\documentclass[a4paper]{article}

\usepackage[pages=all, color=black, position={current page.south}, placement=bottom, scale=1, opacity=1, vshift=5mm]{background}
\SetBgContents{
	\textit{Preprint submitted to Mathematical Geosciences}
}      

\usepackage[margin=1in]{geometry} 

\usepackage{amsmath}
\usepackage{amsthm}
\usepackage{amssymb}
\usepackage{booktabs}

\usepackage{graphicx} 
\usepackage{float}
\usepackage{algorithm}  
\usepackage{algpseudocode}
\usepackage{color}
\usepackage{setspace}

\usepackage{hyperref}
\usepackage{amsmath}
\usepackage{algorithm}
\usepackage{algpseudocode}
\usepackage{emptypage} 
\usepackage{multicol} 
\usepackage{graphicx}
\usepackage{caption} 
\usepackage{float}
\usepackage{nomencl}
\usepackage{amsmath}
\usepackage{amsthm}
\usepackage{amssymb}
\usepackage{amsfonts}
\usepackage{bm}
\usepackage{tabularx}
\usepackage{longtable}
\usepackage{algorithm}
\usepackage{algpseudocode}
\usepackage{subcaption}
\usepackage{multirow} 

\usepackage[utf8]{inputenc}
\usepackage{hyperref}
\hypersetup{
	unicode,
	pdfauthor={Guido Di Federico, Louis J. Durlofsky},
	pdftitle={3D latent diffusion models for parameterizing and history matching facies systems with hierarchical uncertainty},
	pdfsubject={3D latent diffusion models for parameterizing and history matching facies systems with hierarchical uncertainty},
	pdfkeywords={Geological parameterization, diffusion models, latent diffusion, data assimilation, history matching, deep learning},
	pdfproducer={LaTeX},
	pdfcreator={pdflatex}
}

\usepackage[authoryear]{natbib}

\theoremstyle{plain}

\theoremstyle{definition}

\usepackage{graphicx, color}
\graphicspath{{fig/}}

\usepackage{algorithm, algpseudocode} 
\usepackage{mathrsfs} 

\usepackage{lipsum}

\title{3D latent diffusion models for parameterizing and
history matching facies systems under hierarchical
uncertainty}
\author{Guido Di Federico$^1$ \and Louis J. Durlofsky$^1$}

\date{
	\textit{\small$^1$Department of Energy Science \& Engineering, Stanford University, Stanford, CA, 94305, USA}\\ \small{\texttt{\{gdifede, lou\}@stanford.edu}}\\
}

\begin{document}
	\maketitle
	\hrule
\begin{abstract}
Geological parameterization procedures entail the mapping of a high-dimensional geomodel to a low-dimensional latent variable. These parameterizations can be very useful for history matching because the number of variables to be calibrated is greatly reduced, and the mapping can be constructed such that geological realism is automatically preserved. In this work, a parameterization method based on generative latent diffusion models (LDMs) is developed for 3D channel-levee-mud systems. Geomodels with variable \textcolor{black}{geological metaparameters (also referred to as hyperparameters or scenario parameters)}, specifically mud fraction, channel orientation, and channel width, are considered. A perceptual loss term is included during training to improve geological realism. For any set of scenario parameters, an (essentially) infinite number of realizations can be generated, so our LDM parameterizes over a very wide model space. New realizations constructed using the LDM procedure are shown to closely resemble reference geomodels, both visually and in terms of one- and two-point spatial statistics. Flow response distributions, for a specified set of injection and production wells, are also shown to be in close agreement between the two sets of models. The parameterization method is applied for ensemble-based history matching, with model updates performed in the LDM latent space, for cases involving uncertain scenario parameters. For three synthetic true models, we observe clear uncertainty reduction in both production forecasts and geological scenario parameters. The overall method is additionally shown to provide posterior geomodels consistent with the synthetic true model in each case.
\end{abstract}


\maketitle

\section{Introduction}
\label{intro}

Parameterization of geological models entails the mapping of a complex, high-dimensional geological description to a low-dimensional latent variable that, ideally, follows a multi-Gaussian distribution. A new realization, consistent with the underlying geological concept, can then be generated via a decoding, i.e., by passing the latent variable through the parameterization method. This process is especially useful for systems involving complex geological features (e.g., fluvial channels, turbidites, deltaic fans, etc.), where geological realism may be difficult to maintain during history matching. In such cases, the latent variables can be adjusted until the corresponding geological models result in a sufficiently close match between simulation outputs and field observations. This reduces the number of variables to determine and maintains geological realism, while allowing the history matching algorithm to operate in normally-distributed variable space, which is often beneficial.

In this work, we introduce a 3D deep learning-based latent diffusion model for the parameterization of facies-based geological systems. We consider relatively large channel-levee-mud geomodels, and introduce a parameterization that can treat \textcolor{black}{variable continuous geological metaparameters/scenario parameters.} This represents a significant extension of our previous latent diffusion model where we considered 2D, fixed-metaparameter systems~\citep{DIFEDERICO2025105755}. We demonstrate the usefulness of the parameterization method by performing history matching with an ensemble smoother with multiple data assimilation (ESMDA) procedure~\citep{EMERICK20133} operating on latent space variables.

In recent years, a number of deep learning (DL) geomodel parameterizations have been introduced. These have been shown to perform better than more mathematically tractable methods, such as those based on principal component analysis (e.g.,~\citet{Sarma2008}) or discrete cosine transform~\citep{Jafarpour}, which are not always able to reproduce complex geological patterns~\citep{Chan_2019}. DL methods learn the mapping between geological model space and latent space from a training dataset, and are then able to generate new realizations that exhibit the same spatial features. Architectures for this application include CNN-PCA~\citep{LIU2021104676}, variational autoencoders (VAEs)~\citep{Canchumuni_2019, mo_caae, grana_co2}, and generative adversarial networks (GANs)~\citep{Laloy_2018, chan2019parametrization, Canchumuni_2020, HU2023105290, Feng2024, Merzoug2025}. Existing DL parameterizations have proved successful both for standalone geomodeling and history matching for various subsurface flow applications. These include, for example, oil and gas production~\citep{Tang_2021}, groundwater flow~\citep{Song_Mukerji_Hou_2021}, and geological carbon storage~\citep{BAO2024213294}. These methods do, however, have some limitations. Namely, VAEs may provide overly smoothed or discontinuous geological patterns~\citep{Canchumuni_2019, chan2019parametrization, abbate}, while GANs often require computationally intensive (potentially unstable) adversarial training in conjunction with careful hyperparameter tuning. Improved treatments have been presented, including VAE-GAN hybrids~\citep{ZHAO2023105239, FENG2024104833}, progressive training~\citep{Song_Mukerji_Hou_2021}, StyleGAN~\citep{stylegan}, and the use of the Wasserstein loss~\citep{Chan_2019}. Although these developments are promising, the limitations of existing methods have motivated the application of the more recent diffusion models (DMs), which currently represent the state-of-the-art in generative modeling.

DMs, first introduced by~\citet{ho2020denoising}, are a class of generative models trained to learn a denoising process. During denoising, a neural network is used to predict the noise contained in a sample, which is then iteratively subtracted for a number of steps. In this way, the DM transforms an initial noise input into a clean, ``denoised'' image. To accomplish this, the DM learns the mapping between a Gaussian noise realization (latent space) and the corresponding geological model (model space). DMs have been applied for geomodeling purposes in a few studies, and they have been shown to provide high-quality unconditional and conditional realizations for complex geological patterns. These include fluvial channel~\citep{mosser, Xu2024DiffSimDD, DIFEDERICO2025105755}, Gaussian~\citep{alnasser}, and discrete fracture systems~\citep{fractures_ldm}.

Among DMs, latent diffusion models (LDMs, introduced by~\citet{rombach2022highresolution}), appear to be a particularly promising method for geomodel generation and history matching in subsurface flow applications. These methods achieve computational efficiency and scalability by performing the diffusion process in a lower-dimensional latent space rather than in the pixel or voxel space.~\citet{lee2023latent} and~\citet{Ovanger2025} used LDMs to generate realistic conditional realizations of 2D multifacies marine depositional environments. These studies highlighted the superior performance of diffusion-based generation relative to GANs. In~\citet{fractures_ldm}, an LDM was used to parameterize 2D discrete fracture networks (DFNs). LDM generated models exhibited visual improvement over those generated using VAEs, and they preserved property distributions in agreement with reference models. LDMs have also been used for direct generation or post-processing of simulation forecasts. For example,~\citet{zhan} applied an LDM for 2D binary facies models with Gaussian permeability distributions within the two facies. In that study an LDM was trained to generate facies structures conditioned on observed pressure and concentration fields, and another LDM with the same architecture was trained to perform the opposite task.~\citet{exxon} developed an LDM model to post-process upscaled simulation results to provide fine-scale pressure and saturation maps for a 2D waterflooding problem.

To date, many subsurface modeling applications involving DL parameterizations -- and in particular most DM/LDM implementations -- have been for 2D or relatively small 3D domains, containing fewer than $\sim$100,000~cells. Given their powerful generative capabilities and efficient representation of high-dimensional spaces, the extension and application of these methods to larger 3D models is of great interest. In addition, with the exception of the work in \cite{Song2021_multiscenario} for 2D geomodeling with GANs, \textcolor{black}{the global features characterizing the geological scenario are typically fixed} in studies involving flow and history matching. This could represent a key limitation because \textcolor{black}{the scenario parameters themselves are} uncertain in practical applications, and this should be considered during history matching. 

In this work, we address both of these important issues by (1) developing \textcolor{black}{a 3D-LDM applicable over a range of continuous scenario parameters} and (2) applying this treatment to  history matching for cases where \textcolor{black}{these parameters are uncertain.}
The 3D models considered in this work, of grid dimensions 128 $\times$ 128 $\times$ 32 (total of 524,288~cells), are larger than many of the models treated in previous studies. In addition, the three-facies geomodels entail uncertain scenario parameters, namely mud fraction, channel width, and channel orientation.

New samples generated by the 3D-LDM method are shown to exhibit high visual quality in terms of geological features, along with spatial statistics that are consistent with (reference) training samples from Petrel geomodeling software~\citep{Schlumberger}. Significant computational savings are achieved in the geomodeling process. Flow statistics for prior models confirm that the parameterization provides median results, and the correct amount of variability, relative to reference results over a range of metaparameters. History matching is then performed for waterflooding problems using the 3D-LDM representation in combination with an ESMDA procedure~\citep{EMERICK20133}. Results for cases involving synthetic true models corresponding to different scenario parameters are presented.

This paper proceeds as follows. In Section~\ref{case_study}, we describe the channel-levee-mud geomodels considered in this work. In Section~\ref{method}, after reviewing the basics of DMs/LDMs and their use for geomodeling, we describe the 3D-LDM parameterization method and the new treatments introduced in this work. Next, in Section~\ref{results_models}, we display LDM geomodel realizations and compare spatial and flow statistics for an ensemble of LDM generated models to those for reference Petrel models. Latent-space history matching results, for different synthetic true models, are provided in Section~\ref{results_hm}. Finally, in Section~\ref{conclusion}, we summarize this work and suggest directions for future research. Additional algorithmic details are provided in the appendices.

\section{\textcolor{black}{3D fluvial channel systems with variable scenario parameters}}
\label{case_study}

\textcolor{black}{In this work, we define a geological scenario as a realization of a set of continuous geological scenario parameters/metaparameters that characterize the depositional environment.} This approach allows for variation within a general geological concept. For example, in the systems considered here, the metaparameters quantify the geometries of channel bodies in fluvial channel systems. It is important to emphasize that, with the scenario parameters considered here, we are not attempting to treat widely different depositional environments corresponding to distinct geological concepts. For example, these parameters are not meant to characterize settings ranging from fluvial channel systems to eolian sand-dune systems to fractured carbonate systems. Nonetheless, even within a particular geological concept, an infinite number of realizations can be constructed for any set of scenario parameters. In this sense the geomodel uncertainty is hierarchical, as we have uncertain higher-level scenario parameters and, for each set of parameters,
uncertain geobody realizations. \textcolor{black}{We will refer to this as hierarchical uncertainty rather than multiscenario uncertainty, since the latter term could also refer to cases with widely different depositional environments (which are not treated here).}

\textcolor{black}{We can express a particular geomodel $\mathbf{m}$ as $\mathbf{m}(\mathbf{s},\mathbf{r})$, where $\mathbf{s}$ denotes the set of scenario parameters and $\mathbf{r}$ indicates the particular realization. In our case, $\mathbf{r}$ corresponds to the cell-by-cell facies indicator.} The fluvial channel systems considered here contain channel (high permeability), levee (intermediate permeability), and mud/shale (low permeability) facies. The levee facies surrounds the extensive channel bodies, with both deposited within a mud background. Mud, levee, and channel facies are identified in the geomodels by cell values of 0, 0.5, and 1, respectively. \textcolor{black}{The geological scenario parameters $\mathbf{s}$ include mud fraction (denoted $f_m$), channel orientation ($\theta_{\rm ch}$), and channel width ($w_{\rm ch}$). Thus, $\mathbf{s}=[f_m, \theta_{\rm ch}, w_{\rm ch}]$.} The geomodel realizations used for training, and as the reference against which the LDM models will be compared, are constructed using the Petrel geomodeling software~\citep{Schlumberger}. Object-based modeling (OBM) is applied.

The geomodel construction procedure is as follows. A total of 3000 conditional realizations, each containing 128 $\times$ 128 $\times$ 32 cells ($N_x \times N_y \times N_z = N_c = 524,288$), comprise the training dataset. The geological scenario parameters are characterized by uniform distributions, with minimum and maximum values reported in Table~\ref{table:orientation_width_number}. Other features, such as channel depth, wavelength, and amplitude, are the same for all realizations. To construct a realization, we first sample each scenario parameter from its uniform distribution, and then generate a Petrel OBM using these metaparameters, \textcolor{black}{i.e., $\mathbf{r} = \mathbf{r}(\mathbf{s})$.} This acts to provide a diverse set of facies models. For each realization, the specified metaparameters are allowed to vary slightly within Petrel, with a drift of $\pm 10$\% from the specified input value. This means, for example, that for an input channel orientation of 30$^\circ$, the channel bodies within that realization can actually vary from 27$^{^\circ}$ to 33$^{^\circ}$.
\begin{table}[h]
\centering
\caption{Ranges for scenario parameters used as inputs for OBM in Petrel.}
\label{table:orientation_width_number}
\begin{tabular}{@{}lcc@{}}
\toprule
\textbf{Scenario parameter} & \textbf{Min.} & \textbf{Max.} \\
\midrule
Mud fraction, $f_m$ & 0.72 & 0.87 \\
Channel orientation (avg.~angle), $\theta_{\rm ch}$ & 30$^{\circ}$ & 60$^{\circ}$  \\
Channel width (avg.~\# of blocks), $w_{\rm ch}$ & 4 & 6 \\
\bottomrule
\end{tabular}
\end{table}

\begin{figure}[h]
    \centering
    \begin{subfigure}[b]{0.3\textwidth}
        \centering
        \includegraphics[width=\textwidth, trim=50 50 50 50, clip]{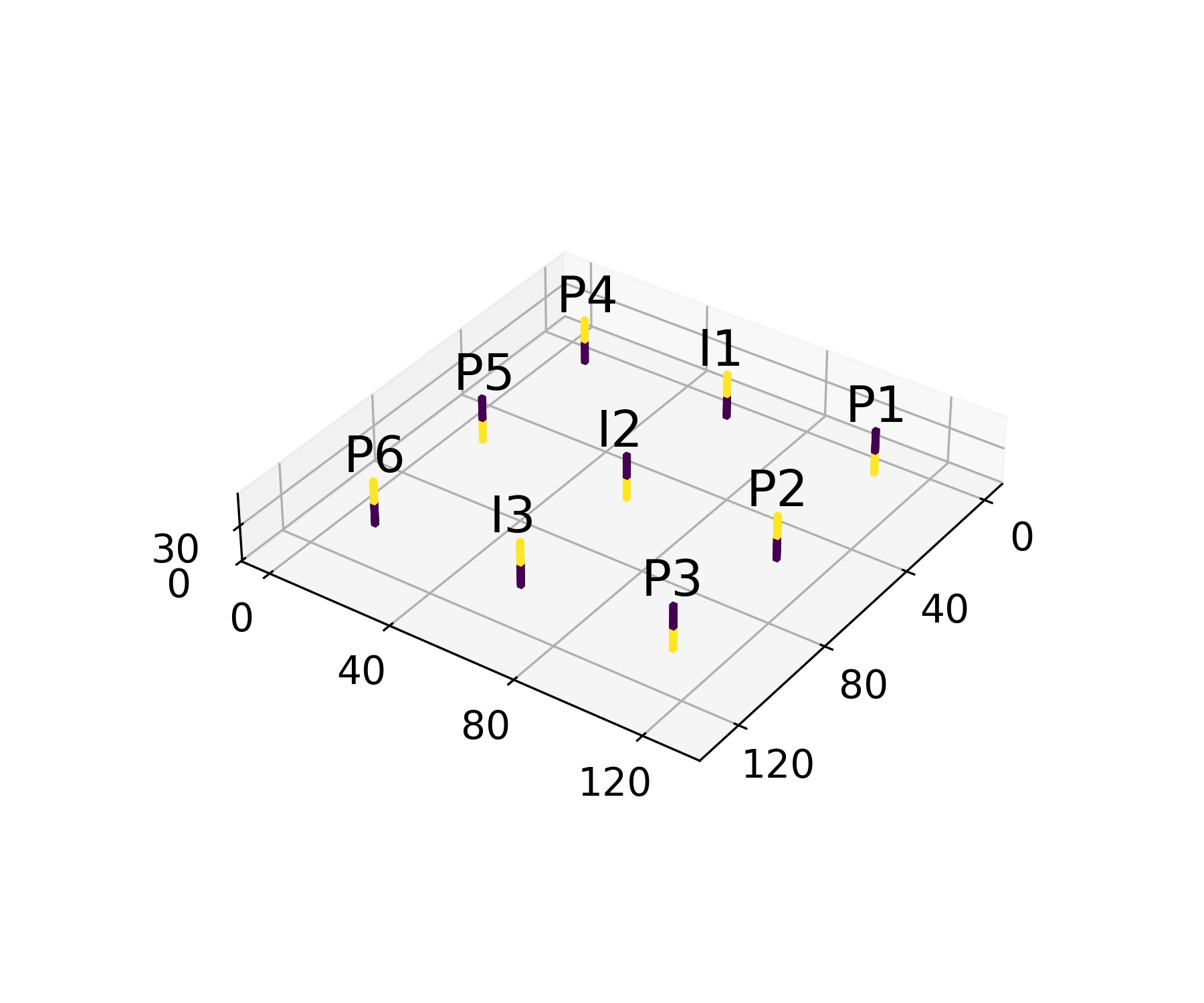}
        \caption{\centering \\ \centering Conditioning locations \\
        P = production well \\
        I = injection well}
    \end{subfigure}
    \hfill
    \begin{subfigure}[b]{0.3\textwidth}
        \centering
        \includegraphics[width=\textwidth, trim=50 50 50 50, clip]{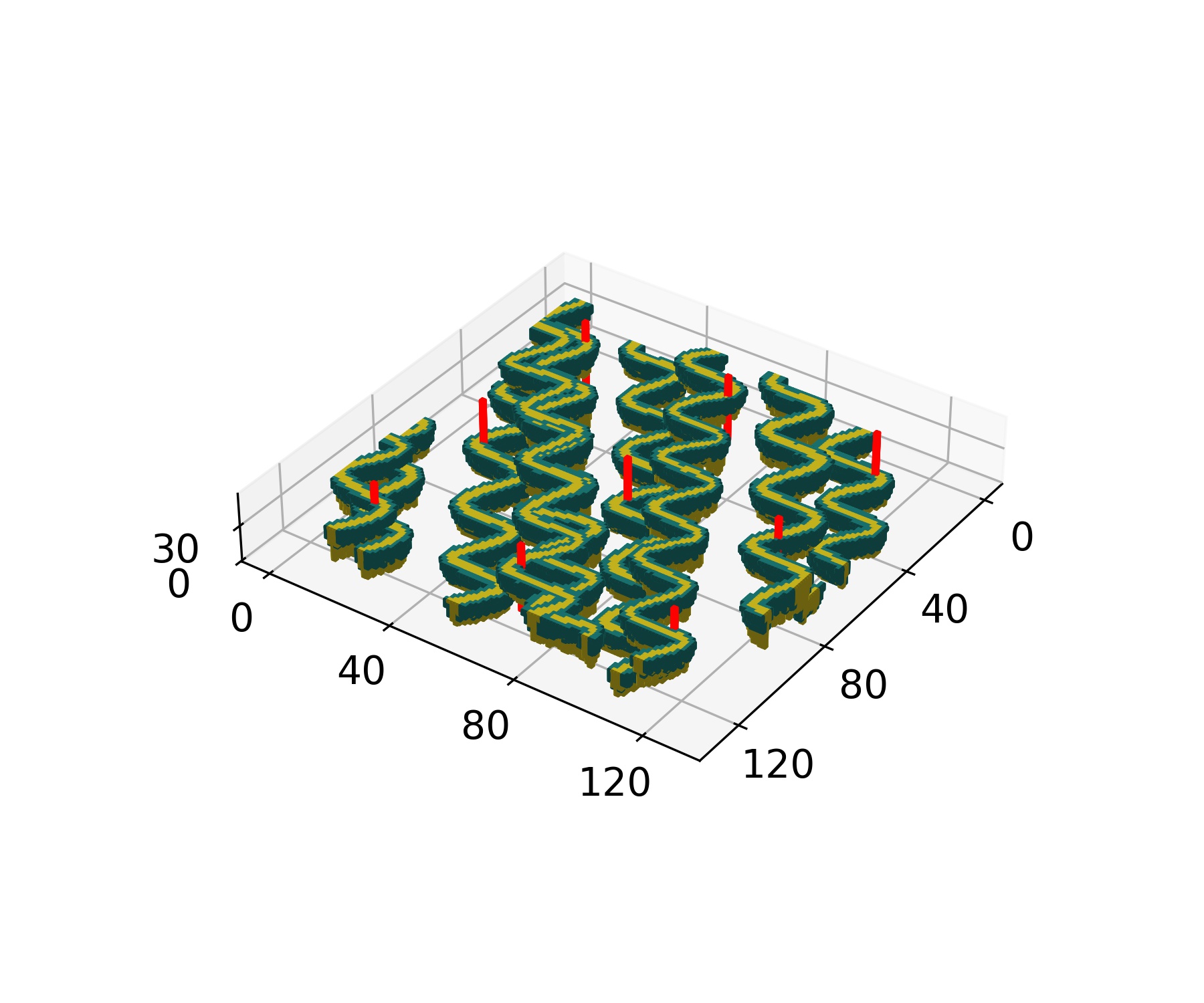}
        \caption{\centering \\ \centering Mud fraction: 0.84\\Channel orientation: 55$^\circ$\\Channel width: 4.2}
    \end{subfigure}
    \hfill
    \begin{subfigure}[b]{0.3\textwidth}
        \centering
        \includegraphics[width=\textwidth, trim=50 50 50 50, clip]{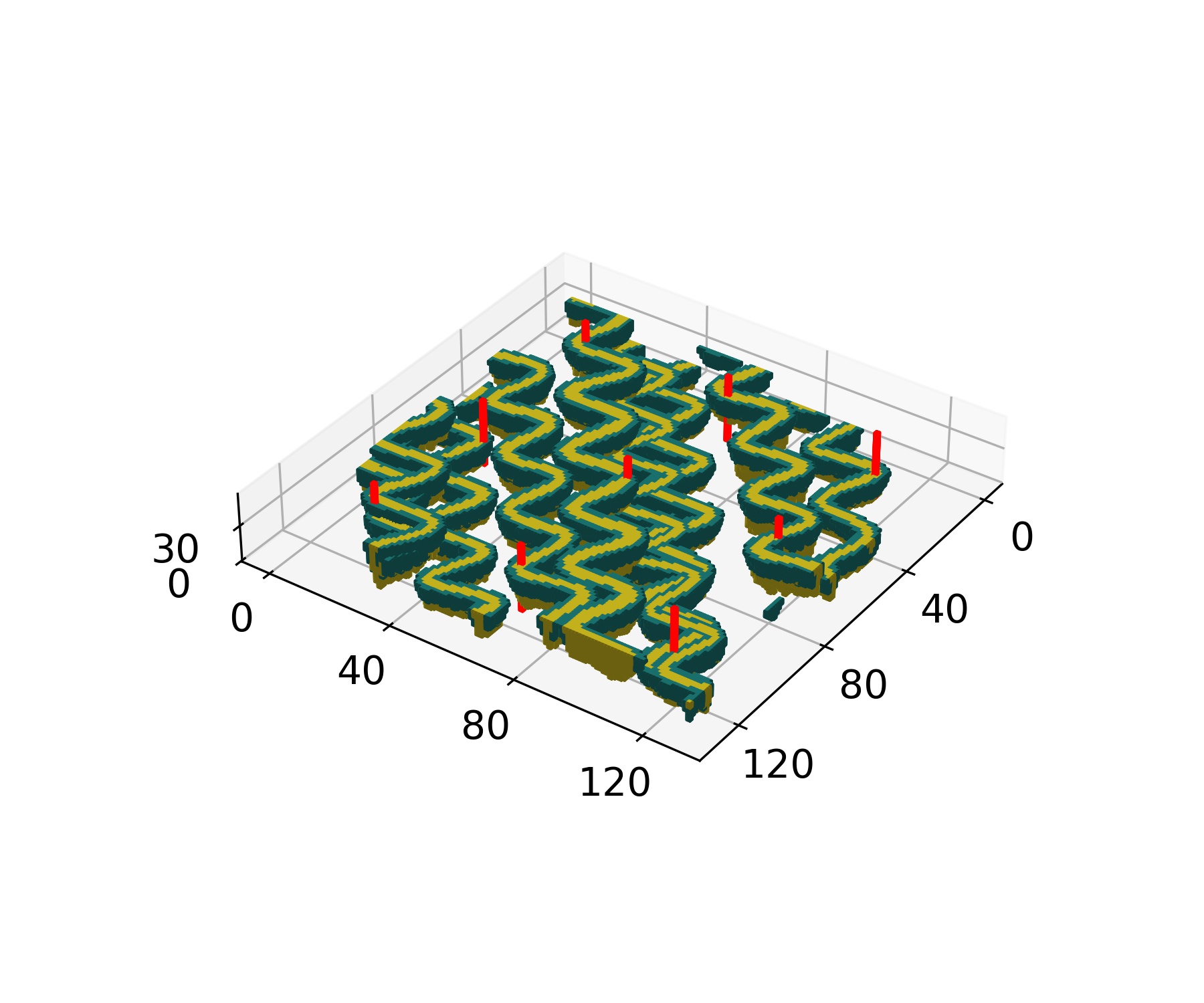}
        \caption{\centering \\ \centering Mud fraction: 0.75\\Channel orientation: 42$^\circ$\\Channel width: 5.1}
    \end{subfigure}
\vfill
    \begin{subfigure}[b]{0.3\textwidth}
        \centering
        \includegraphics[width=\textwidth, trim=50 50 50 50, clip]{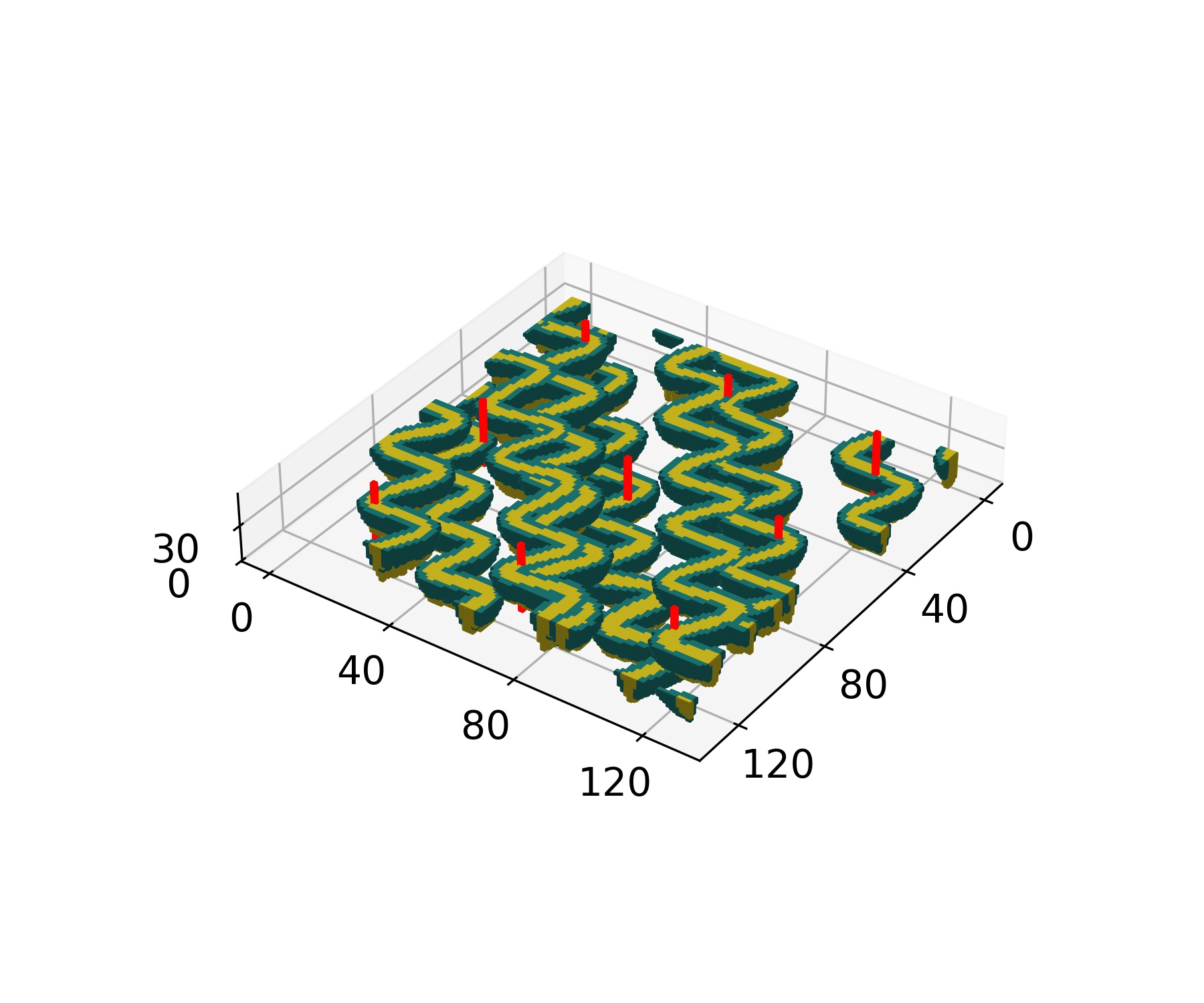}
        \caption{\centering \\ \centering Mud fraction: 0.77\\Channel orientation: 49$^\circ$\\Channel width: 5.9}
    \end{subfigure}
    \hfill
    \begin{subfigure}[b]{0.3\textwidth}
        \centering
        \includegraphics[width=\textwidth, trim=50 50 50 50, clip]{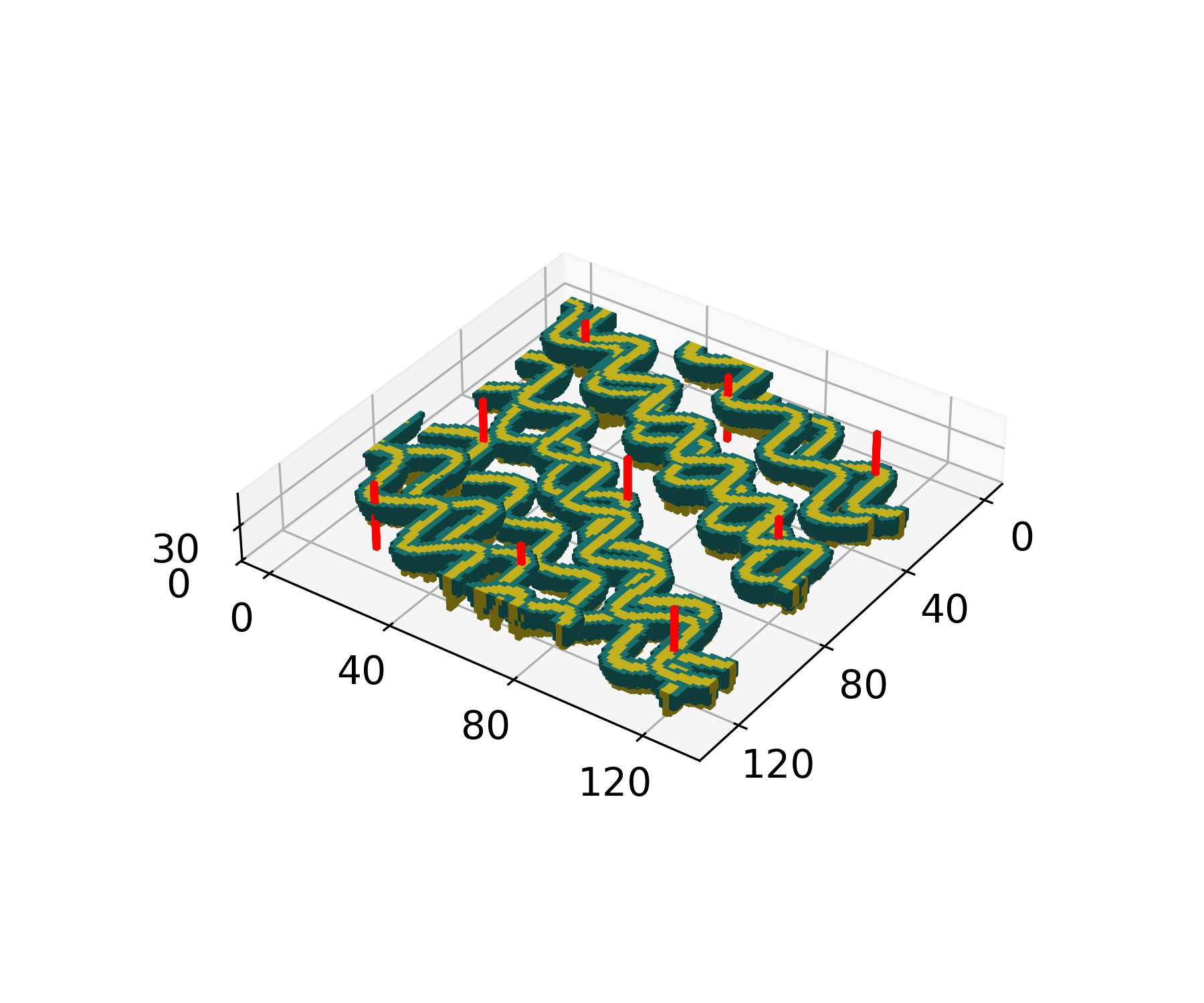}
        \caption{\centering \\ \centering Mud fraction: 0.81\\Channel orientation: 33$^\circ$\\Channel width: 4.4}
    \end{subfigure}
    \hfill
    \begin{subfigure}[b]{0.3\textwidth}
        \centering
        \includegraphics[width=\textwidth, trim=50 50 50 50, clip]{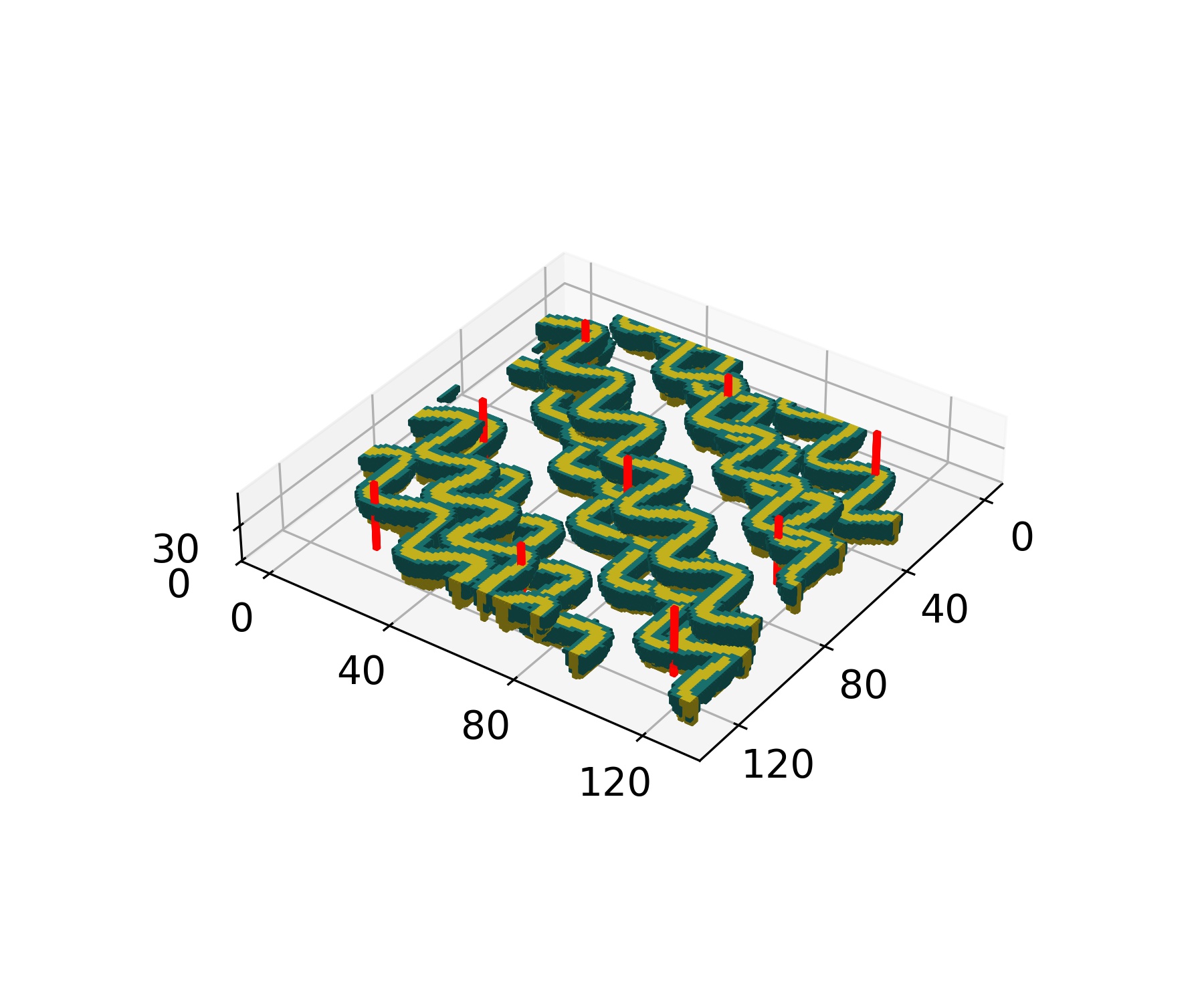}
        \caption{\centering \\ \centering Mud fraction: 0.79\\Channel orientation: 38$^\circ$\\Channel width: 5.5}
    \end{subfigure}

    \caption{(a) Conditioning locations for all geomodels and well labels for (later) flow simulations. (b–f) Representative realizations corresponding to different scenario parameters from the Petrel-generated training set (scenario parameters are given for each geomodel, channel width in units of grid cells). }
    \label{fig:examples_petrel}
\end{figure}

There are $N_w = 9$ wells in the model, as shown in Figure~\ref{fig:examples_petrel}a, that provide conditioning data. At these locations, the models are conditioned to either channel or mud facies along the full vertical thickness. This results in a total of $N_h = N_w \times N_z = 288$ hard data locations. These are depicted in Figure~\ref{fig:examples_petrel}a, where conditioning to channel facies is shown in yellow and conditioning to mud facies in purple.  The conditioning locations are labeled with injection (I) or production (P) well names. These will be used later when we present flow simulation results. 

Sample Petrel realizations are presented in Figure~\ref{fig:examples_petrel}b-f, with the conditioning locations indicated in red. The scenario parameters, which vary over the set, are reported for each model. The orientation angle is relative to the $x$-axis, which is the edge parallel to the line connecting wells P4, I1 and P1. Clear variation in channel orientation and channel density (mud fraction) over the five realizations is evident in Figure~\ref{fig:examples_petrel}. We note that, in 3D visualizations of channel systems, the mud facies is often omitted to better show the spatial configuration of the channel bodies. This type of representation is used in Figure~\ref{fig:examples_petrel}. We will show the mud facies in some later figures in order to better illustrate cross sectional views of the channels, as well as to depict models that have been averaged over multiple realizations.

\section{Latent diffusion models for representing geological systems}
\label{method}

In this section, we first present an overview of the general DM~\citep{ho2020denoising} and LDM~\citep{rombach2022highresolution} procedures. Our 3D-LDM method for 
\textcolor{black}{geological parameterization with variable metaparameters} is then described. From a geomodeling perspective, scenario parameters act as a form of conditioning to global features, as opposed to conditioning to particular values at specific spatial locations, i.e., hard data. Strategies to implement scenario conditioning in a deep learning parameterization context include the use of additional input channels or separate parameter-encoding pipelines. These approaches enable direct control on the generation process, and have proved effective for geomodeling purposes~\citep{Song_Mukerji_Hou_2021}. In this study, however, we adopt a simpler approach, where the method is trained to parameterize for a range of scenario parameters without additional constraints or input features. 

\subsection{Diffusion model (DM)}

Diffusion models are based on the forward process of noising and the (learned) inverse process of denoising. Both processes are performed over a number of discrete steps (denoted by $T$). Noising entails the addition of random Gaussian perturbations to a sample using a function referred to as a scheduler. Let $\mathbf{m}_0 \in \mathbb{R}^{N_x\times N_y\times N_z}$ indicate a sample from a set of geomodels, represented on a discrete grid of $N_x\times N_y\times N_z = N_c$ cells. The $0$ subscript represents the gemodel without any noise. The generic noising step $t$ is defined as
\begin{equation}
\mathbf{m}_t = \sqrt{\bar \alpha_t} \mathbf{m}_0 + \sqrt{1-\bar \alpha_t} \boldsymbol{\epsilon}_t, \text{ with } \boldsymbol{\epsilon}_t = \mathcal{N} (\mathbf{0}, \mathbf{I}_{N_c}),   \label{eqn:noising_step_dm}
\end{equation}
where $\bar{\alpha}_t = \prod_{s=1}^{t} \alpha_s$ and $\mathbf{I}_{N_c}$ is the identity matrix. The scheduler defines the values of the perturbation through the $\alpha_s$ parameter at each step. This is typically a linearly decreasing function from $\alpha_1$ to $\alpha_T$. Step $T$ refers to the last noising step, when the sample is indistinguishable from pure noise, such that $\mathbf{m}_T \approx \mathcal{N} (\mathbf{0}, \mathbf{I}_{N_c})$.
Denoising refers to the opposite process, where noise is progressively subtracted from a perturbed geomodel until a clean (denoised) model is obtained. A neural network with learnable parameters $\theta$ (typically a U-net,~\citep{unet}) is trained to predict the noise $\boldsymbol{\epsilon}_\theta (\mathbf{m}_t,t)$ given a randomly perturbed input $\mathbf{m}_t$ at each step $t$. As such, the training loss function is
\begin{equation}
        L(\theta) = || \boldsymbol{\epsilon}_t - \boldsymbol{\epsilon}_{\theta}(\mathbf{m}_t,t) ||_2 ^2.
        \label{eqn:loss_dm}
\end{equation}
At inference (generation), the DM starts from Gaussian noise $ \mathbf{m}_{T}$ and progressively generates a new realization $ \mathbf{m}_{0}^{\text{DM}}$ by reversing the noising process, i.e.,
\begin{equation}
    \mathbf{m}_{t-1}=\sqrt{\bar{\alpha}_{t-1}}\left(\frac{\mathbf{m}_t-\sqrt{1-\bar{\alpha}_t} \boldsymbol{\epsilon}_\theta\left(\mathbf{m}_t,t\right)}{\sqrt{\bar{\alpha}_t}}\right)+\sqrt{1-\bar{\alpha}_{t-1}} \boldsymbol{\epsilon}_\theta (\mathbf{m}_t,t), \text{ for } t = T,...,1.
    \label{eqn:denoising_step_dm}
\end{equation}
The specific DM method described above, and Eq.~\ref{eqn:denoising_step_dm} in particular, correspond to the denoising diffusion implicit model, DDIM~\citep{song2021denoising}. This model differs from the original denoising diffusion probabilistic model, DDPM~\citep{ho2020denoising}. As discussed in~\citet{DIFEDERICO2025105755}, the DDIM variant is advantageous for our geomodeling application because it enables deterministic sampling in fewer steps than the DDPM. Only the DDIM will be considered throughout the rest of this paper.

Standard DMs operate in a multi-Gaussian latent space of the same dimension as the geomodel space, i.e., there is no dimension reduction. This makes the extension to large geomodels computationally expensive and provides no reduction in the number of variables to be determined during history matching. In LDMs, by contrast, noising and denoising are accomplished by a U-net in a reduced-dimension latent space. A variational autoencoder (VAE) performs the mapping between the latent space and the full geomodel space. We will now discuss the LDM procedure for 3D geomodel parameterization. Our modifications to the 2D-LDM method presented in~\citet{DIFEDERICO2025105755} include the use of 3D neural network layers, as well as a new perceptual loss term during VAE training.

\subsection{3D latent diffusion model (3D-LDM)}

The 3D-LDM utilizes a latent variable $\boldsymbol{\xi} \in \mathbb{R}^{n_x \times n_y \times n_z}$, where $n_x = N_x / f$,  $n_y = N_y / f$, $n_z = N_z / f$, with $f$ being the downsampling ratio~\citep{rombach2022highresolution}. The overall dimension reduction ratio from the geomodel to latent space is $f^3$. The latent space, although lower-dimensional, is still 3D.  The VAE comprises an encoder, such that $\mathcal{E}(\mathbf{m}_0): \mathbb{R}^{N_x \times N_y \times N_z} \rightarrow \mathbb{R}^{n_x \times n_y \times n_z}$, and a decoder, $\mathcal{D}(\boldsymbol{\xi}_0): \mathbb{R}^{n_x \times n_y \times n_z} \rightarrow \mathbb{R}^{N_x \times N_y \times N_z}$. Note the use of the subscript $0$, referring to the encoder operating only on a clean geomodel, $\mathbf{m}_0$, and the decoder only on a fully denoised latent variable, $\boldsymbol{\xi}_0$. The encoder outputs the parameters of a Gaussian distribution, $(\mathcal{E}_\mu, \mathcal{E}_\sigma)$, and the latent variable is sampled via the reparameterization $\boldsymbol{\xi}_0 = \mathcal{E}_\mu + \mathcal{E}_\sigma \cdot \boldsymbol{\epsilon}, \quad \text{where} \quad \boldsymbol{\epsilon} \sim \mathcal{N}(\mathbf{0}, \mathbf{I}_{n_c})$. Overall, we denote $\boldsymbol{\xi}_0 = \mathcal{E}(\mathbf{m}_0)$. 

The VAE, with parameters $\theta$, is trained on the weighted loss function
\begin{equation}
    L_{\text{VAE}}(\theta) = L_{\text{recon}} + \lambda_{\text{KL}} L_{\text{KL}} + \lambda_{\text{h}} L_{\text{h}} + \lambda_{\text{perc}} L_{\text{perc}}.
    \label{eqn:loss_vae}
\end{equation}
Here $ L_{\text{recon}}  = || \mathbf{m}_0 -  \hat{\mathbf{m}}_0 ||_2^2$ is the reconstruction loss computed between input $\mathbf{m}_0$ and reconstructed geomodel $\hat{\mathbf{m}}_0 = \mathcal{D}(\mathcal{E}(\mathbf{m}_0))$. This term ensures the decoded spatial structure is consistent with training samples. The second loss term, $L_{\text{KL}} = D_{\text{KL}}(\mathcal{N}(\mathcal{E}_{\mu},\mathcal{E}_{\sigma}^2) || \mathcal{N}(\mathbf{0}, \mathbf{I}_{n_c}))]$, is the Kullback-Leibler (KL) divergence loss. This term forces the latent variable $ \boldsymbol{\xi}$ to follow a (near) multi-Gaussian distribution with mean $\mathcal{E}_{\mu}$ and variance $\mathcal{E}_{\sigma}^2$. The hard-data loss term, $L_{\text{h}} = \frac{1}{N_{\text{h}}} ||\mathbf{H}( \mathbf{m}_0 - \hat{\mathbf{m}}_0) ||_2^2$, enforces conditioning on decoded outputs. Here $\mathbf{H}$ is a matrix of zeros and ones that extracts the $N_{\text{h}}$ hard data locations. Finally, $L_{\text{perc}}$ is the perceptual loss term, often used in the computer vision field to improve the quality of generated images~\citep{johnson2016perceptuallossesrealtimestyle} (more details are provided below). The variables $\lambda_{\text{KL}}$, $\lambda_{\text{h}}$, and $\lambda_{\text{perc}}$ denote weights. In both the VAE and U-net components, 3D convolutional, upsampling, downsampling, and pooling layers are used.

The perceptual loss term $L_{\text{perc}}$ leverages a pretrained network that compares the high-level feature representations of deep feature maps between training samples $\mathbf{m}_0$ and decoded (reconstructed) samples $\hat{\mathbf{m}}_0$. The operation is represented as $\phi(\cdot)$. For the 3D geomodels considered here, we generate a series of 2D sections by slicing the model along each of the three orthogonal axes. For a given axis \( d \in \{x, y, z\} \), let \( \mathbf{m}_{0,i}^{(d)} \) and \( \hat{\mathbf{m}}_{0,i}^{(d)} \) denote the corresponding 2D slice $i$ of the input and reconstruction, respectively. To reduce the computational cost, a subset of \( N_d^{\text{sub}} = \lfloor 0.2 N_d \rfloor \) slices is randomly selected for each axis. The perceptual loss along axis \( d \) is then computed as
\begin{equation}
L_{\text{perc}}^{(d)} = \frac{1}{N_d^{\text{sub}}} \sum_{i=1}^{N_d^{\text{sub}}} || \phi(\mathbf{m}_{0,i}^{(d)}) - \phi(\hat{\mathbf{m}}_{0,i}^{(d)})||_2^2.
\label{eqn:loss_perc}
\end{equation}
The total perceptual loss aggregates the contributions from the three axes,
\begin{equation}
L_{\text{perc}} = L_{\text{perc}}^{(x)} + L_{\text{perc}}^{(y)} + L_{\text{perc}}^{(z)}.
\label{eqn:loss_perc_tot}
\end{equation}

The perceptual loss addresses some of the limitations of pixel-wise losses (like the reconstruction loss), thus ensuring that decoded models (and, by extension, generated models) are perceptually similar to training models. Established pretrained networks used in this context include VGG16~\citep{vgg}, SqueezeNet~\citep{squeezenet2016}, AlexNet~\citep{alexnet}, and 
ResNet~\citep{resnet50}. For this study, we use ResNet. 

\begin{figure}[h]
  \centering
  \begin{subfigure}[b]{0.65\textwidth}
    \centering
    \includegraphics[width=\textwidth]{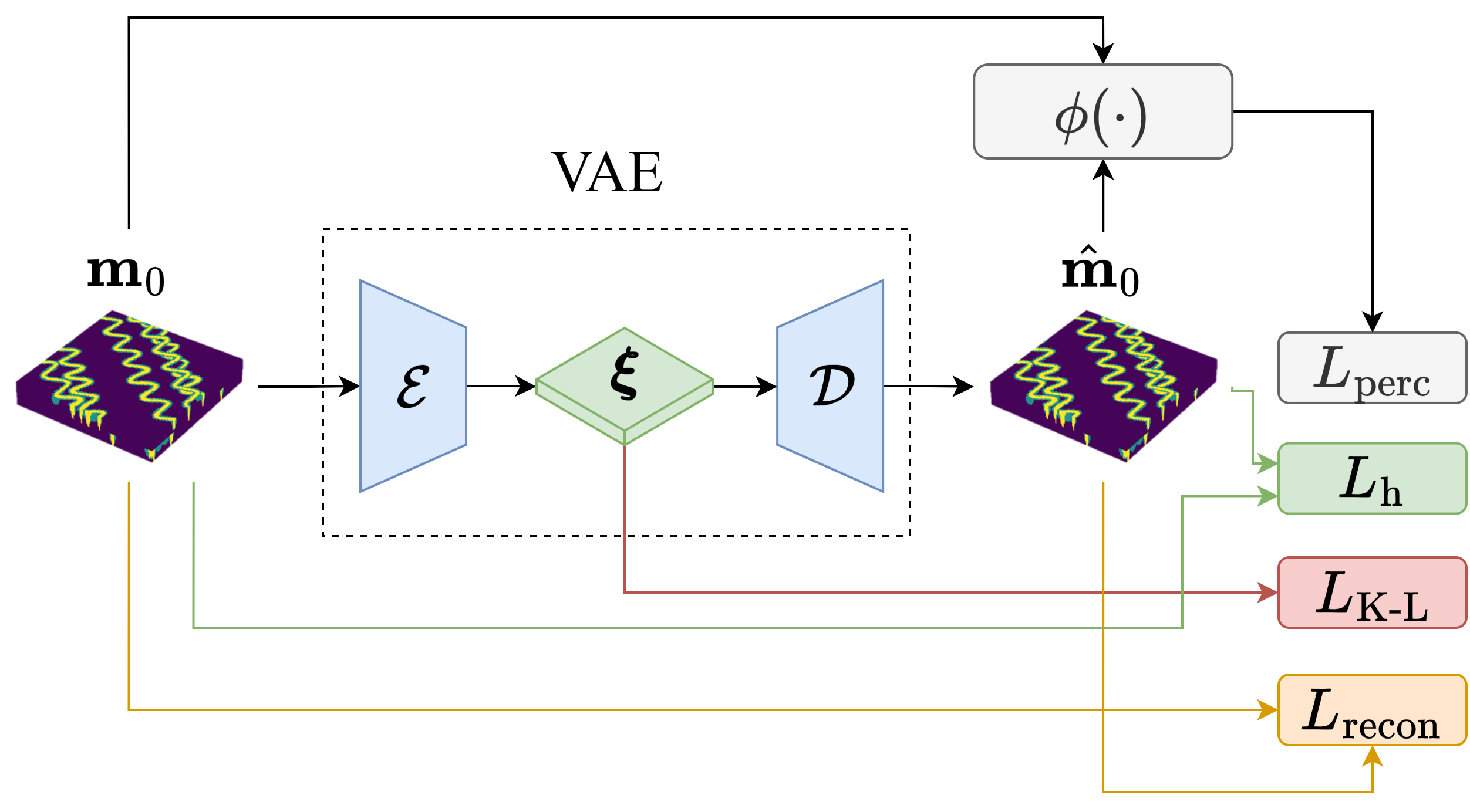}
    \caption{VAE component}
    \label{fig:vae_training}
  \end{subfigure}
  \hfill
  \begin{subfigure}[b]{0.85\textwidth}
    \centering
    \includegraphics[width=\textwidth]{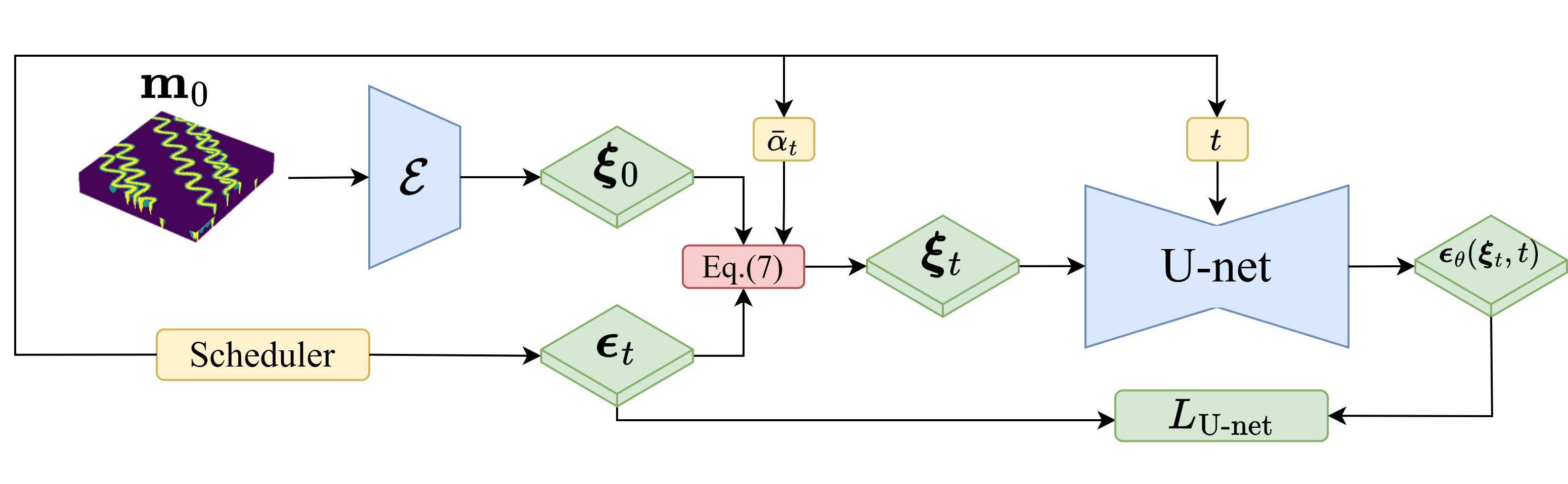}
    \caption{U-net component}
    \label{fig:unet_training}
  \end{subfigure}
  \caption{Schematic of the training procedure for the 3D-LDM method and the computation of loss terms. The VAE component is shown in (a) and the U-net component in (b).}
  \label{fig:ldm_training}
\end{figure}

A schematic of the VAE training procedure and a high-level overview of its architecture are presented in Figure~\ref{fig:ldm_training}a.  The input model $\mathbf{m}_0$ is encoded into the latent variable $\boldsymbol{\xi}$, and decoded back into its reconstruction $\hat{\mathbf{m}}_0$. Reconstruction, perceptual, and hard data loss are computed as a function of $\mathbf{m}_0$ and $\hat{\mathbf{m}}_0$, while the K-L loss involves only $\boldsymbol{\xi}$. \textcolor{black}{Pseudocode is provided in Algorithm~\ref{alg:vae_train} in Appendix~\ref{appendix_a}.} A 80-10-10\% split for the training, validation, and testing sets is applied. Training requires approximately 12~h (2500~epochs) on one NVIDIA A100 GPU. We apply a learning rate of $10^{-4}$ with the Adam optimizer \citep{adam}, and a batch size of 16. Following \citet{rombach2022highresolution}, we set $\lambda_{\text{KL}} = 10^{-6}$. Based on numerical experimentation, we specify $\lambda_{\text{h}}=10^{-2}$ and $\lambda_{\text{perc}}=10^{-3}$.

Once the VAE is trained, its weights are frozen. The U-net, which operates on $\boldsymbol{\xi}$, is then trained.
Since noising and denoising now occur in the latent space, Eqs.~\ref{eqn:noising_step_dm} and~\ref{eqn:denoising_step_dm} are rewritten as
\begin{equation}
    \boldsymbol{\xi}_t = \sqrt{\bar \alpha_t} \boldsymbol{\xi}_0 + \sqrt{1-\bar \alpha_t} \boldsymbol{\epsilon}_t,  \text{ with } \boldsymbol{\epsilon}_t = \mathcal{N} (\mathbf{0}, \mathbf{I}_{n_c}),
    \label{eqn:noising_step_ldm}
\end{equation}

\begin{equation}
    \boldsymbol{\xi}_{t-1}=\sqrt{\bar{\alpha}_{t-1}}\left(\frac{\boldsymbol{\xi}_t-\sqrt{1-\bar{\alpha}_t} \boldsymbol{\epsilon}_\theta\left(\boldsymbol{\xi}_t,t\right)}{\sqrt{\bar{\alpha}_t}}\right)+\sqrt{1-\bar{\alpha}_{t-1}} \boldsymbol{\epsilon}_\theta (\boldsymbol{\xi}_t,t).
    \label{eqn:sampling_ddim_ldm}
\end{equation}
The loss in Eq.~\ref{eqn:loss_dm} now becomes
\begin{equation}
        L_{\text{U-net}}(\theta) = || \boldsymbol{\epsilon}_t - \boldsymbol{\epsilon}_{\theta}(\boldsymbol{\xi}_t,t)||_2 ^2.
        \label{eqn:loss_ldm}
\end{equation}
Figure~\ref{fig:ldm_training}b depicts the U-net training procedure. The encoded geomodel, $\boldsymbol{\xi}_0$, which here represents a ``clean'' latent variable, is perturbed by a scheduler-added noise at a random discrete step $t$, using Eq.~\ref{eqn:noising_step_ldm}, to obtain $\boldsymbol{\xi}_t$. The U-net is then used to predict this noise from the inputs $\boldsymbol{\xi}_t$ and $t$, as $\boldsymbol{\epsilon}_{\theta}(\boldsymbol{\xi}_t,t)$, and finally to compute $L_{\text{U-net}}$. \textcolor{black}{Pseudocode for the U-net training is given in Algorithm~\ref{alg:unet_train} in Appendix~\ref{appendix_a}.}

As noted in \cite{DIFEDERICO2025105755}, LDMs require a compromise between dimension reduction ratio $f$ and generation quality. \textcolor{black}{In this study, we performed numerical experimentation with $f=4$, 8 and 16. We found the use of $f=8$ (overall $f^3=512$) to provide the best balance, i.e., a sufficiently low-dimension latent space and fast geomodel generation, while also achieving minimal information loss from encoding. Thus $f=8$ is used in all results in this paper.} The same dataset split, optimizer, learning rate, and batch size applied for the training of the VAE are used here. For the U-net, training requires approximately 8~h (300~epochs) on the same NVIDIA A100 GPU. We adopt a DDIM sampling and a linear scheduler for $T=20$~steps, with $\alpha_1 = 0.9985$ and $\alpha_T = 0.9805$. 

The implementation developed for this study is based on the libraries \texttt{diffusers} and \texttt{monai} (see the ``Code availability" section for additional details). The detailed architectures used in this work are provided in Table~\ref{tab:vae_architecture} for the VAE and Table~\ref{tab:unet_architecture} for the U-net (in Appendix~\ref{appendix_b}).

\begin{figure}[h]
  \centering
  \begin{subfigure}[b]{0.99\textwidth}
    \centering
    \includegraphics[width=\textwidth]{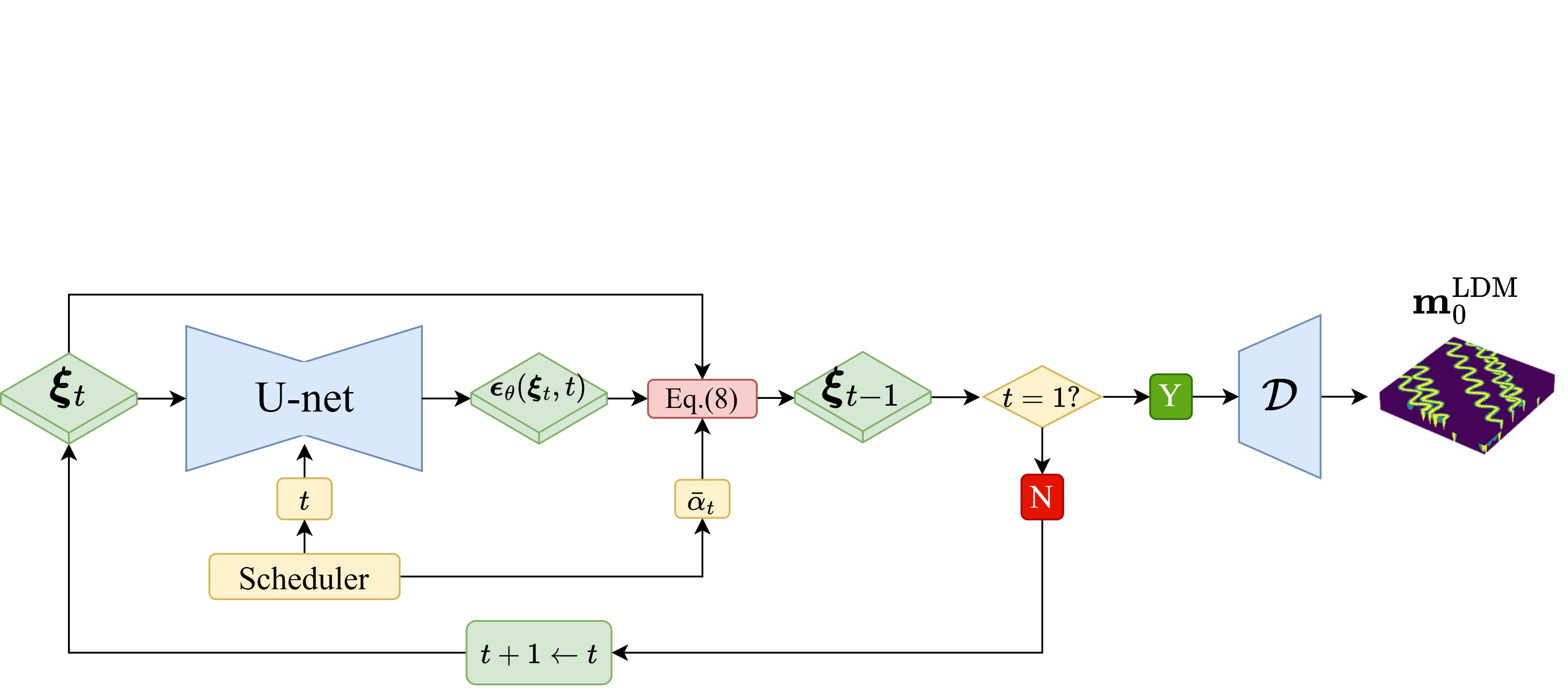}
    \caption{3D-LDM generation process}
    \label{fig:ldm_generation}
  \end{subfigure}
  \hfill
  \begin{subfigure}[b]{0.99\textwidth}
    \centering
    \includegraphics[width=\textwidth, trim=300 0 780 0, clip]{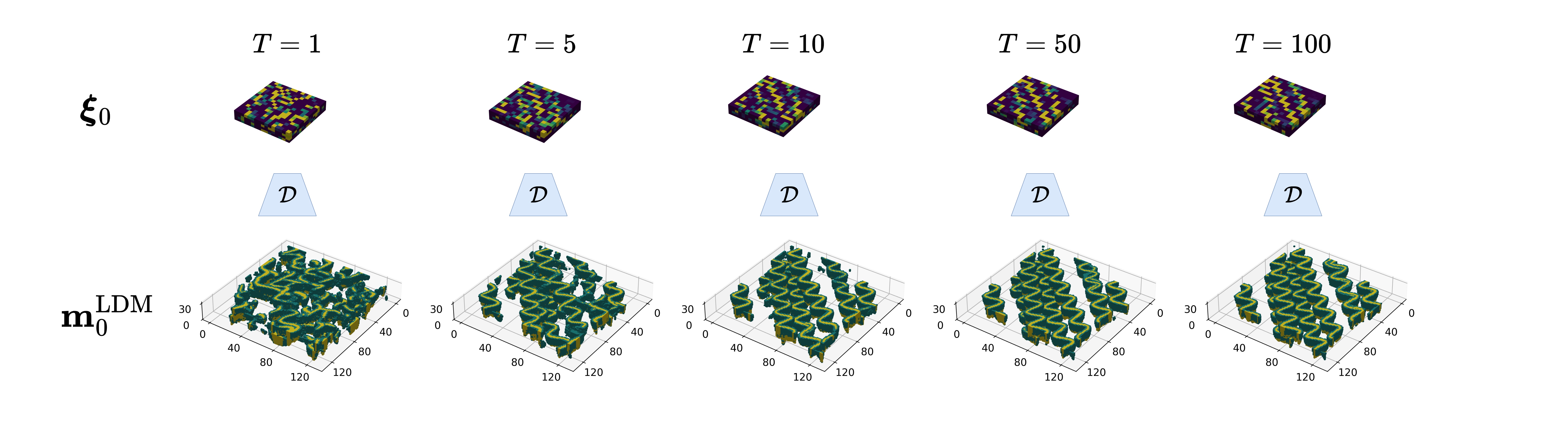}
    \caption{3D denoising visualization}
    \label{fig:denoising_3d}
  \end{subfigure}
  \caption{Illustration of the geomodel generation process with the 3D-LDM method. In (a) the discrete-step denoising performed by the U-net in the latent space is shown, while in (b) the denoised latent variables and corresponding decoded geomodels with different numbers of total denoising steps are depicted.}
  \label{fig:combined_diffusion}
\end{figure}

After both the VAE and the U-net are trained, a new realization ${\mathbf{m}}_0^{\text{LDM}}$ can be generated by iteratively denoising a random sample $\boldsymbol{\xi}_T \sim \mathcal{N}(\mathbf{0}, \mathbf{I}_{n_c})$. A visual representation of this process is given in Figure~\ref{fig:combined_diffusion}a. The generic latent variable $\boldsymbol{\xi}_{t}$ is gradually denoised using the predicted noise from the U-net, $\boldsymbol{\epsilon}_{\theta}(\boldsymbol{\xi}_t,t)$, together with the step $t$ and parameter $\bar{\alpha}_t$ from the scheduler, through Eq.~\ref{eqn:sampling_ddim_ldm}. The denoised latent variable $\boldsymbol{\xi}_{t-1}$ is obtained at each step. When the last step is reached, i.e., $t=1$, the latent variable is decoded into the generated geomodel through application of $\mathbf{m}_0^{\text{LDM}} = \mathcal{D}(\boldsymbol{\xi}_0)$. \textcolor{black}{Pseudocode for the generation (inference) procedure is provided in Algorithm~\ref{alg:ldm_generate} in Appendix~\ref{appendix_a}.} 

Figure~\ref{fig:combined_diffusion}b shows an example of geomodels generated by decoding the fully denoised latent variable $\boldsymbol{\xi}_0$, for different numbers of total DDIM steps ($T$), starting from the same input $\boldsymbol{\xi}_T$. It can be observed that, for $T>10$, the overall quality remains stable, and extending the denoising process only provides a cleaner, more noise-free geomodel. The images here represent
the generated geomodel for different $T$ values. This is for illustration purposes only because, in the actual implementation, the latent variable is decoded only at the final step.

\section{Geomodel generation and flow simulations with 3D-LDM}
\label{results_models}

We now employ the 3D-LDM to generate realizations of the three-facies channelized system. To evaluate the similarity between LDM generated models and reference models, we utilize both qualitative visual assessments and quantitative metrics. The latter include spatial statistics and flow simulation results. Though the main purpose of our LDM procedure is to provide a parameterized representation that can be used for history matching, the approach also leads to significant computational savings in the model generation step itself. Specifically, using the same NVIDIA A100 GPU, only 2~minutes are required to generate 1000, 128 $\times$ 128 $\times$ 32 realizations, compared to several hours using the Petrel OBM tool on an Intel Core i7-7700 CPU.

\subsection{Qualitative assessments}

\begin{figure}[h]
    \centering
    \begin{subfigure}[b]{0.3\textwidth}
        \centering
        \includegraphics[width=\textwidth, trim=50 50 50 50, clip]{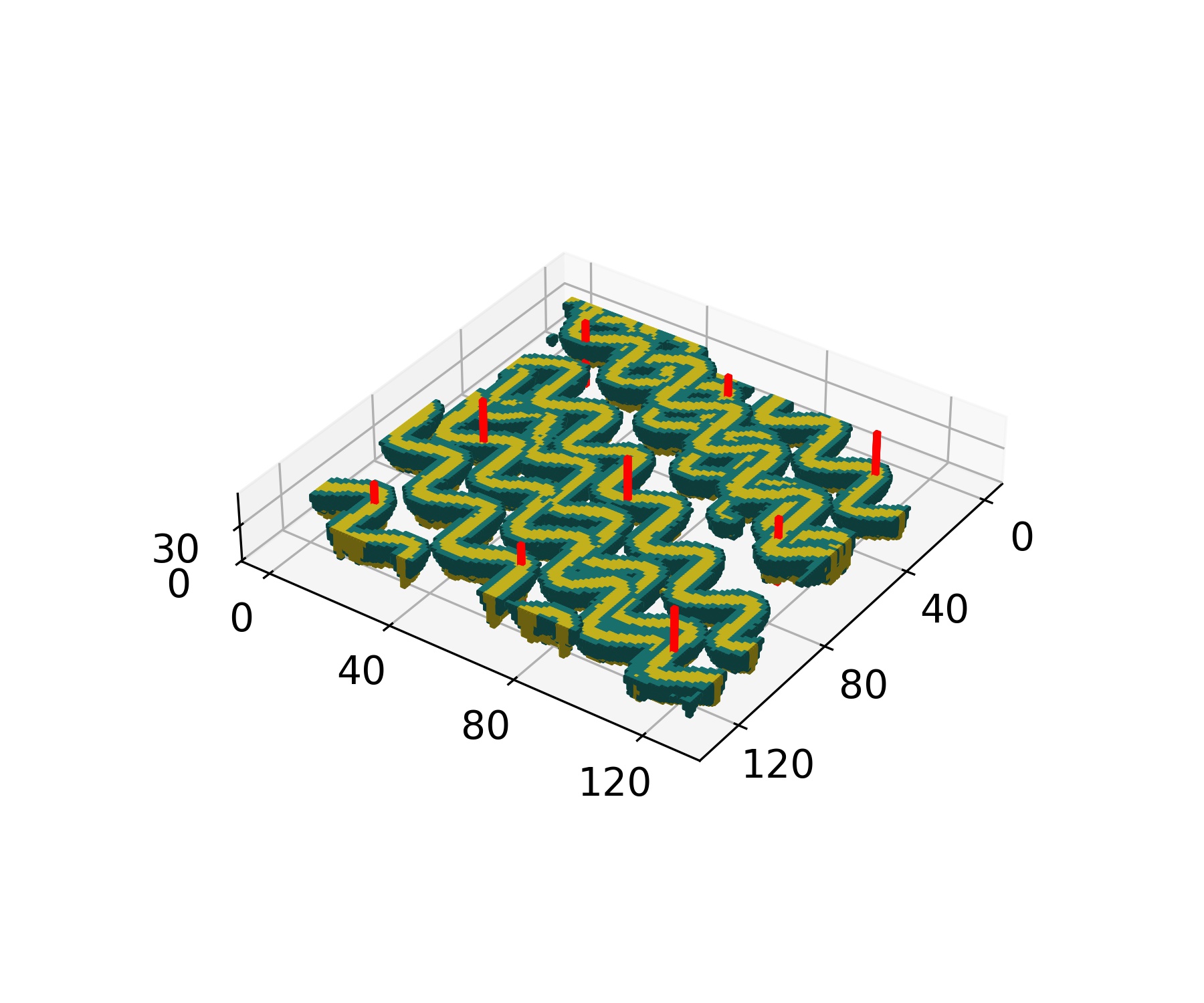}
        \caption{\centering \\ \centering Mud fraction: 0.74\\Channel orientation: 31$^\circ$\\Channel width: 6.0}
    \end{subfigure}
    \hfill
    \begin{subfigure}[b]{0.3\textwidth}
        \centering
        \includegraphics[width=\textwidth, trim=50 50 50 50, clip]{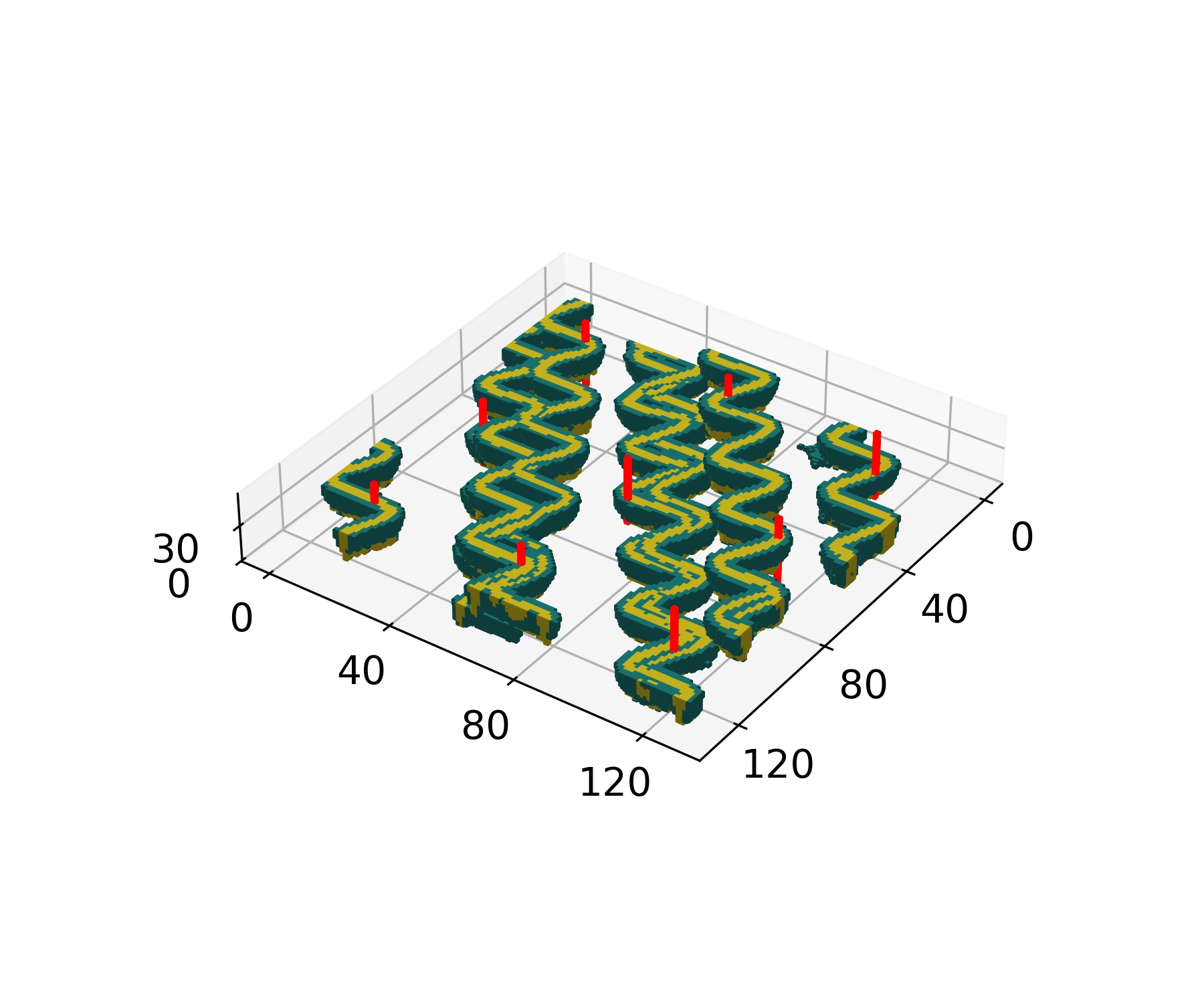}
        \caption{\centering \\ \centering Mud fraction: 0.85\\Channel orientation: 58$^\circ$\\Channel width: 4.4}
    \end{subfigure}
    \hfill
    \begin{subfigure}[b]{0.3\textwidth}
        \centering
        \includegraphics[width=\textwidth, trim=50 50 50 50, clip]{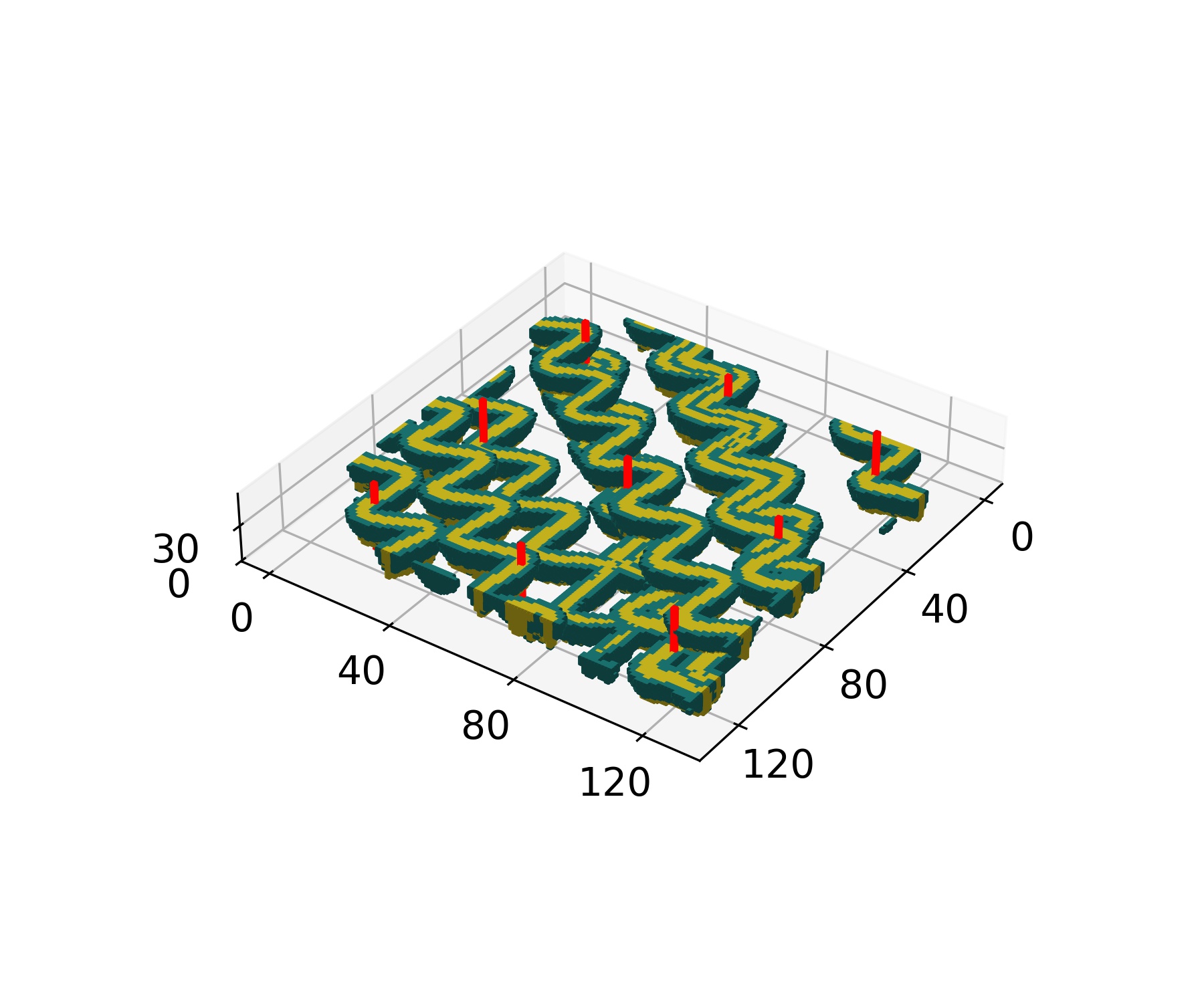}
        \caption{\centering \\ \centering Mud fraction: 0.81\\Channel orientation: 38$^\circ$\\Channel width: 5.4}
    \end{subfigure}

    \begin{subfigure}[b]{0.3\textwidth}
        \centering
        \includegraphics[width=\textwidth, trim=50 50 50 50, clip]{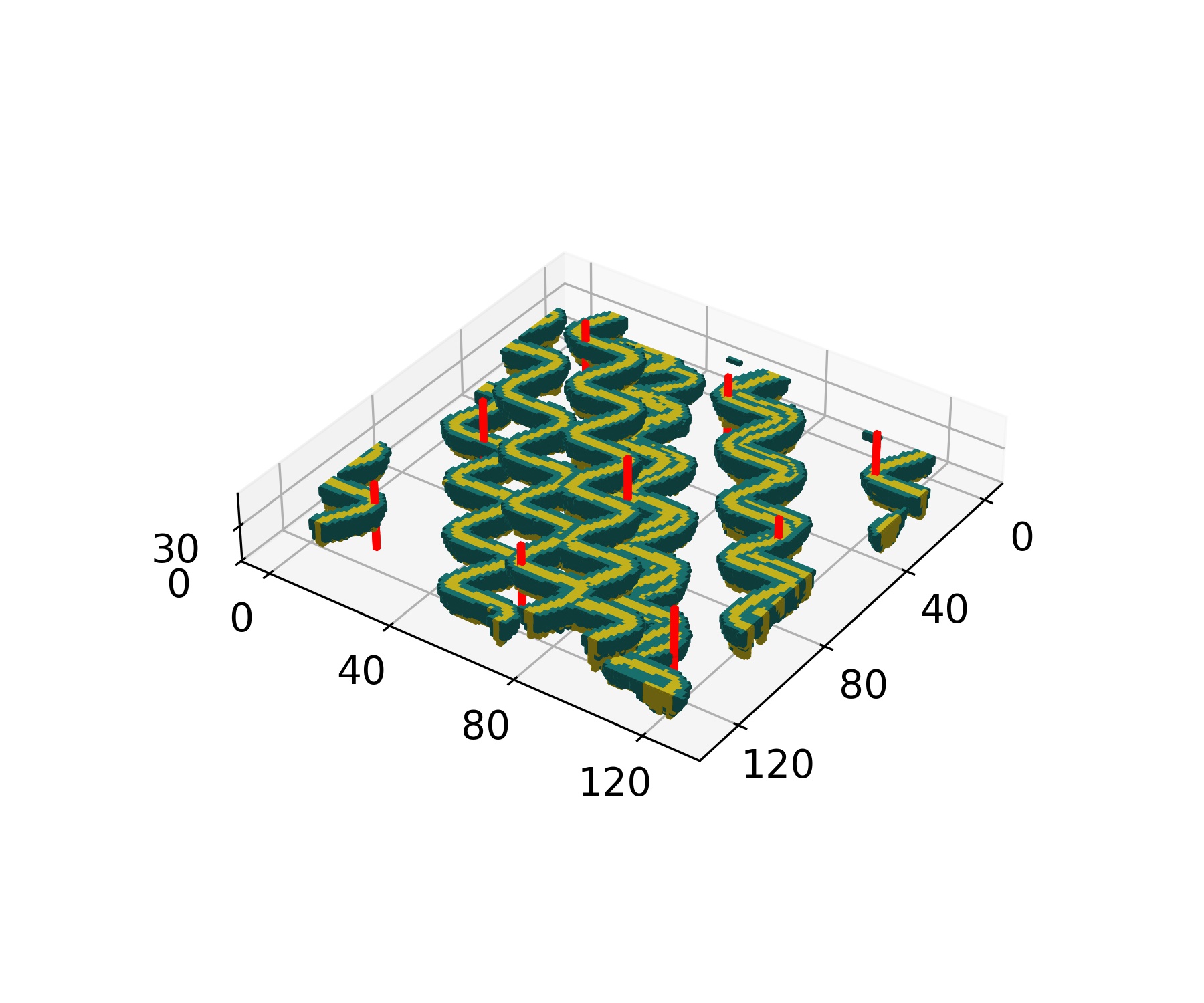}
        \caption{\centering \\ \centering Mud fraction: 0.83\\Channel orientation: 54$^\circ$\\Channel width: 4.6}
    \end{subfigure}
    \hfill
    \begin{subfigure}[b]{0.3\textwidth}
        \centering
        \includegraphics[width=\textwidth, trim=50 50 50 50, clip]{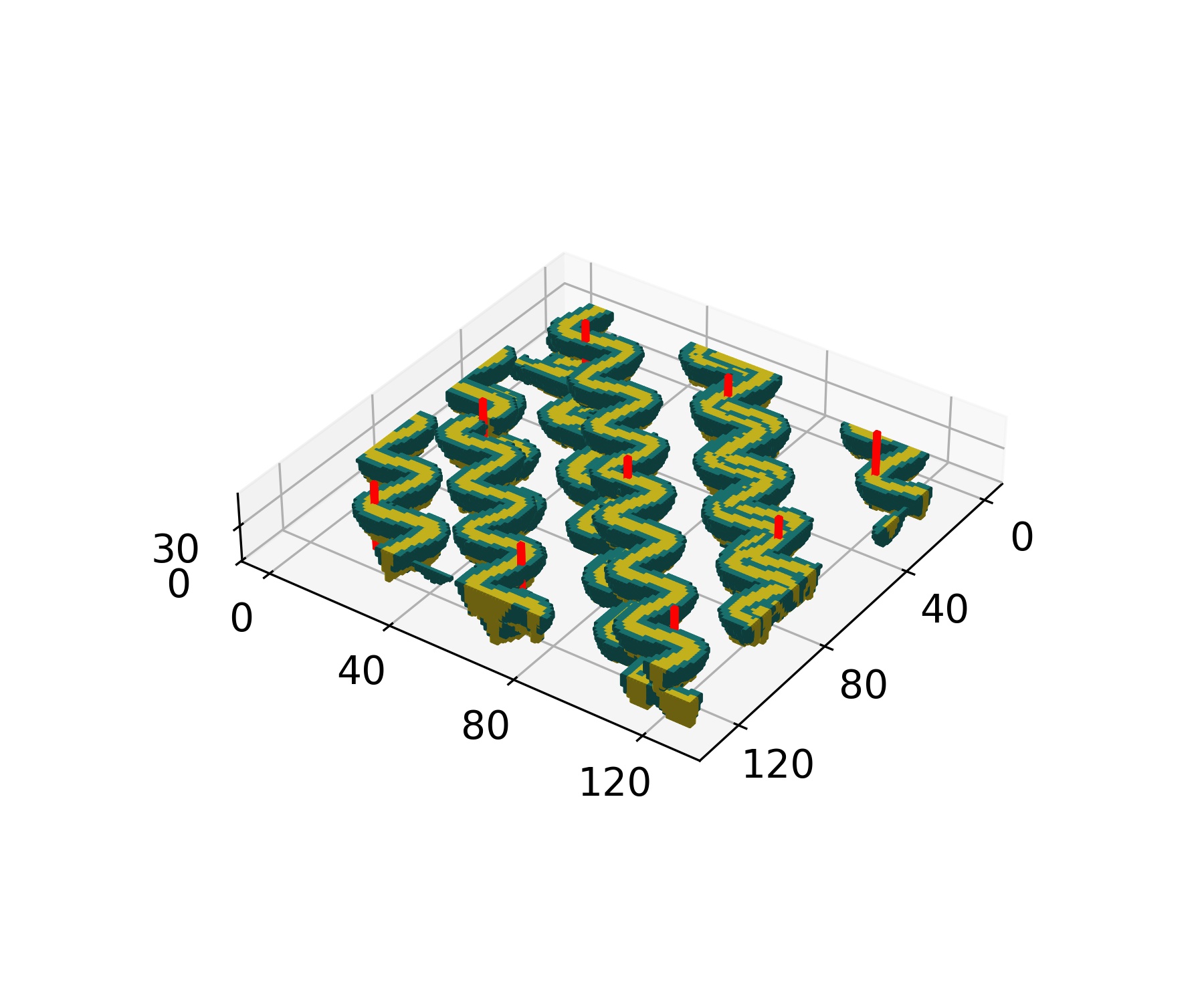}
        \caption{\centering \\ \centering Mud fraction: 0.83\\Channel orientation: 44$^\circ$\\Channel width: 5.5}
    \end{subfigure}
    \hfill
    \begin{subfigure}[b]{0.3\textwidth}
        \centering
        \includegraphics[width=\textwidth, trim=50 50 50 50, clip]{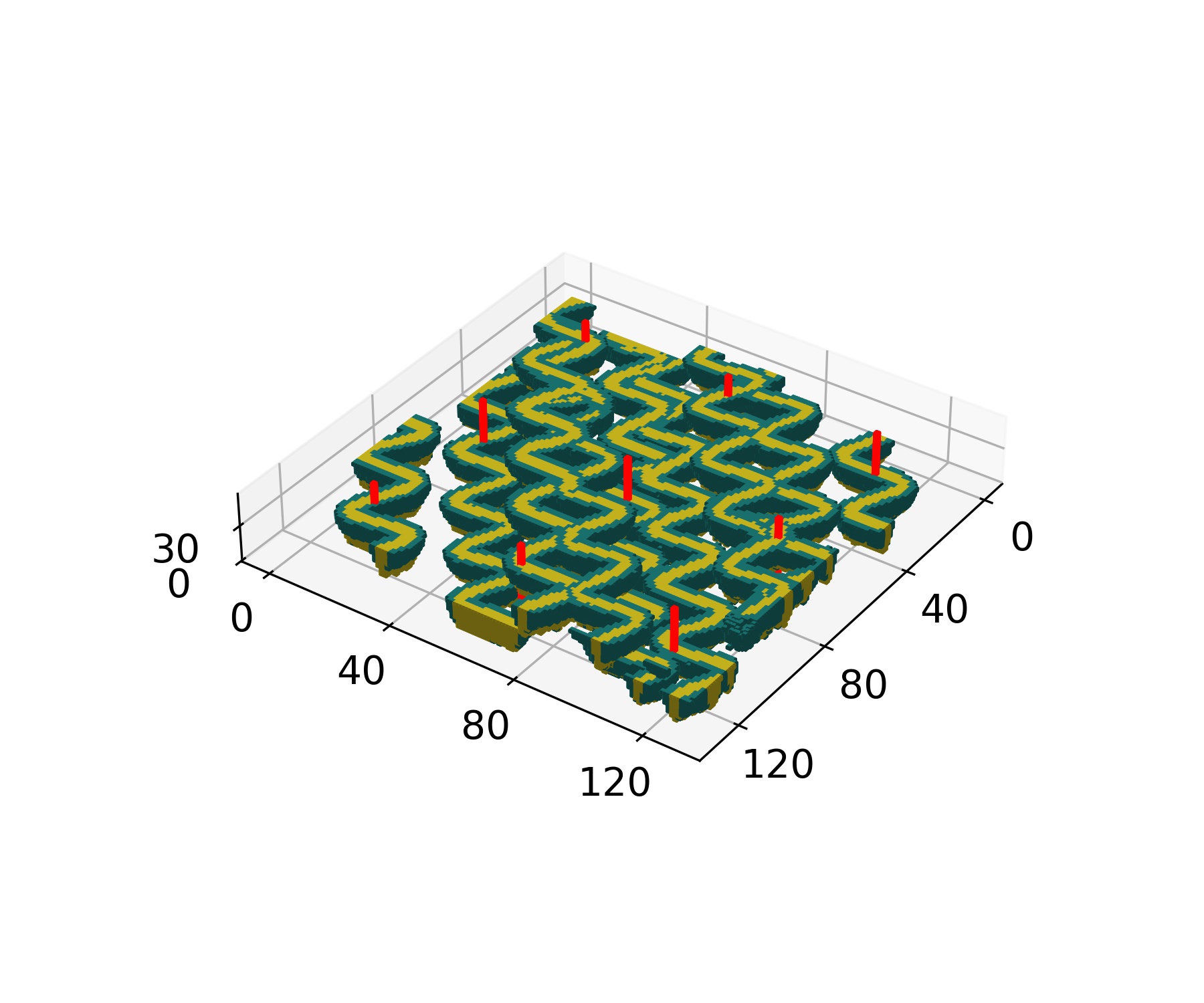}
        \caption{\centering \\ \centering Mud fraction: 0.72\\Channel orientation: 50$^\circ$\\Channel width: 5.2}
    \end{subfigure}

    \caption{Randomly selected 3D-LDM generated realizations corresponding to different scenario parameters. Scenario parameters are reported for each geomodel.}
    \label{fig:examples_ldm}
\end{figure}

Examples of 3D-LDM generated models are shown in Figure~\ref{fig:examples_ldm}. Here we use the same presentation style as in Figure~\ref{fig:examples_petrel}. Visually, the LDM geomodels display geological features consistent with those in the reference Petrel models (Figure~\ref{fig:examples_petrel}). Specifically, correct channel shape, channel continuity, and the location of levees with respect to channels are observed. 

The scenario parameters \textcolor{black}{$\mathbf{s}=[f_m, \theta_{\rm ch}, w_{\rm ch}]$} for Petrel-generated samples are prescribed as OBM inputs, so they are immediately available. This is not the case for 3D-LDM realizations, however, since the only input is a randomly generated $\boldsymbol{\xi}_T$ field. Thus, if scenario parameters are of interest, these must be determined from the LDM-generated geomodels. Mud fraction can be directly computed as the fraction of 0-facies cells. To efficiently determine the channel orientation and channel width for LDM-generated realizations, we train a separate convolutional neural network (CNN) to provide these quantities given the input \textcolor{black}{facies model $\mathbf{r}$. We denote this estimation of $\theta_{\rm ch}$ and $w_{\rm ch}$ as $\tilde{\mathbf{s}} = \text{CNN}(\mathbf{r})$}. This CNN is trained on the same dataset as was used for the 3D-LDM training. To simplify this procedure, only layers $z=1$ and $z=16$ are used, as we have found these to be representative of the overall geomodel scenario parameters. The average testing RMSE with this approach is about 2.3$^{\circ}$ and 0.15~grid blocks for channel orientation and width. These errors are comparable to the variability introduced by the OBM drift discussed in Section~\ref{case_study}, so we believe this treatment is sufficient for current (reporting) purposes.

The scenario parameters computed in this way for the LDM-generated geomodels are provided in Figure~\ref{fig:examples_ldm}. It is evident that the geomodels correspond to a variety of scenario parameters -- high and low mud fractions, different orientations, wide and thin channels -- consistent with the training set. This is an important observation as it suggests that the full range of models comprising the training set can indeed be generated. Another model characteristic worth assessing is the level of accuracy in hard data conditioning. We observe that hard data at conditioning locations are honored with an accuracy of 98\% over the full set of LDM generated models. This could be improved by increasing $\lambda_{\text{h}}$ in Eq.~\ref{eqn:loss_vae}, though that could negatively impact other geomodel features.

\begin{figure}[h]
    \centering
    \begin{subfigure}[b]{0.8\textwidth}
        \centering
        \includegraphics[width=\textwidth]{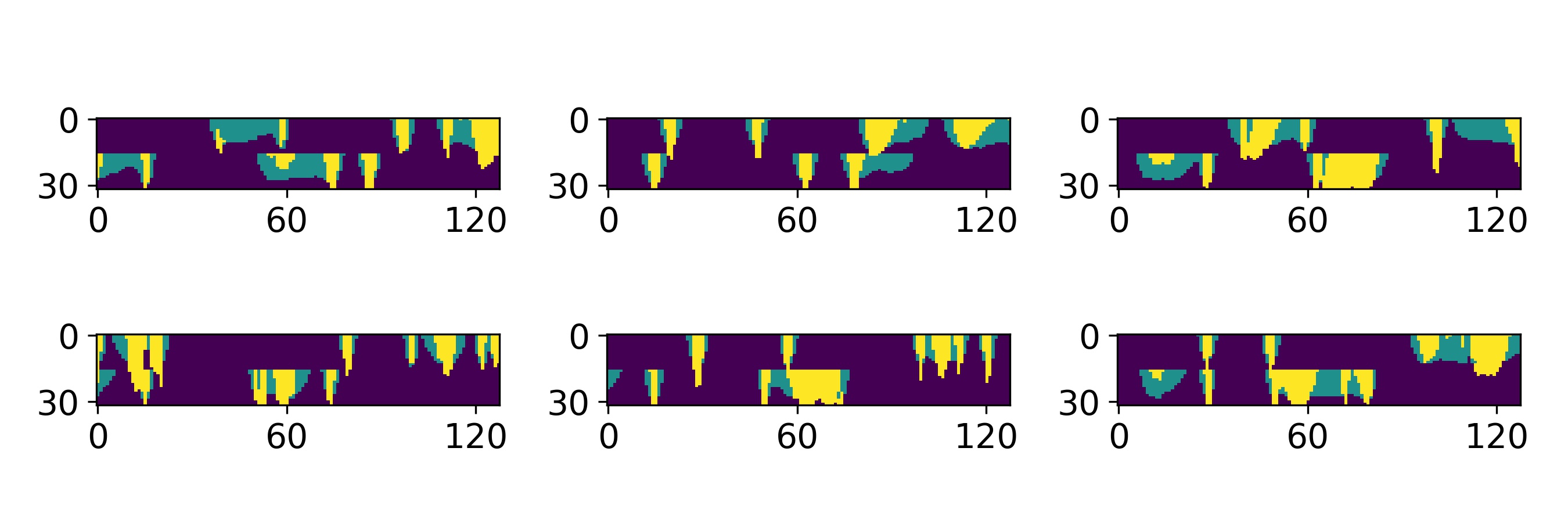}
        \caption{Petrel}
        \label{fig:sections_petrel}
    \end{subfigure}
    \begin{subfigure}[b]{0.8\textwidth}
        \centering
        \includegraphics[width=\textwidth]{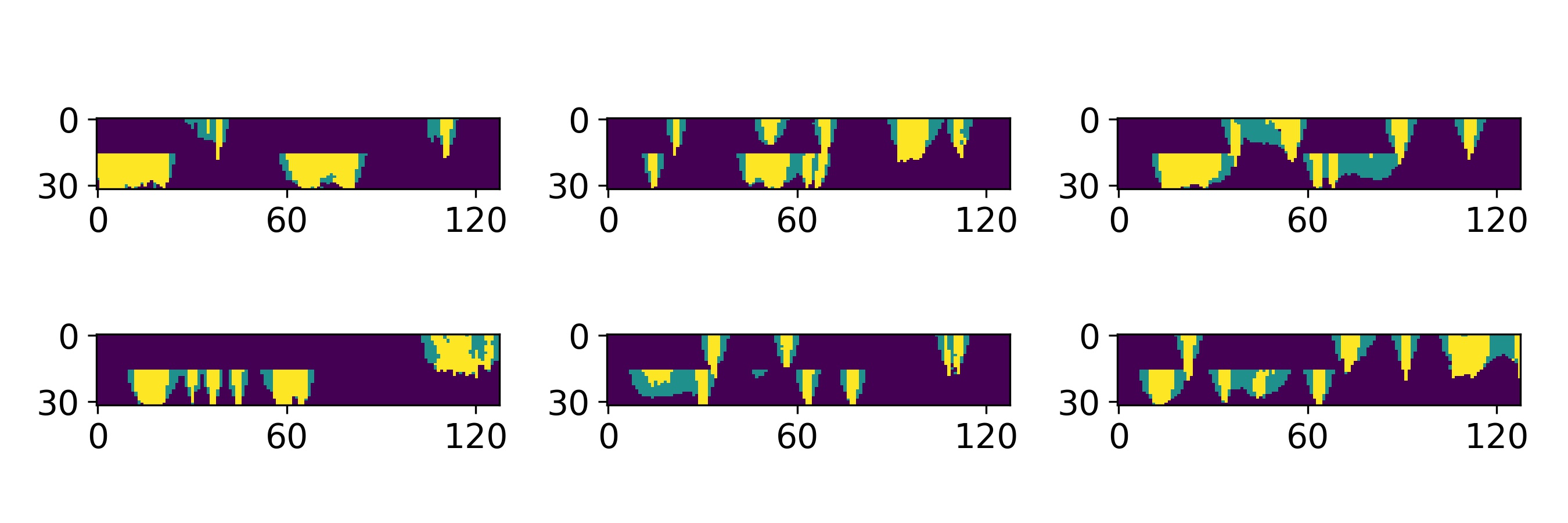}
        \caption{3D-LDM}
        \label{fig:sections_ldm}
    \end{subfigure}
    \caption{Comparison of vertical cross sections ($y-z$ plane, $x=64$) for (a) randomly selected Petrel realizations and (b) 3D-LDM realizations.}
    \label{fig:cross_sections}
\end{figure}

Cross sectional ($y-z$ plane) views of the LDM-generated models appear in Figure~\ref{fig:cross_sections}. Petrel cross sections are shown for comparison purposes. Both sets of models were randomly selected, so there is not a one-to-one correspondence between the Petrel and 3D-LDM realizations. We see that the shapes of the channel cross sections (e.g., decreasing width as a function of depth) are preserved. The levee facies surrounding the upper portion of each channel is also captured. It is important to note that the wider or narrower sections observed within the same realization are a consequence of the sinuous nature of the channels. As a result, the cross sectional cut may be anywhere from nearly perpendicular to nearly parallel to the main channel axis. 

\begin{figure}[h]
  \centering
  \begin{subfigure}[b]{0.4\textwidth}
    \centering
\includegraphics[width=\textwidth, trim=30 400 30 30, clip]{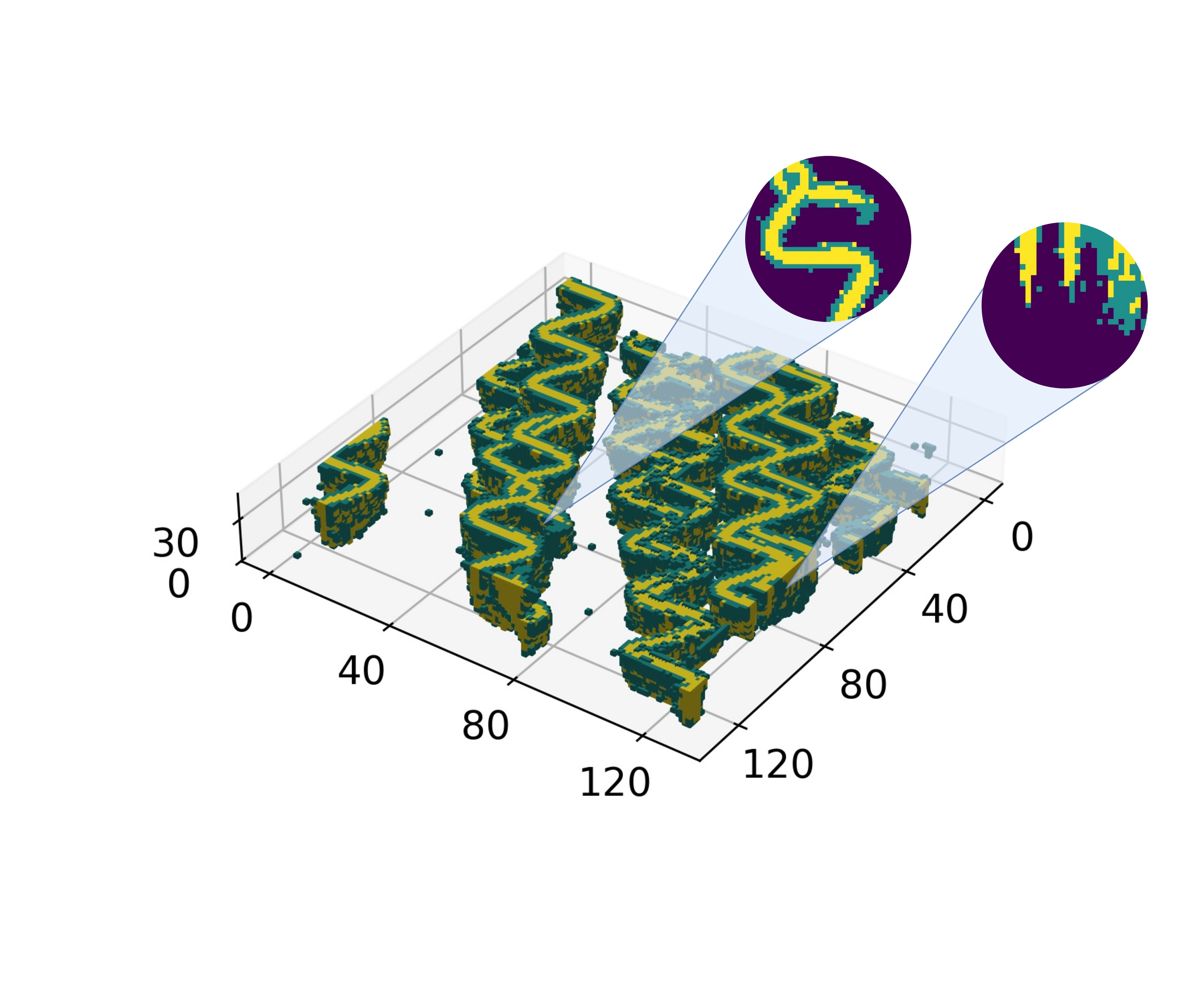}
    \caption{Without perceptual loss}
    \label{fig:perc_loss_a}
  \end{subfigure}
  \begin{subfigure}[b]{0.4\textwidth}
    \centering \includegraphics[width=\textwidth,, trim=30 400 30 30, clip]{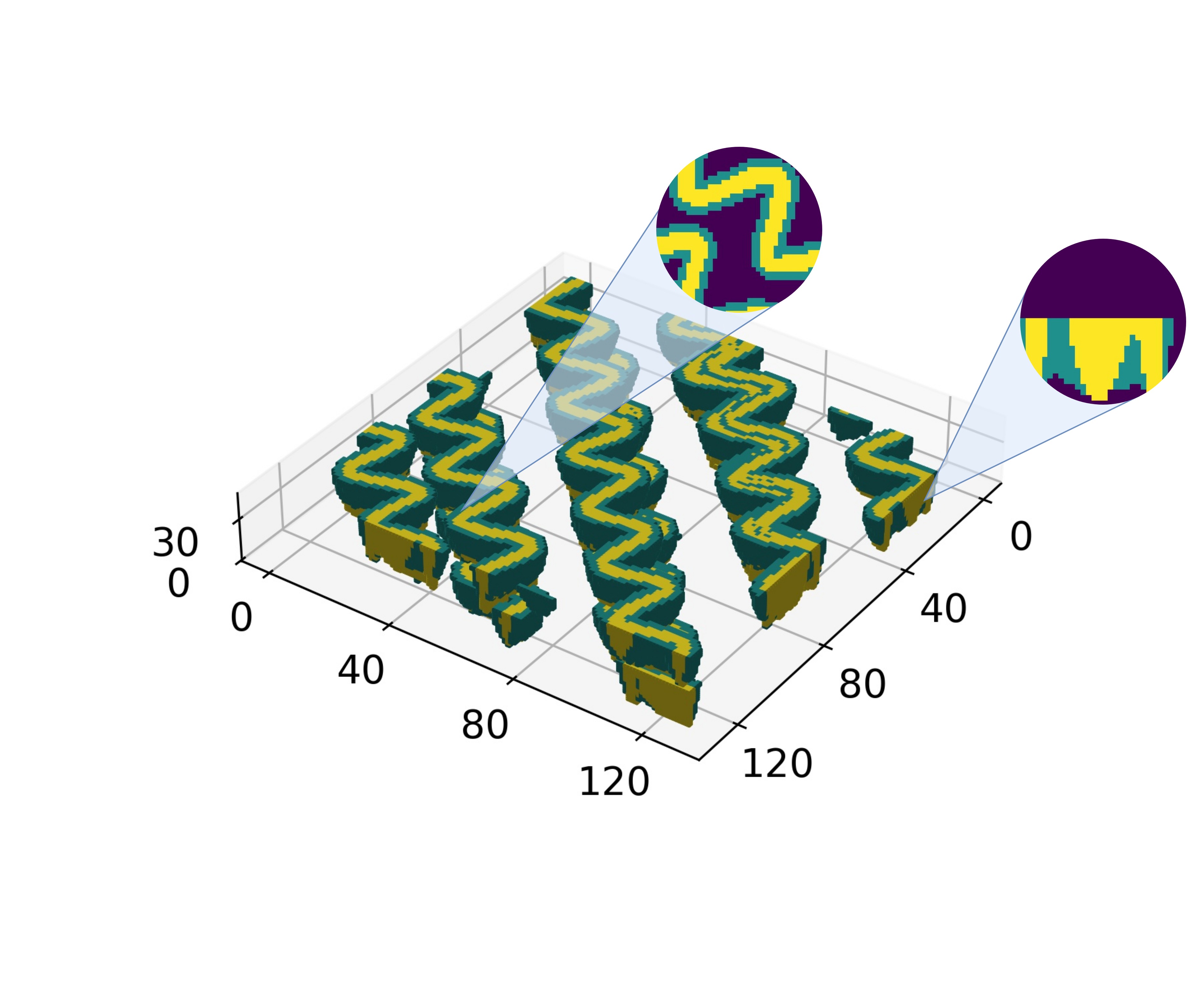}
    \caption{With perceptual loss}
    \label{fig:perc_loss_b}
  \end{subfigure}
  \caption{Effect of the perceptual loss on 3D-LDM geomodel quality: (a)~generated model without $\lambda_{\text{perc}} L_{\text{perc}}$ term, (b)~generated model with $\lambda_{\text{perc}} L_{\text{perc}}$ term.}
  \label{fig:perc_loss_combined}
\end{figure}

Figure~\ref{fig:perc_loss_combined} highlights the impact of the perceptual loss term, $\lambda_{\text{perc}} L_{\text{perc}}$ in Eq.~\ref{eqn:loss_vae}, in the VAE training. This loss term significantly reduces (essentially eliminates) the occurrence of unrealistic features in the decoded geomodels such as broken channel bodies, incorrect facies ordering, or isolated levee cells. Such unwanted features are visible in the magnified (circular) views in the areal and vertical cross section planes in Figure~\ref{fig:perc_loss_combined}a, where the perceptual loss is not included. There we see some ``speckling'' where levee facies appear within channel facies. This issue is not observed in Figure~\ref{fig:perc_loss_combined}b, where the perceptual loss is included. Although a one-to-one comparison is not possible as the samples are generated by LDMs with different loss functions and hence different parameters, the model quality shown in Figure~\ref{fig:perc_loss_combined}b is consistently observed. This demonstrates that the use of perceptual loss is indeed beneficial for our \textcolor{black}{variable scenario} modeling.

\subsection{Quantitative assessments}
\label{quant_assess}

\begin{figure}[h]
  \centering
  \begin{subfigure}[b]{0.3\textwidth}
    \centering
    \includegraphics[width=\linewidth]{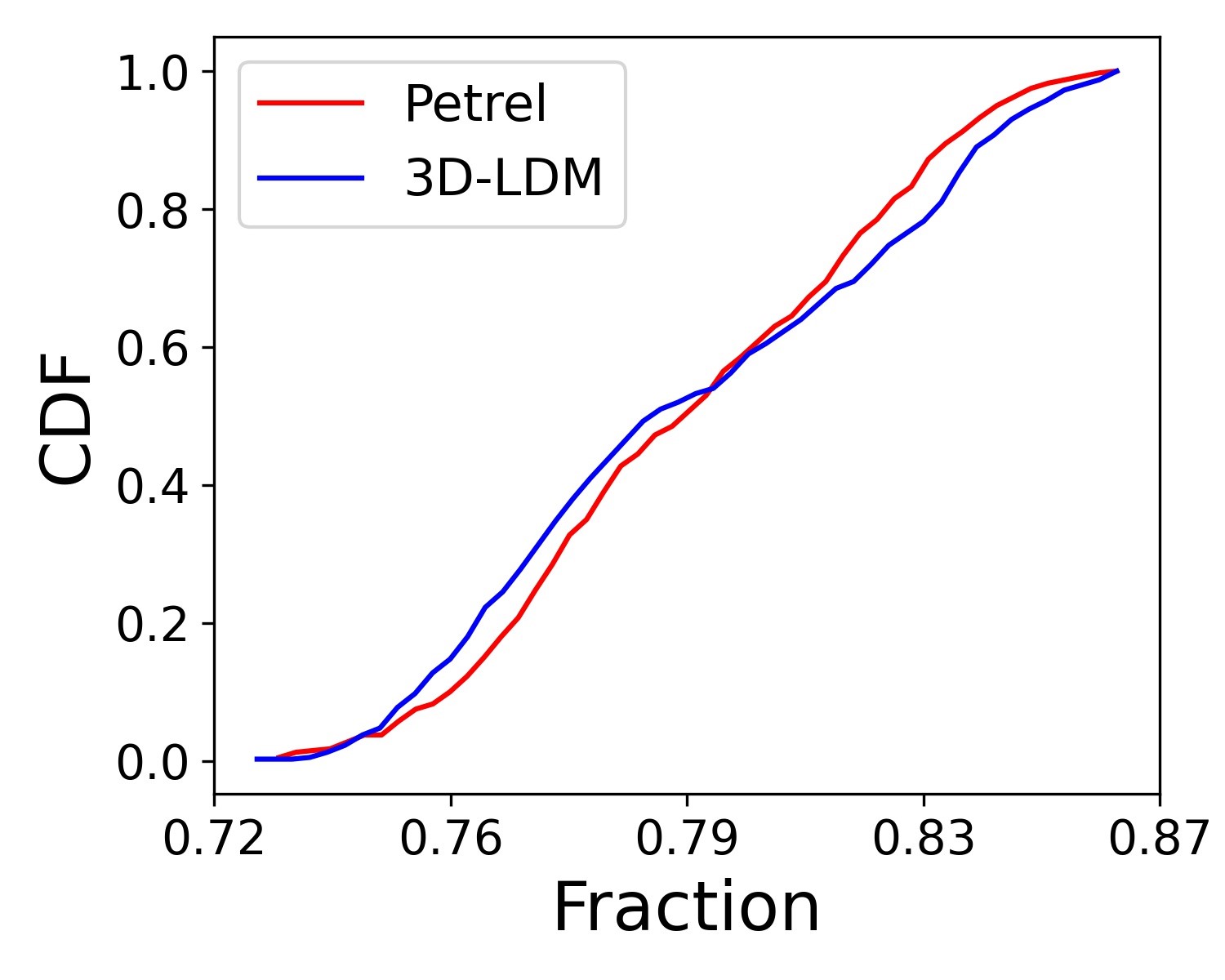}
    \caption{Mud facies}
  \end{subfigure}
  \hfill
  \begin{subfigure}[b]{0.3\textwidth}
    \centering
    \includegraphics[width=\linewidth]{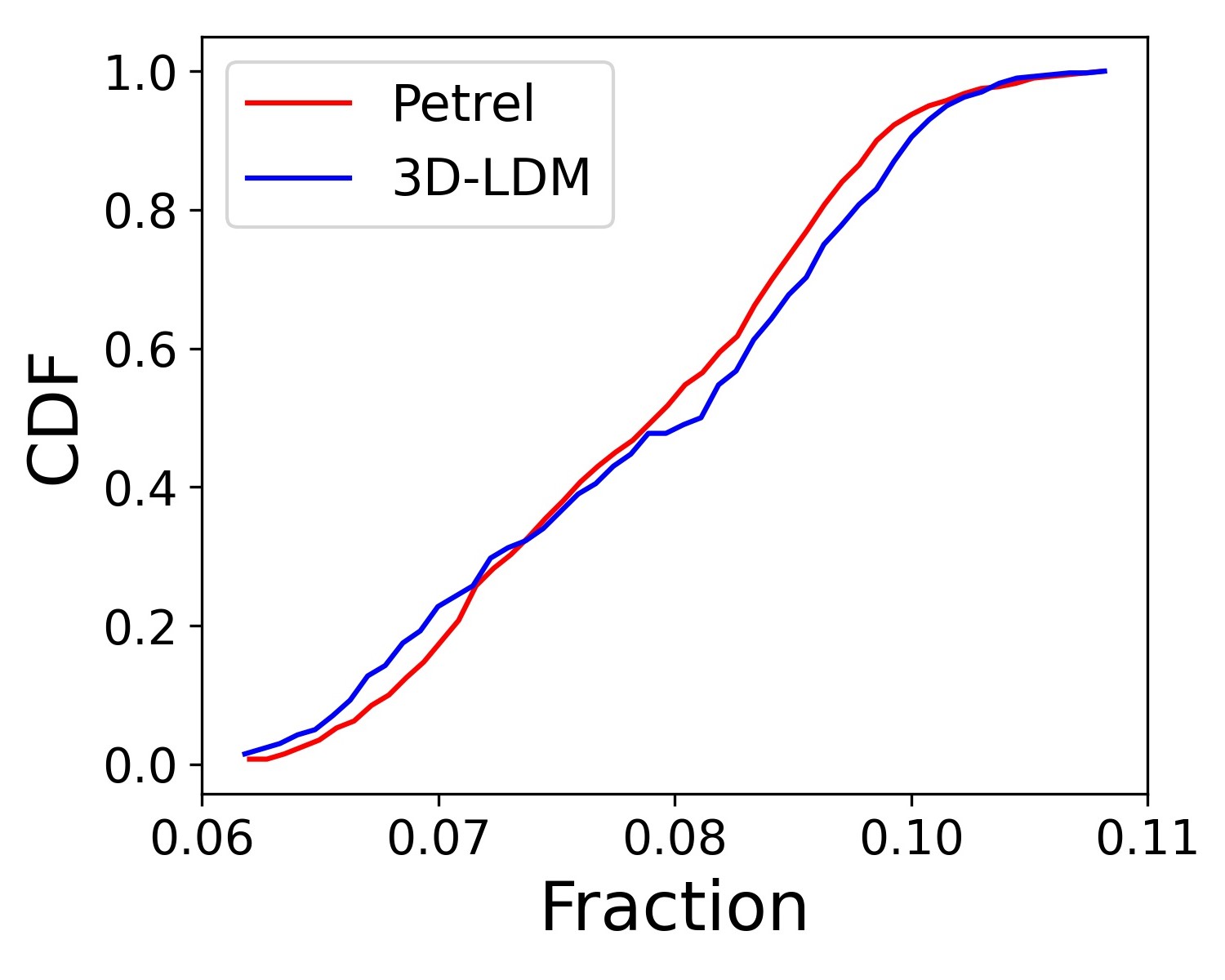}
    \caption{Levee facies}
  \end{subfigure}
  \hfill
  \begin{subfigure}[b]{0.3\textwidth}
    \centering
    \includegraphics[width=\linewidth]{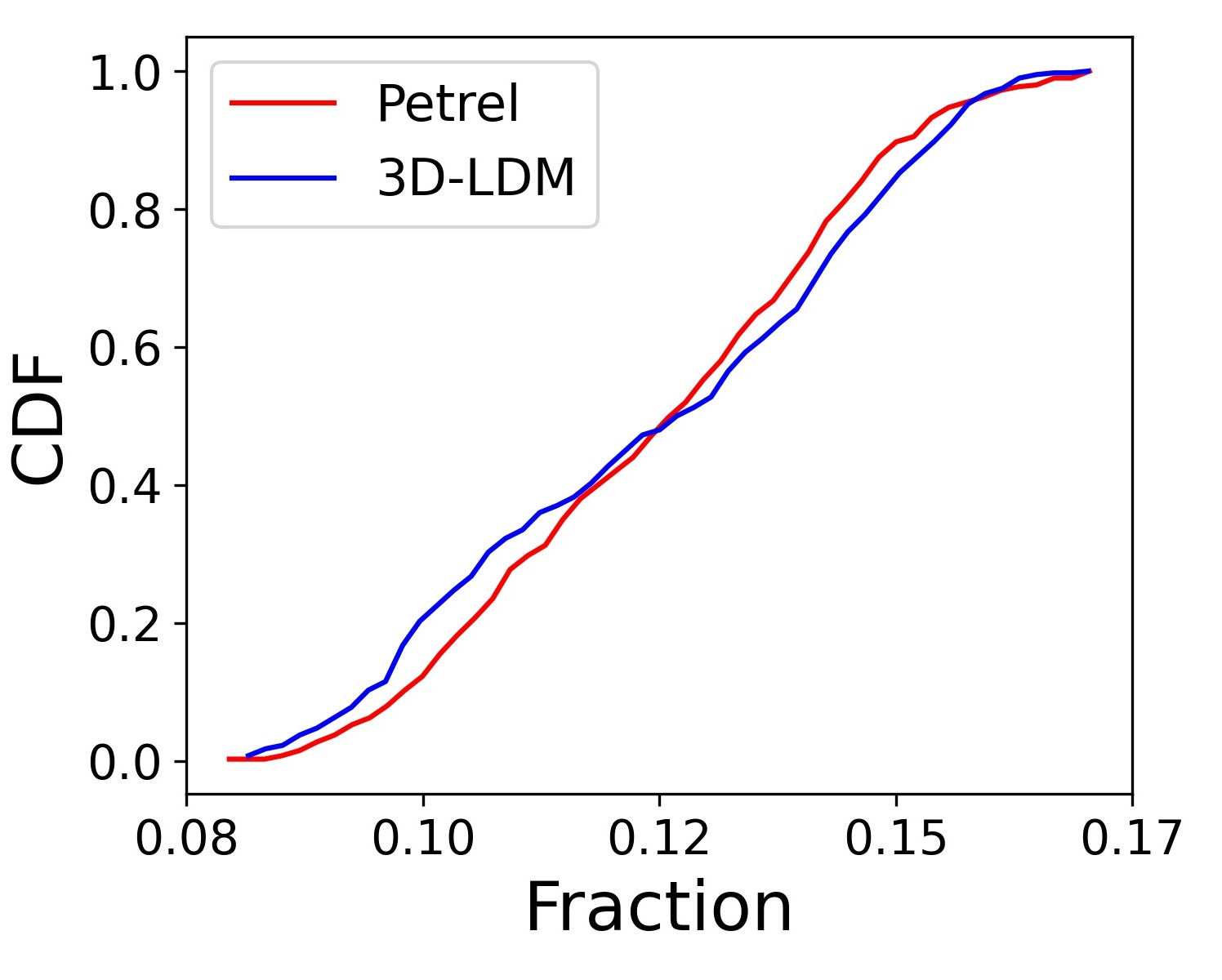}
    \caption{Channel facies}
  \end{subfigure}

  \caption{Comparison of CDFs for the three facies fractions for Petrel realizations (red curves) and 3D-LDM realizations (blue curves). Curves constructed from 200 new Petrel and 3D-LDM geomodels.}
  \label{fig:facies_fractions}
\end{figure}

We now compare facies fractions distributions, variogram functions, and flow response statistics for sets of 200 newly generated 3D-LDM and Petrel models. Figure~\ref{fig:facies_fractions} shows cumulative density functions (CDFs) for the mud, levee, and channel facies fractions. Although slight deviations are observed along the CDFs, there is clearly a high degree of consistency between the distributions for the two sets of models. It is also apparent that 3D-LDM realizations span the same ranges in facies fractions as the Petrel models. 

Variograms are commonly used in geostatistics to quantify two-point spatial correlations. These fully characterize multi-Gaussian fields, and provide useful (though incomplete) metrics for more general multipoint geostatistical models, such as those considered here. For a 2D slice $i$ along axis $d$ of a geomodel, $\mathbf{m}_{0,i}^{(d)}$, the variogram (or ``semi-variance'') $\gamma(h_{\theta})$ is a function of the lag distance $h_{\theta}$ in the direction aligned with angle $\theta$~\citep{pyrcz2014geostatistical}, i.e.,
\begin{equation}
\gamma(h_{\theta}) = \frac{1}{2} \, \mathrm{Var} \left[ \mathbf{m}_{0,i}^{(d)}(l) - \mathbf{m}_{0,i}^{(d)}(l + h_\theta) \right].
\end{equation}
Here $\text{Var}$ is the variance operator across an ensemble of geomodels, and $\mathbf{m}_{0,i}^{(d)}(l)$ indicates the facies value at location $l$. 

\begin{figure}
    \centering
    \begin{subfigure}[b]{0.45\textwidth}
        \centering
        \includegraphics[width=\textwidth]{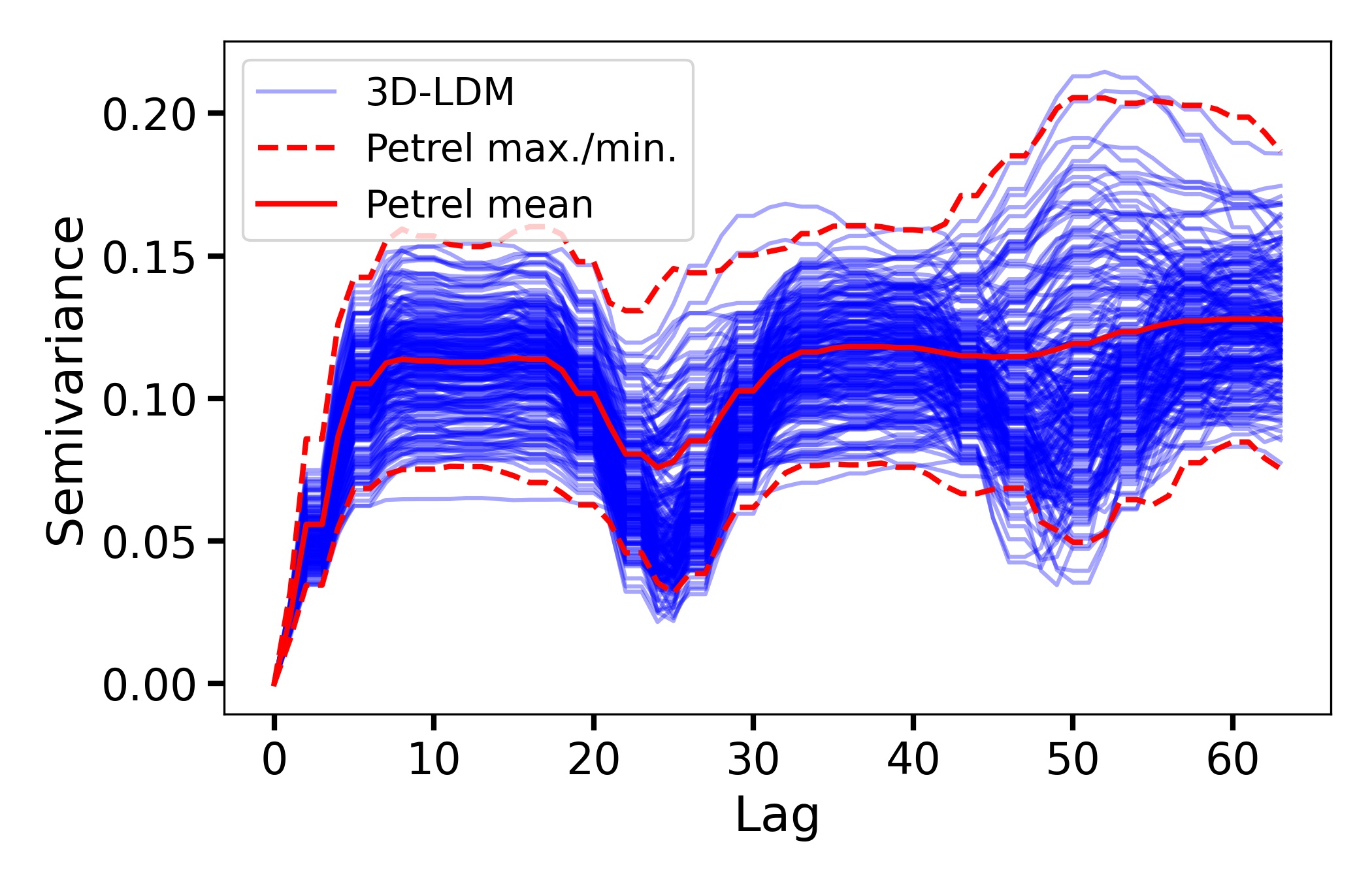}
        \caption{Horizontal, angle $45^{\circ}$, layer $z=1$}
        \label{fig:variogram_45_z1}
    \end{subfigure}
    \begin{subfigure}[b]{0.45\textwidth}
        \centering
        \includegraphics[width=\textwidth]{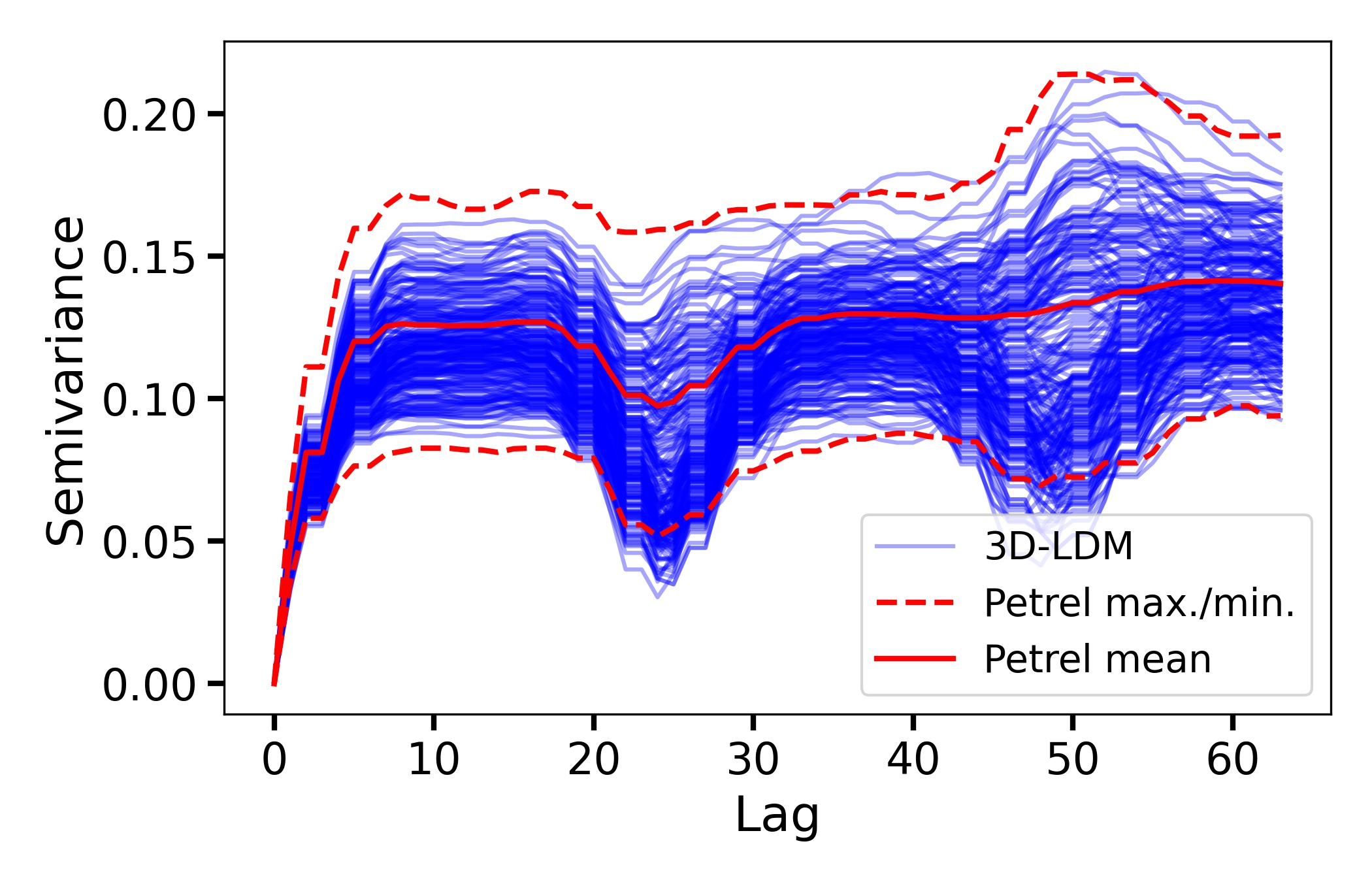}
        \caption{Horizontal, angle $45^{\circ}$, layer $z=16$}
        \label{fig:variogram_45_z18}
    \end{subfigure}
    \begin{subfigure}[b]{0.45\textwidth}
        \centering
        \includegraphics[width=\textwidth]{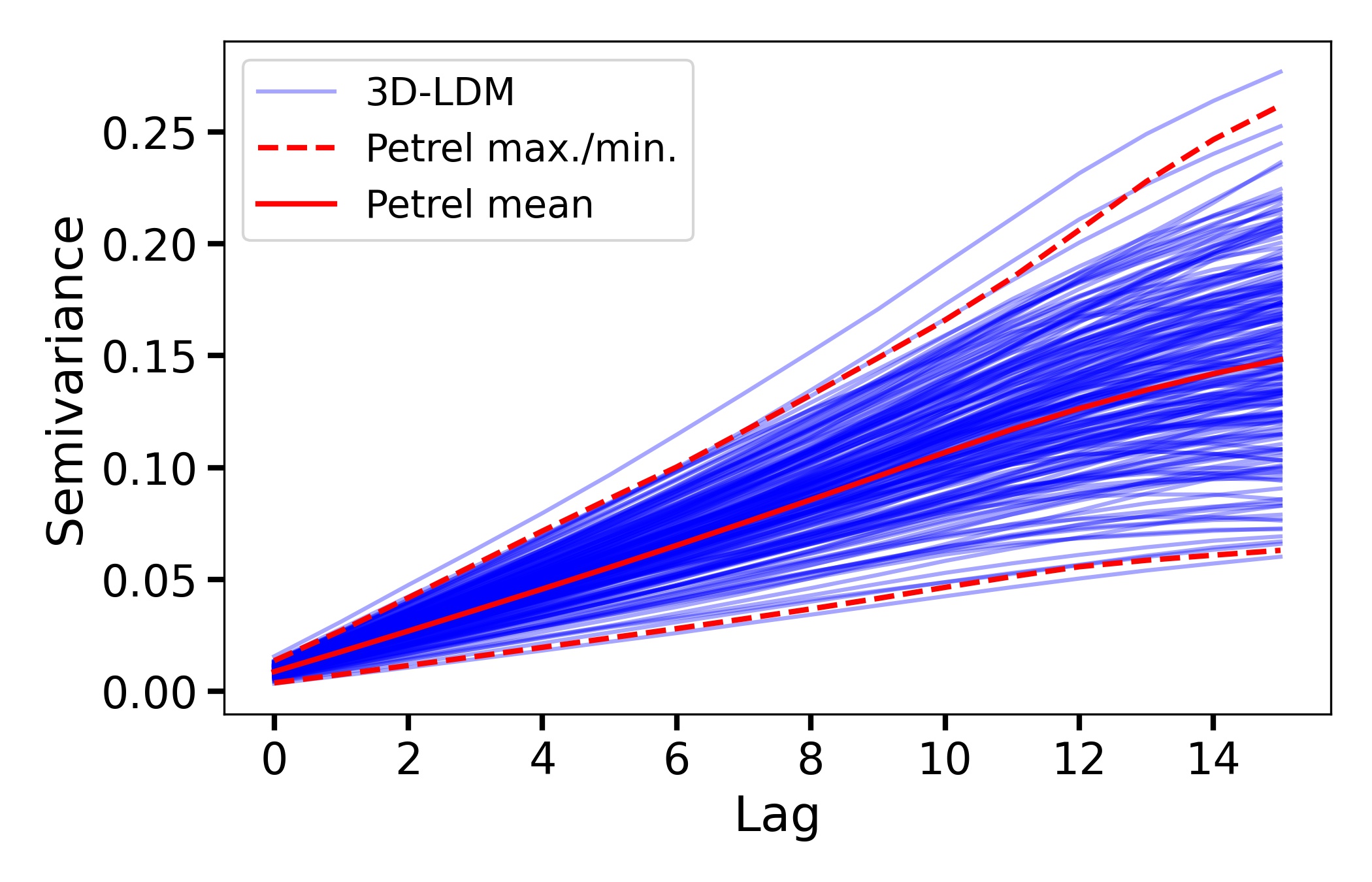}
        \caption{Vertical, angle $90^{\circ}$, section $x=64$}
        \label{fig:variogram_90_x64}
    \end{subfigure}
    \begin{subfigure}[b]{0.45\textwidth}
        \centering
        \includegraphics[width=\textwidth]{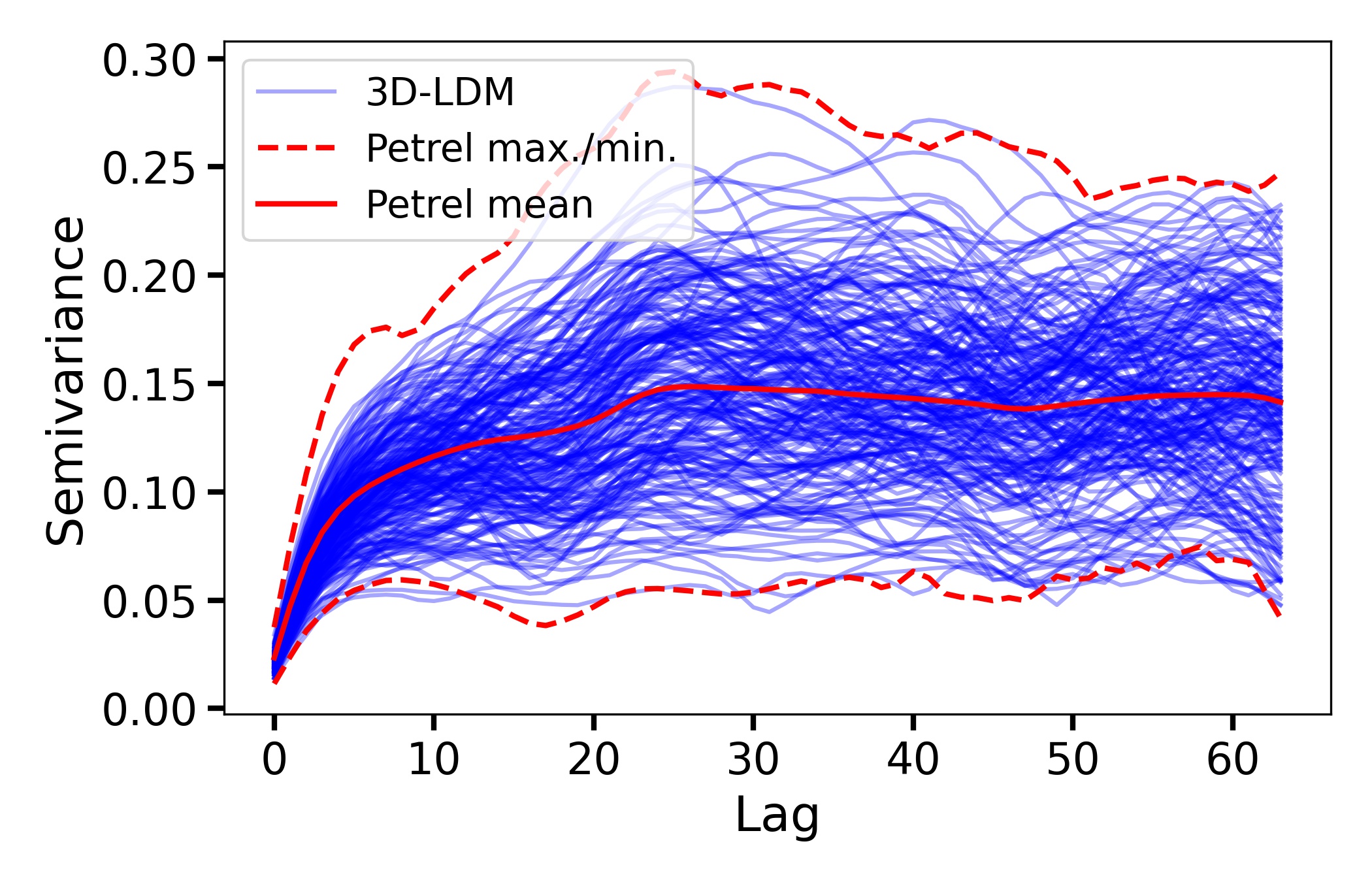}
        \caption{Vertical, angle $0^{\circ}$, section $x=64$}
        \label{fig:variogram_0_x64}
    \end{subfigure}
    \caption{Comparisons of variograms in the horizontal and vertical direction, with lag distance in grid blocks. Red solid and dashed curves denote mean, maximum, and minimum value at each lag distance for 200 new Petrel models, while blue curves show individual results for 200 new 3D-LDM generated models.}
    \label{fig:variograms}
\end{figure}

Variograms are presented in Figure~\ref{fig:variograms}. Plots are shown for the diagonal direction for two representative layers ($x-y$ plane, Figure~\ref{fig:variograms}a,b), and for a vertical cross section ($y-z$ plane, Figure~\ref{fig:variograms}c,d) along both the horizontal and vertical directions. Each blue curve corresponds to the variogram for a particular 3D-LDM generated realization. The red curves indicate the mean, maximum, and minimum values for the ensemble of 200 Petrel geomodels. The upper and lower red curves in each plot are seen to lie near the edges of the set of blue curves, demonstrating that the range of responses is consistent between the two sets of results. In addition, the middle red curve (mean) appears around the middle of the cloud of blue curves, as would be expected. Similar results are observed along other directions and in other layers or cross sections. In total, these results show that fundamental (two-point) statistics quantifying the spatial arrangement of facies are captured in the LDM models. Higher-order or facies-continuity statistics could also be computed if such metrics are important for a particular application. We will instead next consider flow metrics, since our target application is history matching.

We now evaluate the performance of our 3D-LDM parameterization in a flow simulation setting. Flow simulations are performed for a water-flooding problem involving a two-phase immiscible oil-water system. The setup entails three water injection wells and six production wells arranged in a line-drive pattern, as shown in Figure~\ref{fig:examples_petrel}a.  Grid blocks are of physical dimensions 65~ft $\times$ 65~ft $\times$ 6.5~ft (in the $x$, $y$ and $z$ directions, respectively). The reservoir is initially at a pressure of 4500~psi (at a depth of 4160~ft). Initial oil and water saturations are 0.85 and 0.15, and their viscosities are 5~cP and 1~cP, respectively, at reference conditions. The relative permeability curves are the same as those used in \cite{DIFEDERICO2025105755}. Porosity and permeability are constant for each facies. The specific values are 0.2 and 1000~mD for channels, 0.15 and 200~mD for levee, and 0.05 and 25~mD for mud. Wells are all bottom-hole pressure (BHP) controlled, with injectors operating at 4700~psi and producers at 4300~psi. The wells are perforated only at grid blocks corresponding to channel facies (see Figure~\ref{fig:examples_petrel}a). The simulation time frame is 1500~days. All runs are performed using the tNavigator simulator~\citep{rock_flow_dynamics}. \textcolor{black}{Using 16 CPU cores, a single simulation run requires about 10~minutes.}

\begin{figure}
    \centering
    \begin{subfigure}[b]{0.32\textwidth}
        \centering
        \includegraphics[width=\textwidth, trim=15 0 0 0, clip]{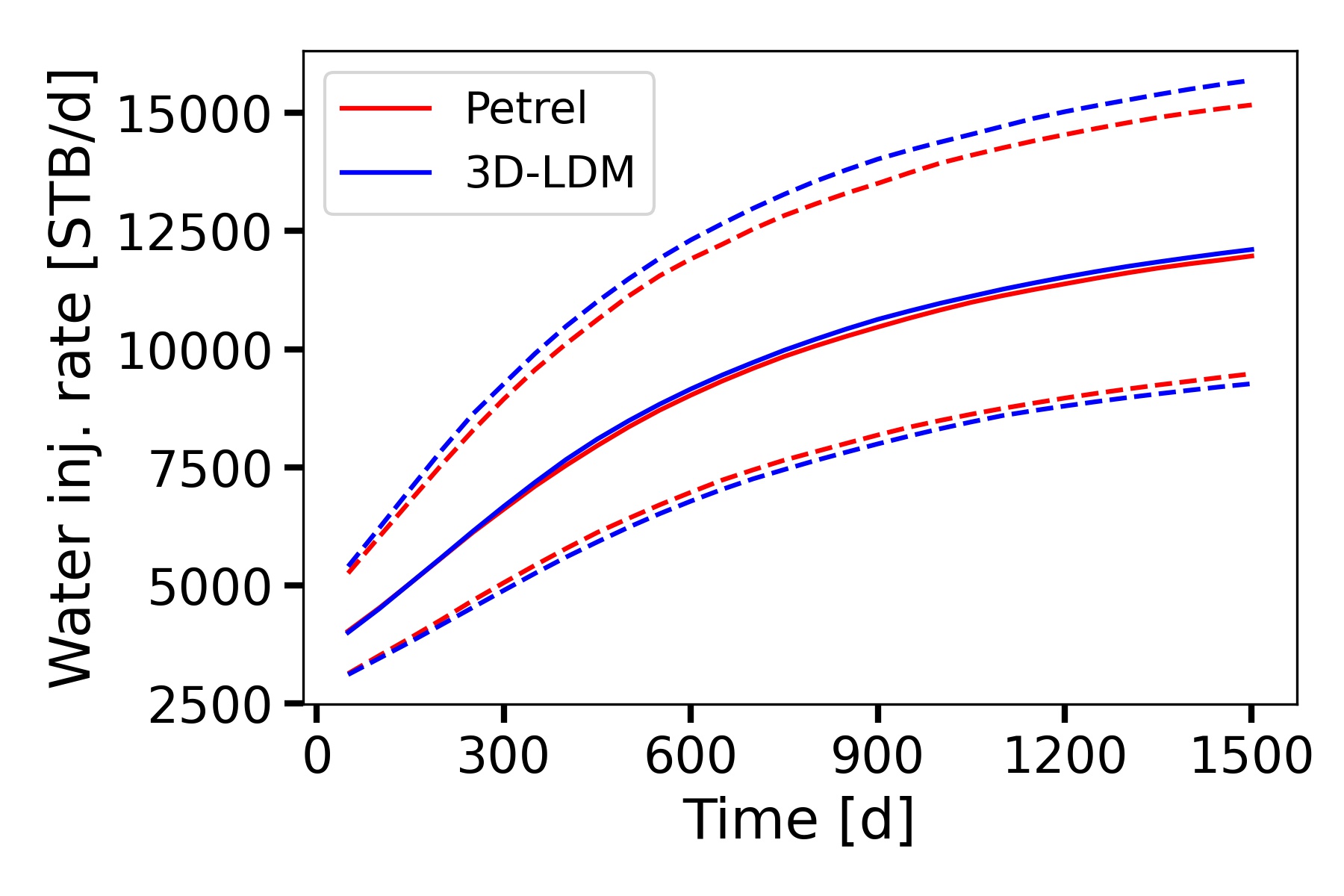}
        \caption{Water injection}
        \label{fig:field_inj}
    \end{subfigure}
    \hfill
    \begin{subfigure}[b]{0.32\textwidth}
        \centering
        \includegraphics[width=\textwidth, trim=15 0 0 0, clip]{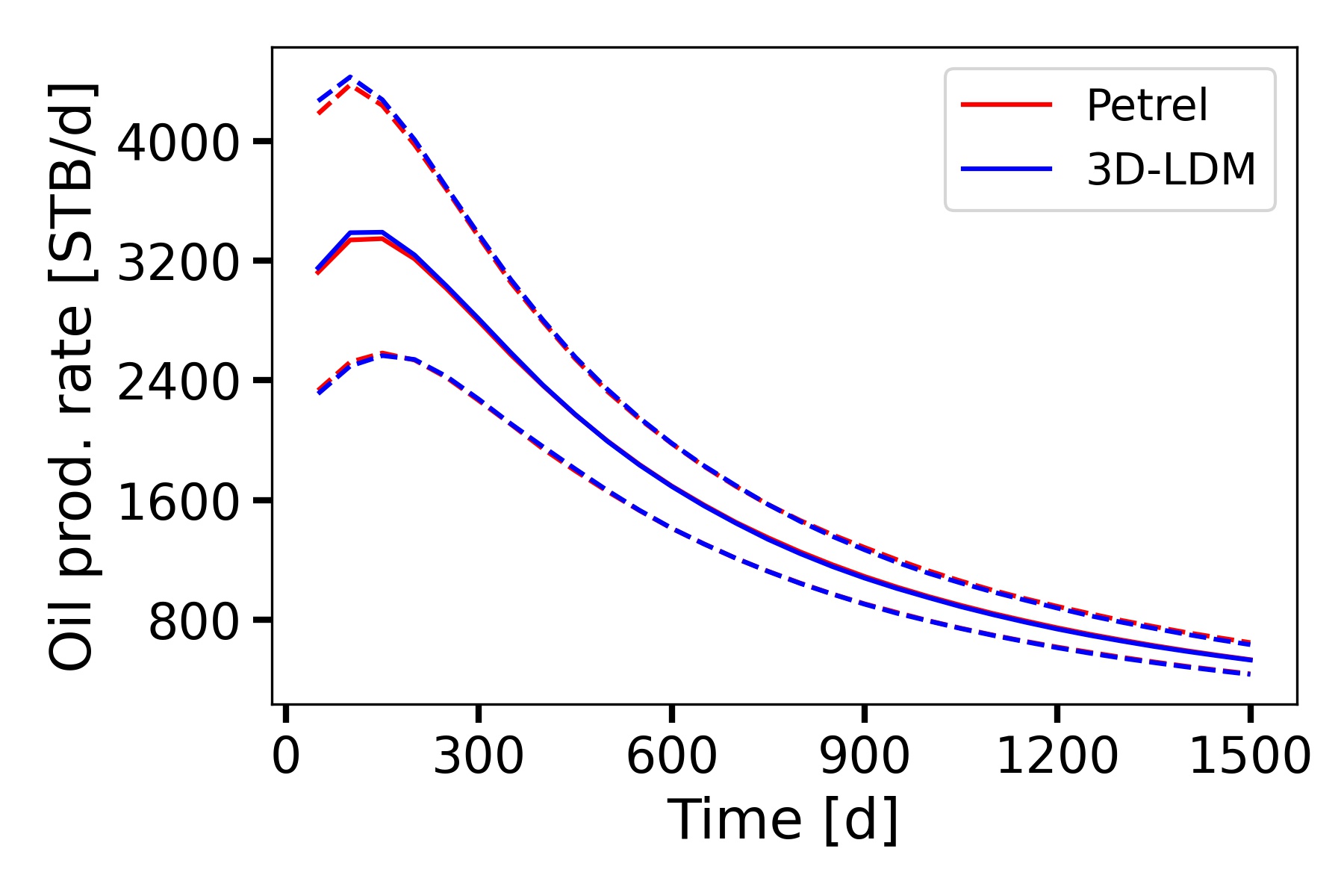}
        \caption{Oil production}
        \label{fig:field_oil}
    \end{subfigure}
    \hfill
    \begin{subfigure}[b]{0.32\textwidth}
        \centering
        \includegraphics[width=\textwidth, trim=15 0 0 0, clip]{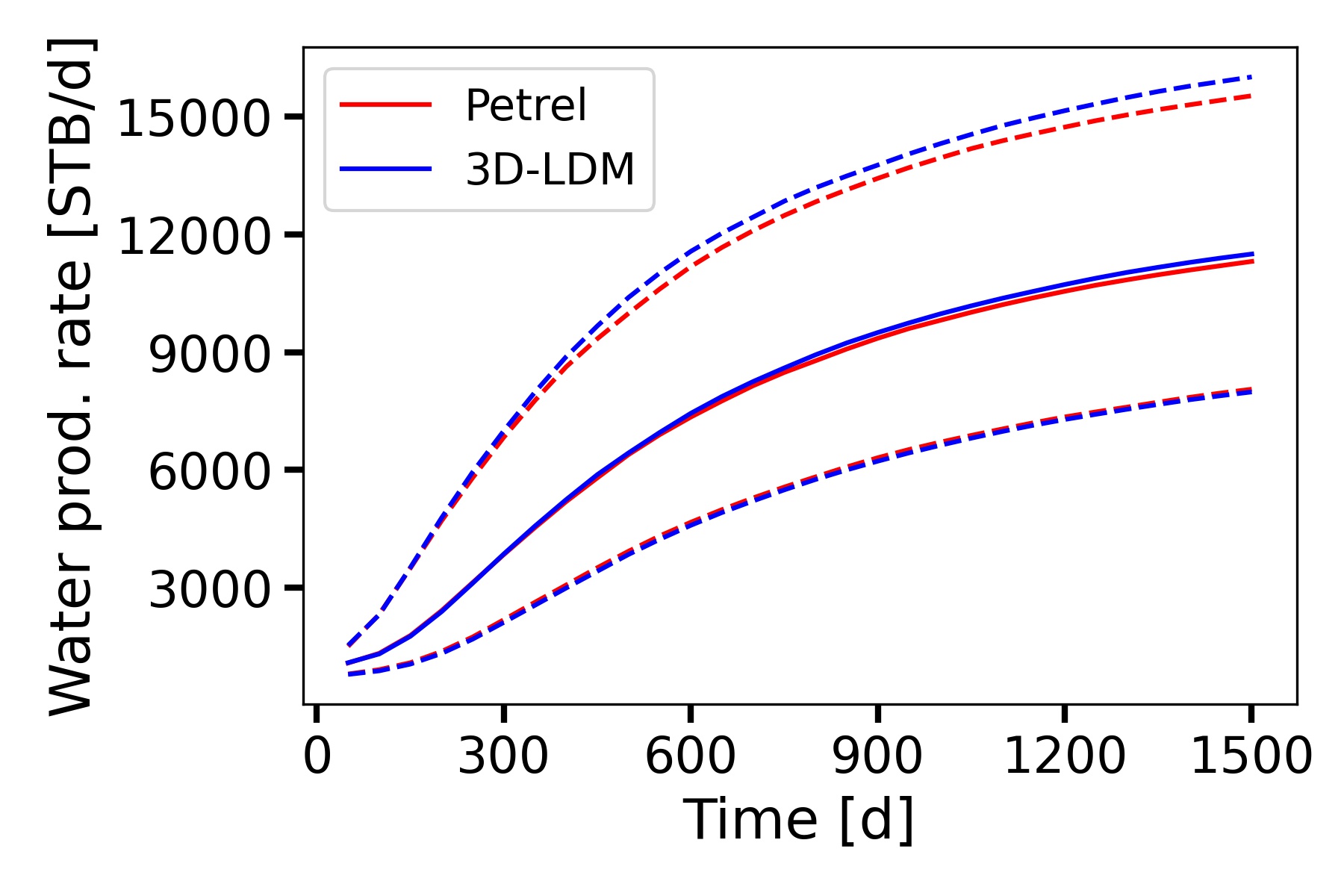}
        \caption{Water production}
        \label{fig:field_wat}
    \end{subfigure}
    \caption{Comparison of field-level flow statistics for Petrel realizations (red curves) and 3D-LDM realizations (blue curves) over ensembles of 200 new models. Solid curves denote P$_{50}$ results, lower and upper dashed curves are P$_{10}$ and P$_{90}$ results.}
    \label{fig:field_rates}
\end{figure}

Results for flow statistics are shown in Figure~\ref{fig:field_rates}. For both the Petrel and 3D-LDM ensembles, we present the 10th, 50th and 90th percentile flow response (P$_{10}$, P$_{50}$ and P$_{90}$) over the full set of runs at each time step. These results are at the field level, i.e., summed over all three injectors (Figure~\ref{fig:field_rates}a) or all six producers (Figure~\ref{fig:field_rates}b,c). These flow statistics display very close agreement -- the only visible discrepancies are slight deviations in the P$_{90}$ water injection  (Figure~\ref{fig:field_rates}a) and water production (Figure~\ref{fig:field_rates}c) curves. We note that consistency is also observed at the well level (these results are not shown for conciseness). The agreement highlighted in Figure~\ref{fig:field_rates} is important because it suggests that the range of spatial features that impact flow are indeed captured in the LDM generated models.

The results in this section demonstrate that our 3D-LDM procedure is accurate for the range of scenario parameters considered (see Table~\ref{table:orientation_width_number}). For somewhat larger parameter ranges (e.g., $f_m$ ranging from 0.6 to 0.9, $\theta_{\rm ch}$ from 0$^{\circ}$ to 60$^{\circ}$, $w_{\rm ch}$ from 3 to 7~grid blocks), the 3D-LDM did not perform as well. For such cases, the generated models still appeared geologically realistic (visually), though they did not span the full range of the training models. Specifically, we observed incomplete coverage of the ranges in the generated geomodels for ensemble metrics such as those shown in Figures~\ref{fig:facies_fractions} and \ref{fig:variograms}.

Given that our target application is history matching, where hard data and geologic interpretations are available, we expect to have prior ranges for the scenario parameters that are not overly broad (i.e., along the lines of those in Table~\ref{table:orientation_width_number}). In this setting, the 3D-LDM developed here should be applicable. If much larger prior uncertainty ranges are required, some modifications to the model may be needed.

\section{Latent space history matching with uncertain geological scenario}
\label{results_hm}

In this section, we apply the 3D-LDM within the context of history matching. The flow problem and simulation setup are as described in Section~\ref{quant_assess}. For discrete facies geomodels such as those considered in this study, the goal of the data assimilation procedure is to assign a facies type to each grid block. In the \textcolor{black}{hierarchical uncertainty} case, this process is more challenging because the latent variables must characterize (indirectly) the specific scenario parameters, which in this case correspond to mud fraction, channel orientation, and channel width. 

\textcolor{black}{For the purposes of history matching, the training dataset (or a statistically representative subset, as assumed in Section~\ref{quant_assess}) defines the prior. In other words, the history matching method applied here assumes the priors have been ``validated'' and that the true model and data lie within the spaces defined by the training dataset. If this is not the case, the priors should be modified accordingly, or a treatment for model error that accounts for this discrepancy should be incorporated. Either or both of these steps must be performed before the history matching procedure described in Section~\ref{hm_esmda} is applied.}

\subsection{History matching cases and procedure}
\label{hm_esmda}

We will apply the history matching procedure for three different synthetic true models, denoted as Cases~1--3. None of these true models was used in training. The true models correspond to different scenario parameters, which are given in Table~\ref{table:true_params}. This allows us to assess the robustness of the overall procedure across different geological scenarios. 

\begin{table}[h]
\centering
\caption{Scenario parameter values for the synthetic true models used in history matching.}
\label{table:true_params}
\begin{tabular}{@{}lccc@{}}
\toprule
\textbf{True scenario parameter} & \textbf{Case~1} & \textbf{Case~2} & \textbf{Case~3} \\
\midrule
Mud fraction, $f_m$ & 0.75 & 0.85 & 0.81 \\
Channel orientation (avg.~angle), $\theta_{\rm ch}$ & 50$^{\circ}$ & 59$^{\circ}$ & 32$^{\circ}$  \\
Channel width (avg.~\# of blocks), $w_{\rm ch}$ & 5.9 & 4.3 & 4.6 \\
\bottomrule
\end{tabular}
\end{table}

History matching updates are performed in the LDM latent space, such that the variables to calibrate correspond to $\boldsymbol{\xi}_{T} \in \mathbb{R}^{n_c}$. This reduces the number of history matching variables from $N_c=524,288$ (i.e., with no parameterization) to $n_c=N_c/f^3=1024$. At every iteration, the 3D-LDM parameterization generates the facies model corresponding to the updated $\boldsymbol{\xi}_{T}$, $\mathbf{m}_0^{\text{LDM}}(\boldsymbol{\xi}_T)$, through the procedure shown in Figure~\ref{fig:ldm_generation}. Grid blocks are then assigned the appropriate (facies-based) porosity and permeability values used in Section~\ref{quant_assess}. Flow simulations are performed in the geomodel space. Note that, as demonstrated for a 2D case in \cite{DIFEDERICO2025105755}, it is straightforward to include the facies properties (porosity and permeability) in the set of unknowns to be determined during history matching, though this is not considered here.

True data $\mathbf{d}_{\text{true}}\in \mathbb{R}^{N_d}$, where $N_d$ is the number of data points, are obtained by simulating the synthetic true model for each case. The observed data used for history matching, denoted $\mathbf{d}^*_{\text{obs}}\in \mathbb{R}^{N_d}$, include Gaussian noise, intended to represent measurement error. This error is modeled with a zero mean and a diagonal (no space or time correlation) covariance matrix $C_d \in \mathbb{R}^{N_d \times N_d}$. Thus the (noisy) observed data are expressed as $\mathbf{d}^*_{\text{obs}} = \mathbf{d}_{\text{true}} + \mathcal{N}(\mathbf{0}, C_d)$. Here we take the standard deviation of the error to be 2\% of the corresponding simulated value. We split the timeline into a historical period (first 500~days), and a forecast period (last 1000~days). Observations consist of water injection rates at the three injection wells (I1--I3), and oil and water production rates at the six production wells (P1--P6), recorded every 50~days, for a total of $N_d = 150$ observations. 

\textcolor{black}{Before describing the specific history matching procedure applied in this study, we discuss the hierarchical data assimilation problem that is being treated. Recall that we can express the geomodel as $\mathbf{m}=\mathbf{m}(\mathbf{s},\mathbf{r})$, where $\mathbf{s}$ and $\mathbf{r}$ refer to scenario parameters and cell-by-cell facies values, respectively. With this representation, and consistent with, e.g.,~\citet{roininen2016hyperpriorsmaternfieldsapplications} and~\citet{TENG2025104961}, the joint posterior density in a Bayesian framework, denoted $p(\mathbf{s}, \mathbf{r}|\mathbf{d}^*_{\text{obs}})$, is given by
\begin{equation}
p(\mathbf{s}, \mathbf{r}|\mathbf{d}^*_{\text{obs}}) = \frac{ p(\mathbf{d}^*_{\text{obs}}|\mathbf{s}, \mathbf{r}) \, p(\mathbf{s}, \mathbf{r})}{p(\mathbf{d}^*_{\text{obs}})}.
\label{eqn:joint_post}
\end{equation}
Here $p(\mathbf{d}^*_{\text{obs}}|\mathbf{s},\mathbf{r})$ is the likelihood function, $p(\mathbf{s}, \mathbf{r})$ is the prior density, and $p(\mathbf{d}^*_{\text{obs}})$ is the evidence, which acts as a normalization constant. Following~\citep{TENG2025104961}, the joint posterior can be decomposed into the marginal posterior of $\mathbf{s}$ and the conditional posterior of $\mathbf{r}$ given $\mathbf{s}$
\begin{equation}
p(\mathbf{s}, \mathbf{r}|\mathbf{d}^*_{\text{obs}}) = p(\mathbf{s}|\mathbf{d}^*_{\text{obs}}) \, p(\mathbf{r}|\mathbf{s}, \mathbf{d}^*_{\text{obs}}).
\label{eqn:joint_post_decomp}
\end{equation}
In turn, $p(\mathbf{s}|\mathbf{d}^*_{\text{obs}})$ is given by Bayes' rule as
\begin{equation}
p(\mathbf{s}|\mathbf{d}^*_{\text{obs}}) = \frac{p(\mathbf{d}^*_{\text{obs}}|\mathbf{s}) \, p(\mathbf{s})}{p(\mathbf{d}^*_{\text{obs}})},
\label{eqn:bayes_marginal}
\end{equation}
where $p(\mathbf{s})$ is the prior of the scenario parameters. The likelihood in Eq.~\ref{eqn:bayes_marginal} is difficult to compute because it entails marginalization over $\mathbf{r}$, i.e.,
\begin{equation}
p(\mathbf{d}^*_{\text{obs}}|\mathbf{s}) = \int p(\mathbf{d}^*_{\text{obs}}|\mathbf{r}, \mathbf{s}) \, p(\mathbf{r}|\mathbf{s}) \, d\mathbf{r}.
\label{eqn:int_marginal}
\end{equation}
In our parameterization, the variables $(\mathbf{s}, \mathbf{r})$ are replaced with the latent variable $\boldsymbol{\xi}_T$. The data assimilation problem is thus simplified considerably, as we now only need to estimate the posterior distribution $p(\boldsymbol{\xi}_T|\mathbf{d}^*_{\text{obs}})$ instead of $p(\mathbf{s}, \mathbf{r}|\mathbf{d}^*_{\text{obs}})$. In addition, in contrast to \textbf{s} and \textbf{r}, $\boldsymbol{\xi}_T$ is normally distributed.} 

\textcolor{black}{We apply the widely used ensemble smoother with multiple data assimilation (ESMDA) procedure, developed by Emerick and Reynolds~\cite{EMERICK20133}, for history matching, i.e., to sample the posterior $p(\boldsymbol{\xi}_T|\mathbf{d}^*_{\text{obs}})$. ESMDA operates directly on the latent variables. No specialized treatments such as localization are used. Because the relationship between $\boldsymbol{\xi}_T$ and spatial locations is indirect, distance-based localization could be challenging to apply. Correlation-based localization~\cite{LACERDA2019690}, which leverages the sample cross-correlation between latent variables and observed data, could be considered. This did not appear to be necessary for the examples in this paper.} 

In ESMDA, an ensemble of $\boldsymbol{\xi}_T$ fields is updated (superscript $u$) at every iteration based on the mismatch between observed data $\mathbf{d}_{\text{obs}}^*$ and simulated data $\mathbf{d}$, obtained by performing flow simulation for each geomodel in the ensemble. The update equation is
\begin{equation}
    \boldsymbol{\xi}^{u,j}_{T} = \boldsymbol{\xi}^{j}_{T} + C_{\xi d} (C_{dd} + \alpha C_d)^{-1}(\mathbf{d}^{j} - \mathbf{d}^*_\text{obs} ), \quad \text{ for } j = 1, \dots, N_e,
\end{equation}
where $N_e$ is the number of ensemble members, $C_{\xi d} \in \mathbb{R}^{n_c \times N_d}$ is the cross-covariance matrix between $\boldsymbol{\xi}_{T}$ and $\mathbf{d}$, and $C_{d d} \in \mathbb{R}^{N_d \times N_d}$ is the autocovariance of $\mathbf{d}$. The value $\alpha$ is an inflation coefficient, which must satisfy $\sum_{i=1}^{N_a} {\alpha_i}^{-1} = 1$. Here we set $N_e = 200$, $N_a =10$, and $\alpha_i$ values to [57.017, 35.0, 25.0, 20.0, 18.0, 15.0, 12.0, 8.0, 5.0, 3.0], as suggested by~\citet{EMERICK20133}. \textcolor{black}{Since the ensemble members are independent, simulations can be performed in parallel. We simulate these models in batches of 20, which results in an elapsed time of approximately 1.5~hours per ESMDA iteration, or about 15~hours total for a full history matching run.}

The discrete, non-Gaussian nature of the facies models considered here would present challenges for ESMDA applied directly in the geomodel space (i.e., with no parameterization). For example, directly updating individual grid blocks over data assimilation steps may not preserve the continuity or geological character of the geological structures (though localization can be used to mitigate this effect). The use of the 3D-LDM parameterization addresses this concern since the latent variable $\boldsymbol{\xi}_T$ is normally distributed, consistent with ESMDA assumptions, and the 3D-LDM mapping maintains geological realism after latent variable updates. 

\textcolor{black}{On the other hand, it should be noted that the parameterization introduces an additional nonlinear mapping step between the geological model and simulated data. For discrete-facies models, this can result in a discontinuity between scenario parameters and grid-block facies values. As a consequence, $C_{\xi d}$ may only approximate the relationship between $\boldsymbol{\xi}_T$ and $\mathbf{d}$. Nevertheless, the results presented below confirm the effectiveness of the 3D-LDM-based ESMDA procedure for our cases. This effective performance may be related to the fact that the data used here are well rates, which involve summation over multiple geological layers. Thus, local changes in facies type may have a relatively small impact on \textbf{d} (recall also that we condition to hard data at well locations). In future work, we plan to evaluate the performance of 3D-LDM with other history matching methods, such as Markov chain Monte Carlo.}

\subsection{History matching results for Case~1}

\begin{figure}[h]
    \centering
    \begin{subfigure}[b]{0.3\textwidth}
        \centering
        \includegraphics[width=\textwidth, trim=50 50 50 50, clip]{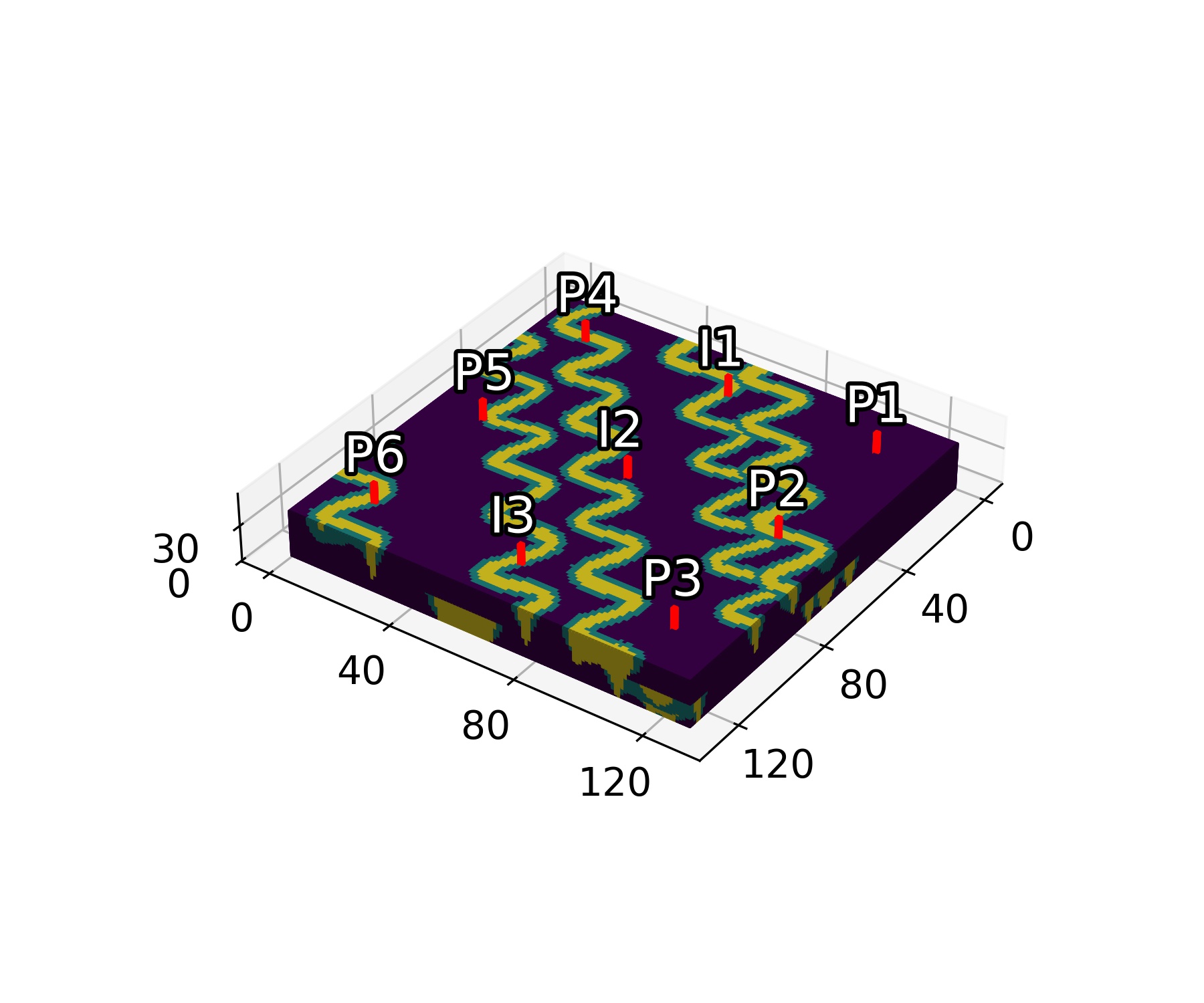}
        \caption{True 1}
        \label{fig:true_top}
    \end{subfigure}
    \hfill
    \begin{subfigure}[b]{0.3\textwidth}
        \centering
        \includegraphics[width=\textwidth, trim=50 50 50 50, clip]{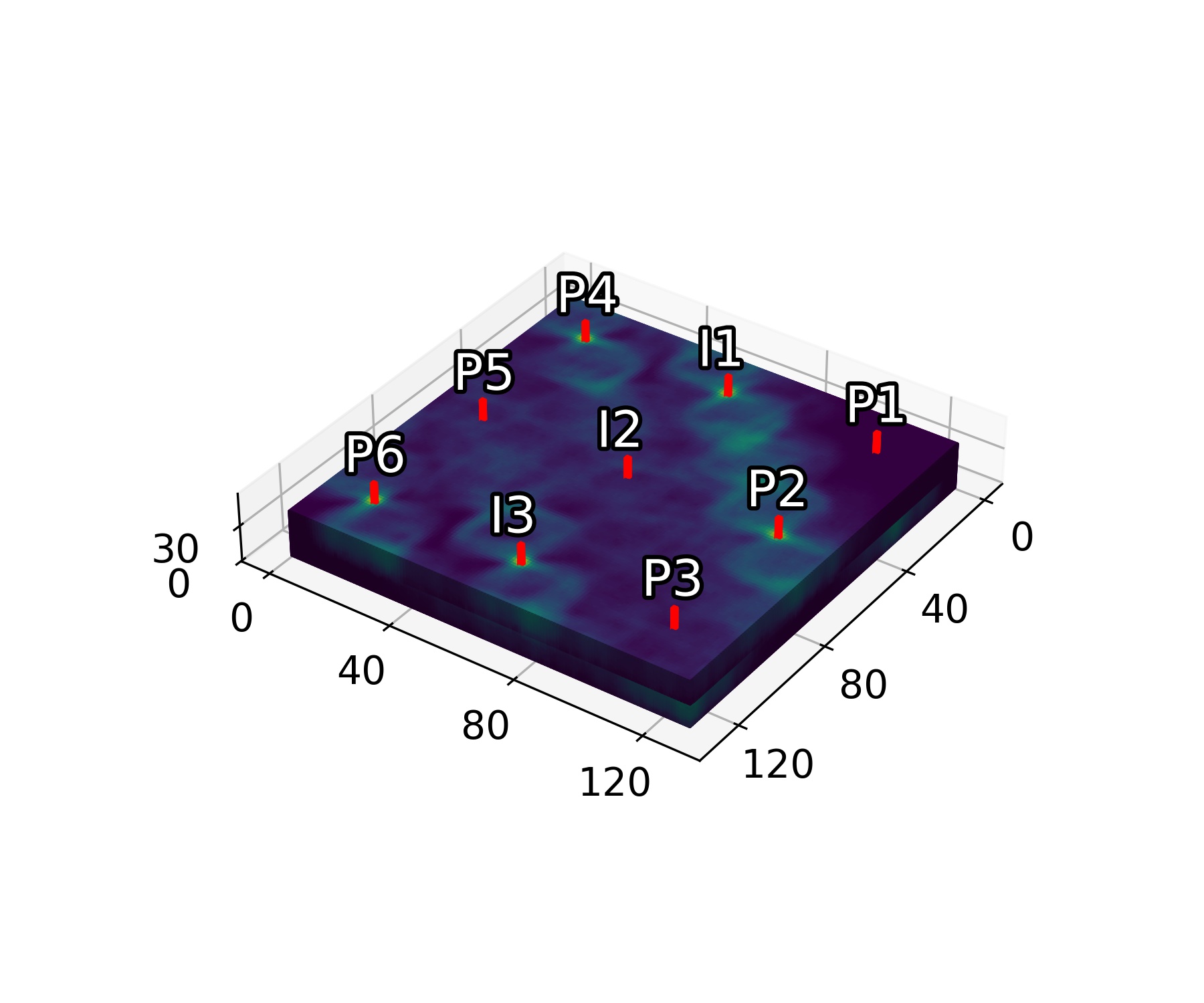}
        \caption{Prior mean}
        \label{fig:prior_mean_top}
    \end{subfigure}
    \hfill
    \begin{subfigure}[b]{0.3\textwidth}
        \centering
        \includegraphics[width=\textwidth, trim=50 50 50 50, clip]{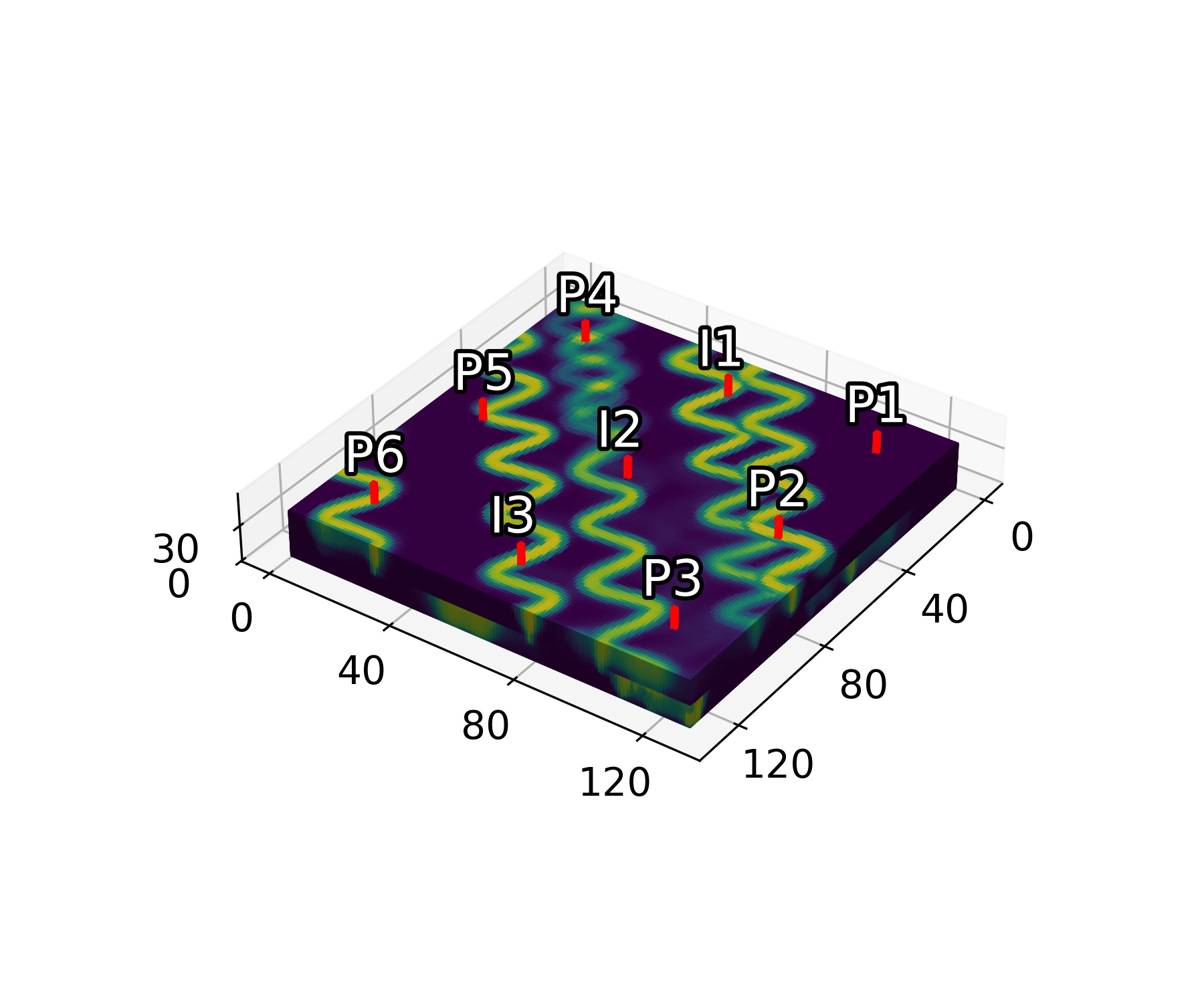}
        \caption{Post.~mean}
        \label{fig:post_mean_top}
    \end{subfigure}

    \begin{subfigure}[b]{0.3\textwidth}
        \centering
        \includegraphics[width=\textwidth, trim=50 50 50 50, clip]{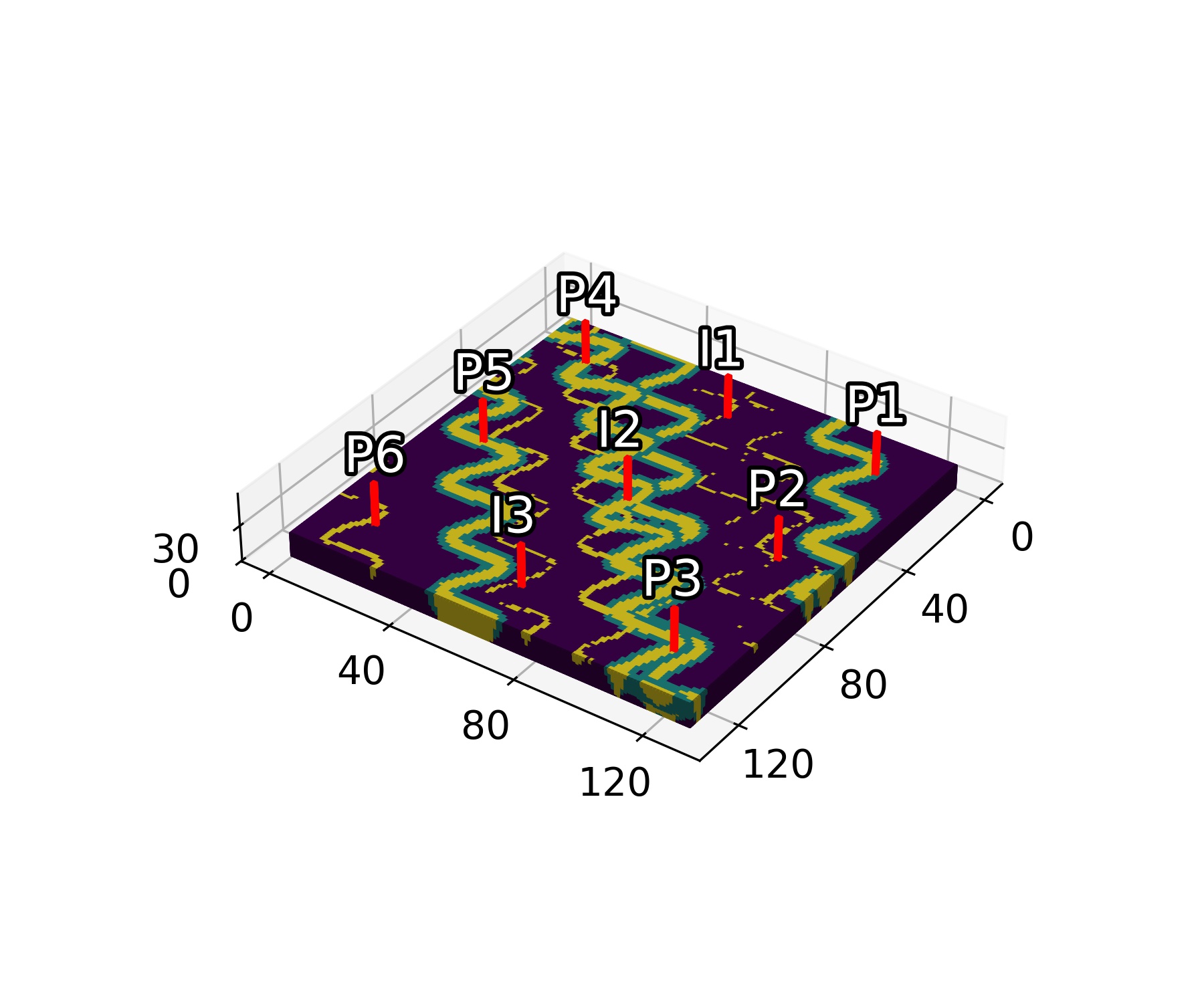}
        \caption{True 1 (lower half)}
        \label{fig:true_bottom1}
    \end{subfigure}
    \hfill
    \begin{subfigure}[b]{0.3\textwidth}
        \centering
        \includegraphics[width=\textwidth, trim=50 50 50 50, clip]{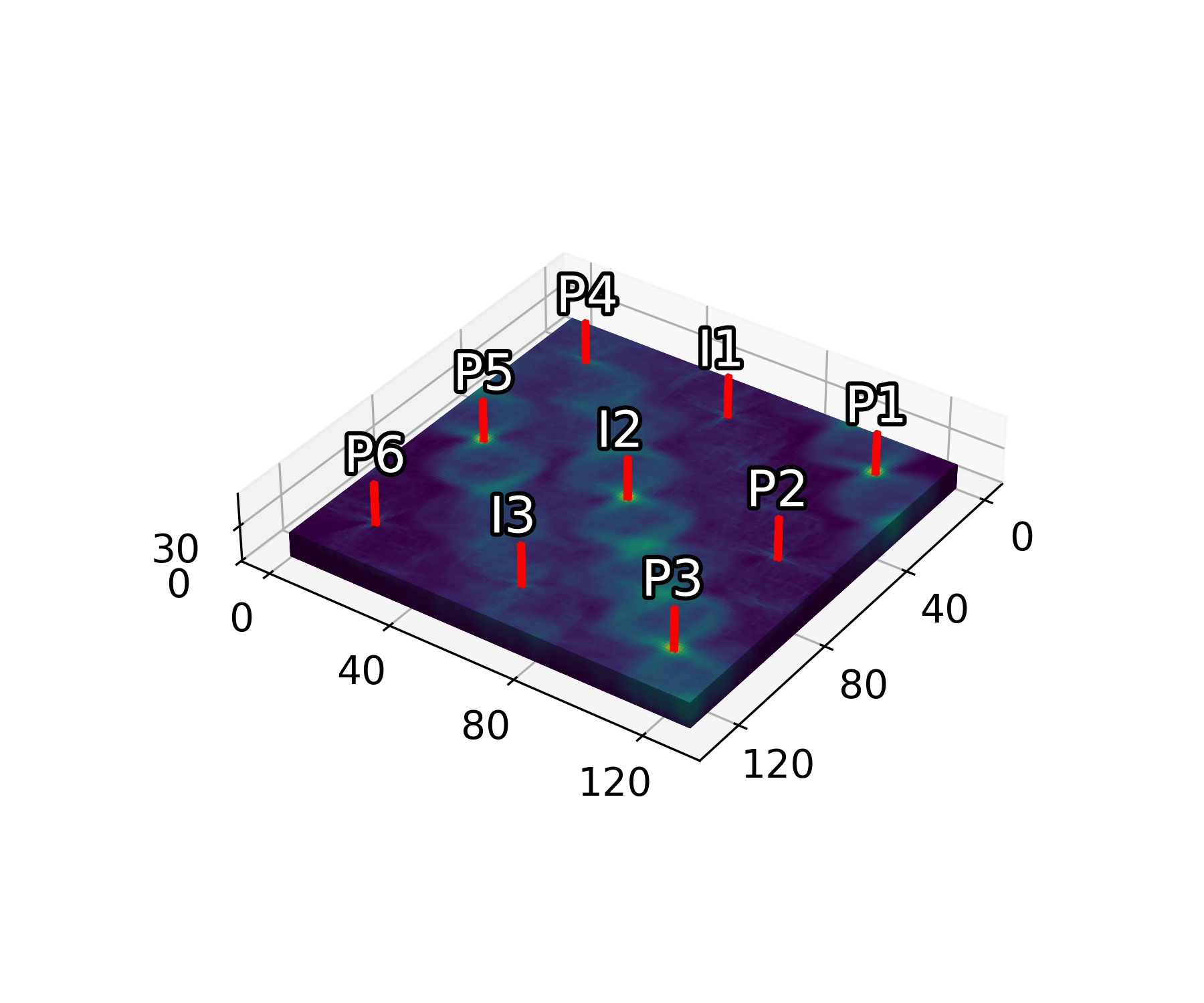}
        \caption{Prior mean (lower half)}
        \label{fig:prior_mean_bottom}
    \end{subfigure}
    \hfill
    \begin{subfigure}[b]{0.3\textwidth}
        \centering
        \includegraphics[width=\textwidth, trim=50 50 50 50, clip]{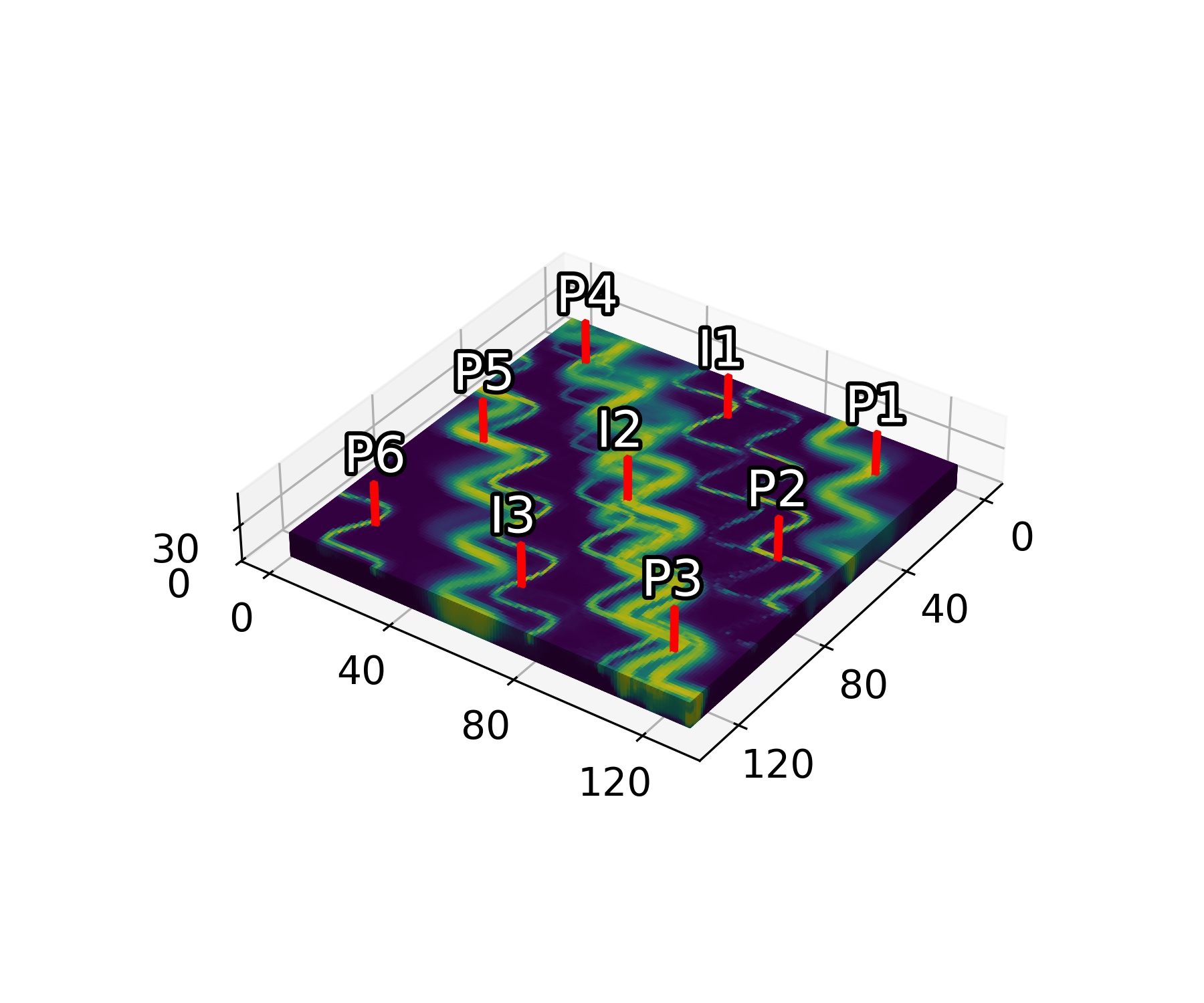}
        \caption{Post.~mean (lower half)}
        \label{fig:post_mean_bottom}
    \end{subfigure}
    \caption{Case 1: (a-c) show the synthetic true model, prior and posterior ensemble means, while (d-f) show the lower halves of the same models.}
    \label{fig:case1_models}
\end{figure}

We now present history matching results for different synthetic true models. The true model for Case~1 is shown in Figure~\ref{fig:case1_models}a and Figure~\ref{fig:case1_models}d. The lower half view in Figure~\ref{fig:case1_models}d, which corresponds to layers~17-32, indicates that some of the features evident in Figure~\ref{fig:case1_models}a do not extend through the full model thickness.

\begin{figure}[h!]
    \centering

    \begin{subfigure}[b]{0.45\textwidth}
        \centering
        \includegraphics[width=\textwidth]{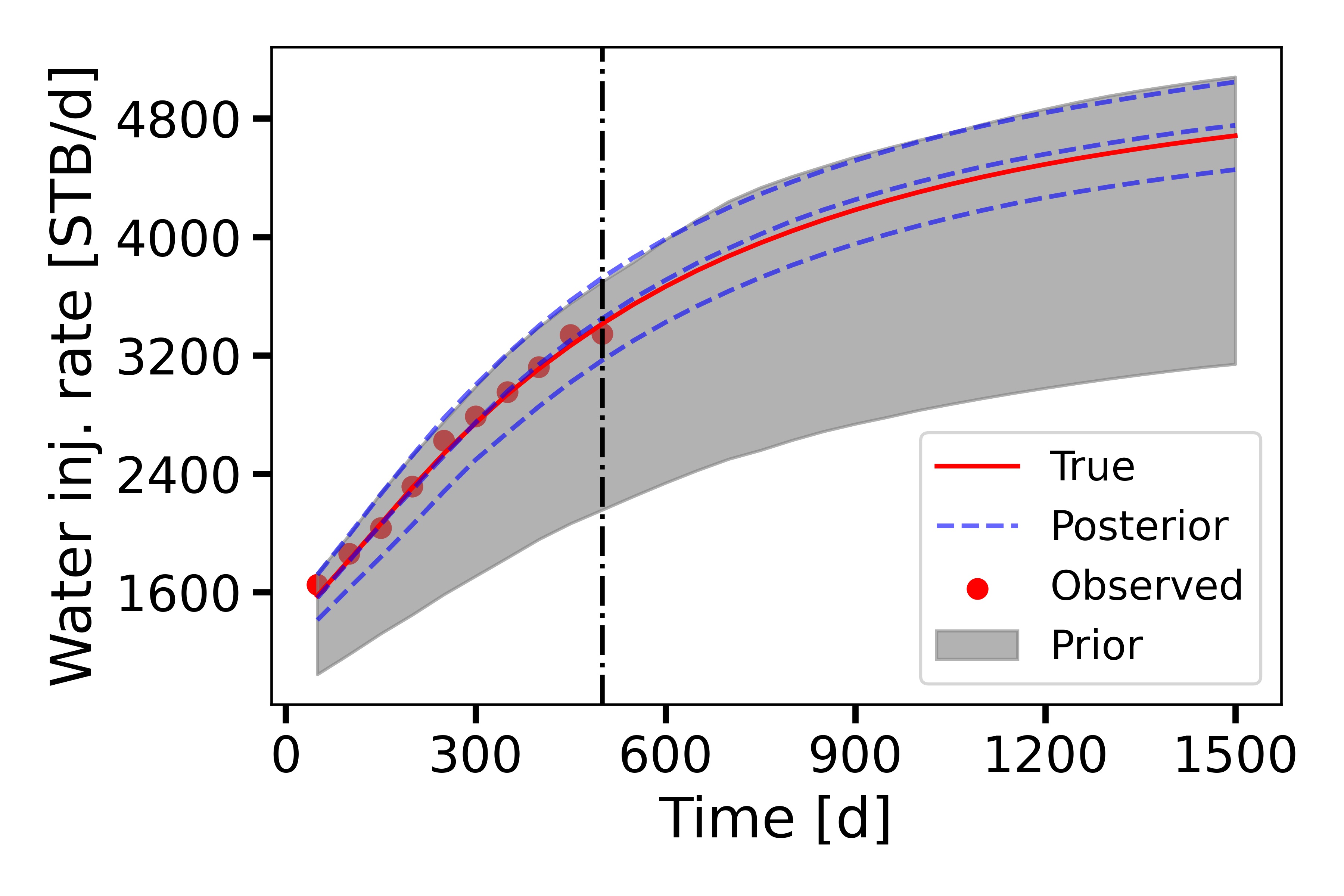}
        \caption{I1 water injection rate}
        \label{fig:I1_WAT}
    \end{subfigure}
    \hfill
    \begin{subfigure}[b]{0.45\textwidth}
        \centering
        \includegraphics[width=\textwidth]{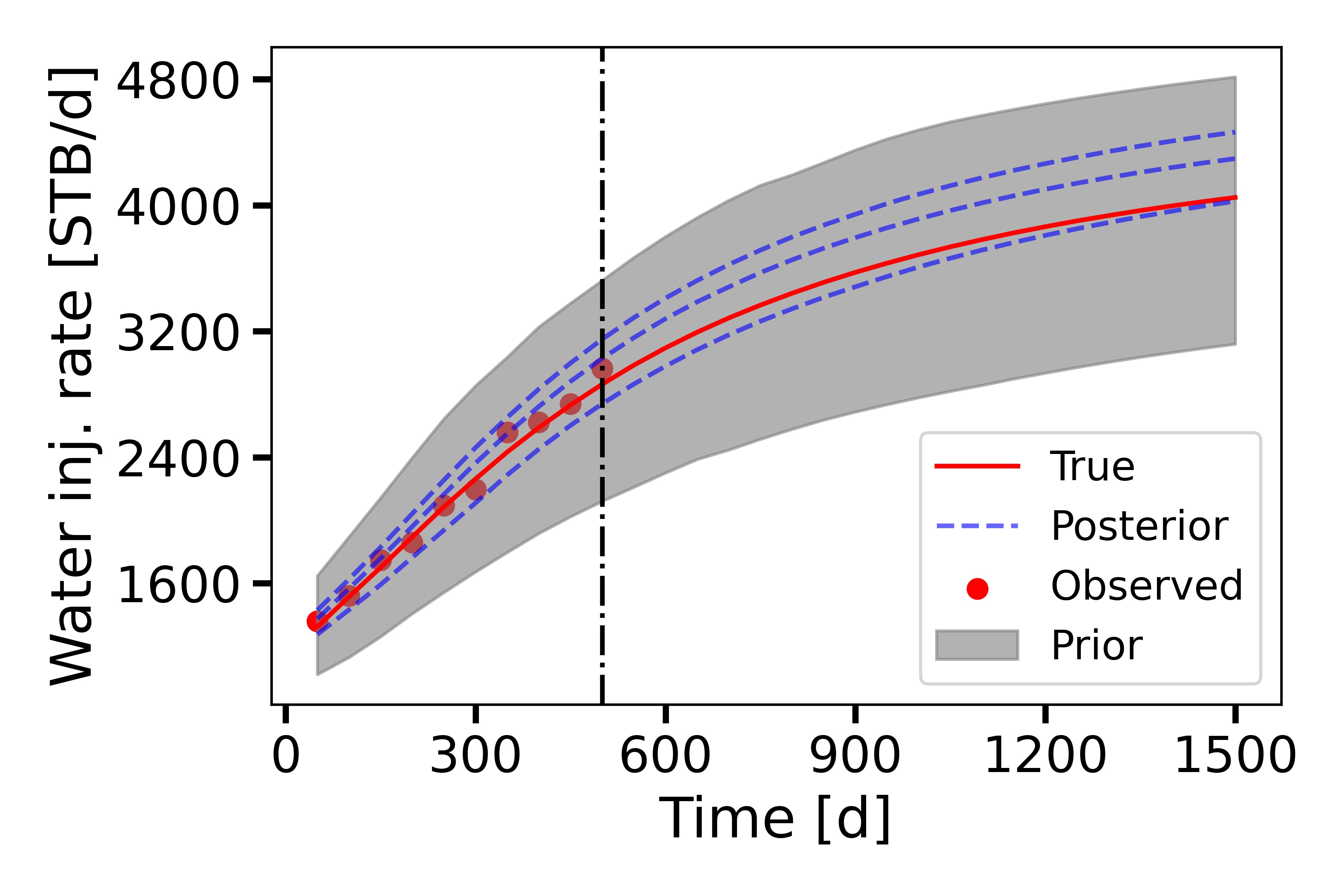}
        \caption{I2 water injection rate}
        \label{fig:I3_WAT}
    \end{subfigure}

    \vspace{1em}

    \begin{subfigure}[b]{0.45\textwidth}
        \centering
        \includegraphics[width=\textwidth]{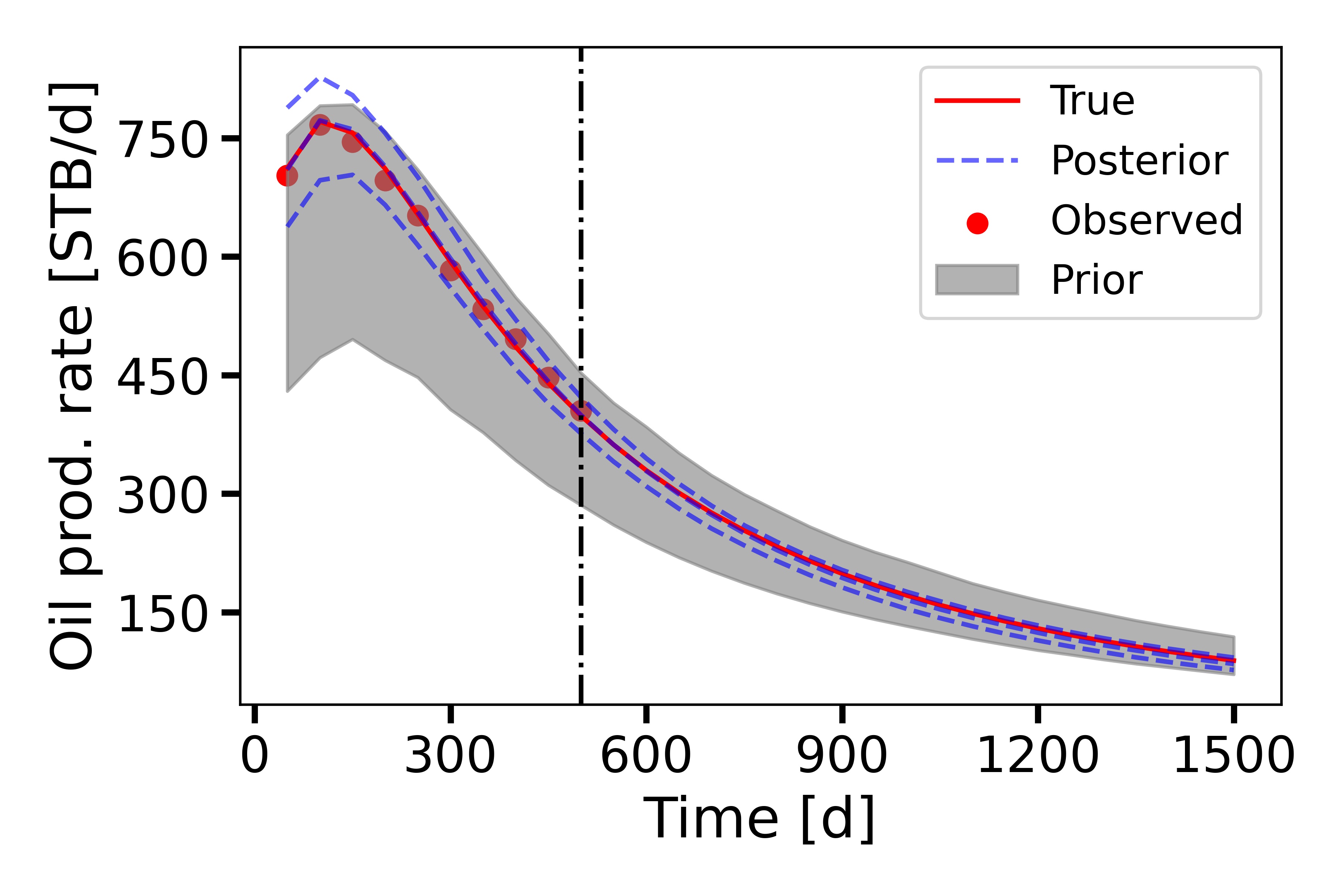}
        \caption{P4 oil production rate}
        \label{fig:P4_OIL}
    \end{subfigure}
    \hfill
    \begin{subfigure}[b]{0.45\textwidth}
        \centering
        \includegraphics[width=\textwidth]{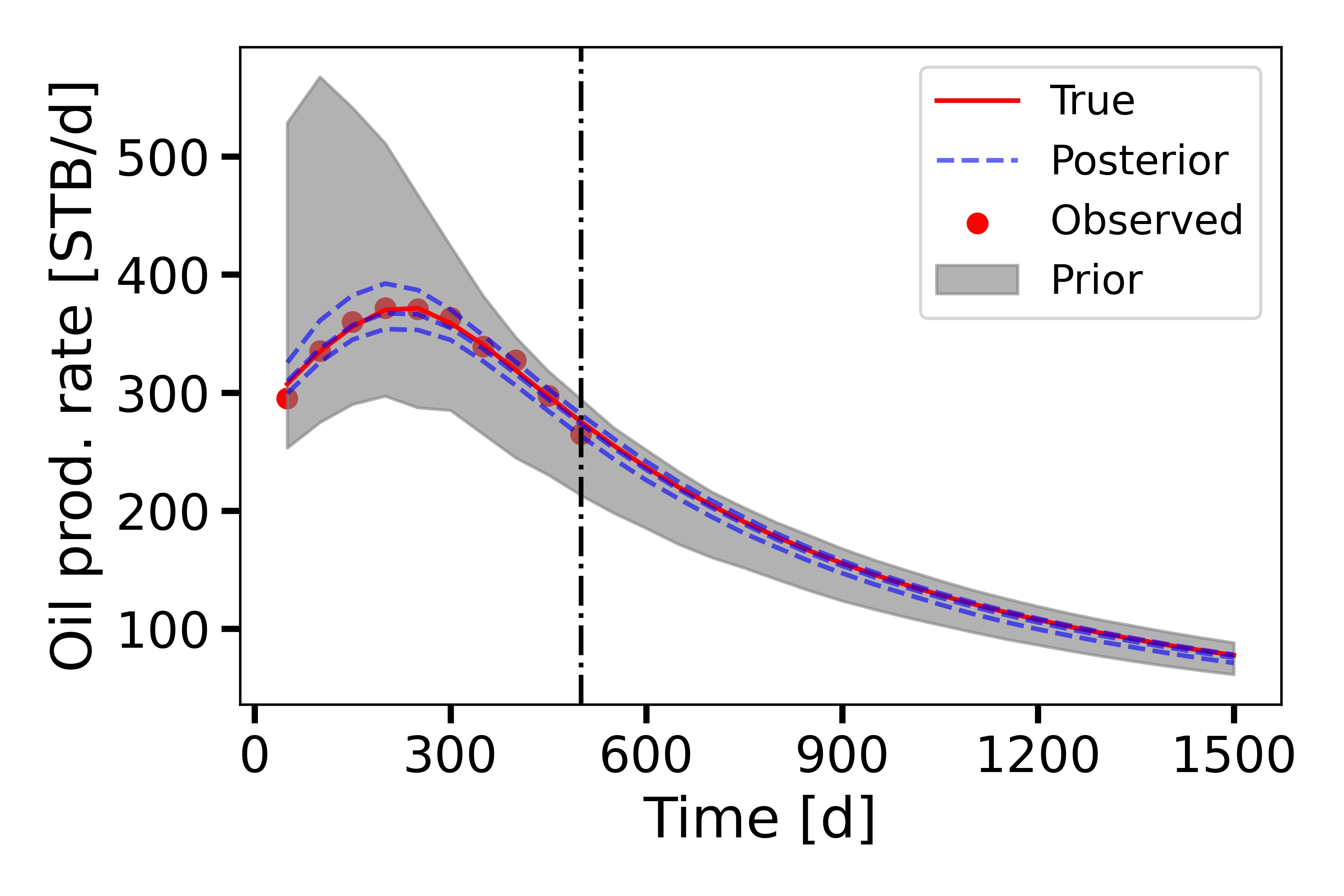}
        \caption{P6 oil production rate}
        \label{fig:P6_OIL}
    \end{subfigure}

    \vspace{1em}

    \begin{subfigure}[b]{0.45\textwidth}
        \centering
        \includegraphics[width=\textwidth]{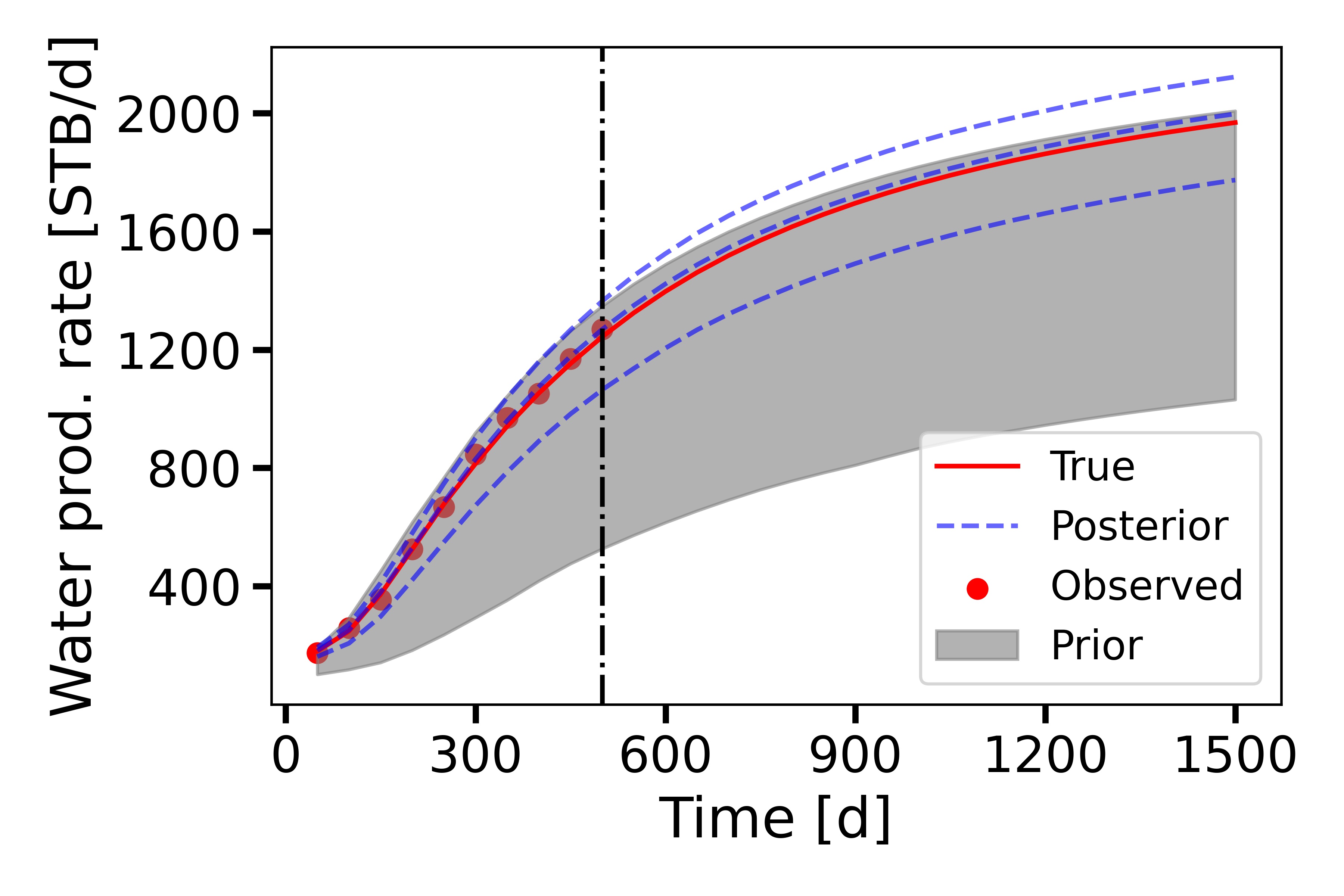}
        \caption{P2 water production rate}
        \label{fig:P1_WAT}
    \end{subfigure}
    \hfill
    \begin{subfigure}[b]{0.45\textwidth}
        \centering
        \includegraphics[width=\textwidth]{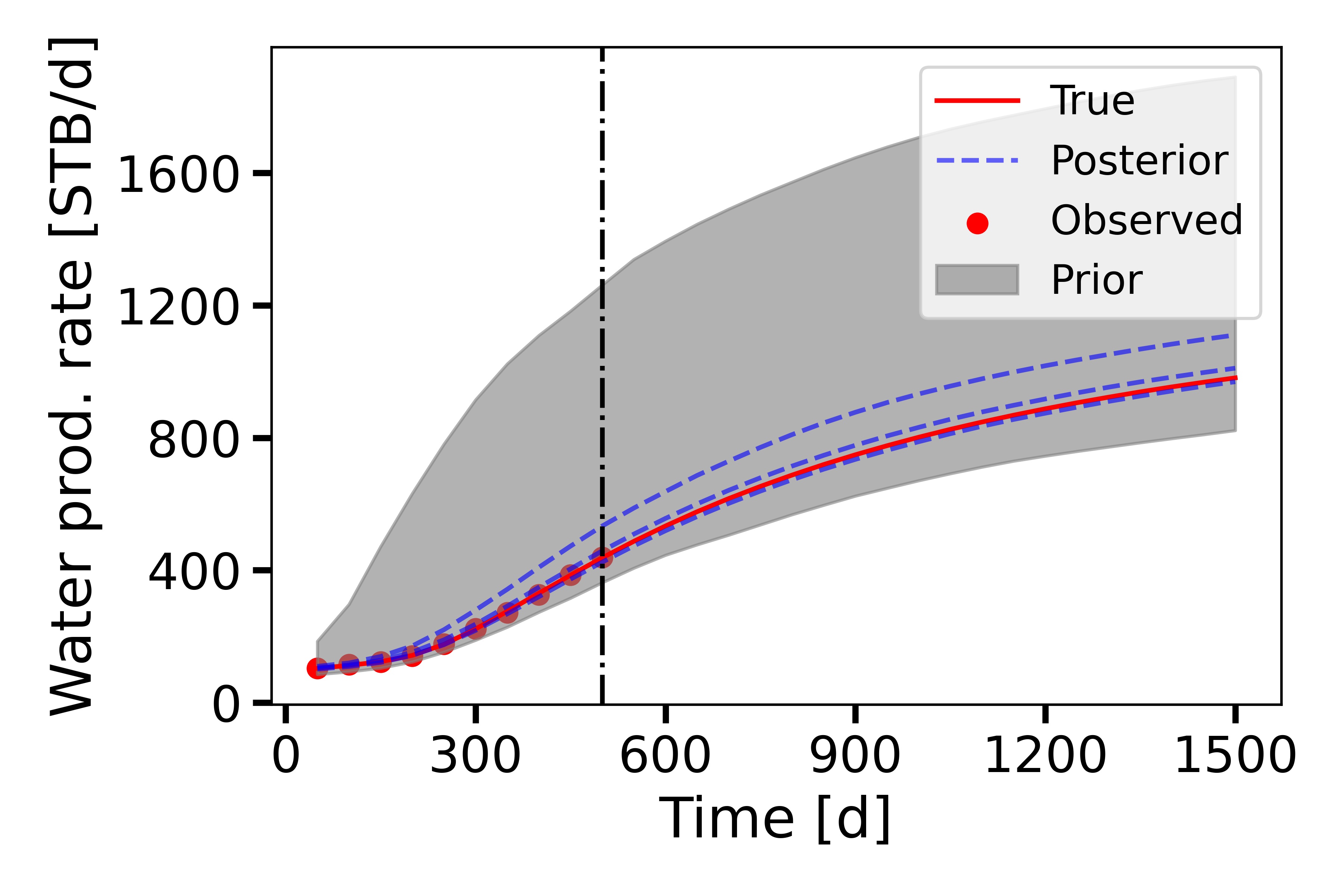}
        \caption{P5 water production rate}
        \label{fig:P5WAT}
    \end{subfigure}

    \caption{Case 1: history matching results for selected wells. Gray regions show the prior P$_{10}$–P$_{90}$ range, blue dashed lines denote the posterior P$_{10}$, P$_{50}$, P$_{90}$ curves, and red points and red curves represent observed and true data. The vertical black dot-dash line indicates the end of the history matching period.}
    \label{fig:hm_rates}
\end{figure}

Prior and posterior flow results, for water injection, oil production, and water production from representative wells, are shown in Figure~\ref{fig:hm_rates}. The gray region in each plot indicates the prior P$_{10}$-P$_{90}$ (percentile) range, and the blue dashed curves show the P$_{10}$, P$_{50}$, and P$_{90}$ curves across the posterior ensemble. The historical and forecast periods are separated by the vertical black line at 500~days. The data points used as observations appear as red dots, while the red curve shows the true model simulation result (without any noise). The narrower ranges from prior to posterior ensemble results reflect the substantial degree of uncertainty reduction achieved in this case. Furthermore, for all wells, the observed data lie within (or very near) the posterior P$_{10}$--P$_{90}$ ranges for the forecast period. These results indicate the improved forecast accuracy after history matching. Analogous behavior is also observed for other flow rate quantities (not shown for conciseness).

We now consider the posterior geomodels themselves. Figure~\ref{fig:case1_models}b and Figure~\ref{fig:case1_models}e show the full model and lower half view of the mean across the prior ensemble of geomodels. High color intensities (yellow, green) indicate high probability of channel/levee facies, while lower intensities (dark blue) indicate high probability of mud facies. Except for the well locations, where conditioning is honored, the prior mean model corresponds to high uncertainty in the channel locations and, correspondingly, in the scenario parameters. The posterior mean, achieved after $N_a$ ESMDA iterations, is shown in Figure~\ref{fig:case1_models}c and Figure~\ref{fig:case1_models}f. Here we see much sharper channel features, which closely resemble the true model. Some ``blurring'' of the posterior mean relative to the true model is evident, however, since a degree of uncertainty remains after history matching. 

\begin{figure}
    \centering

    \begin{subfigure}[b]{0.49\textwidth}
        \centering
        \includegraphics[width=\textwidth]{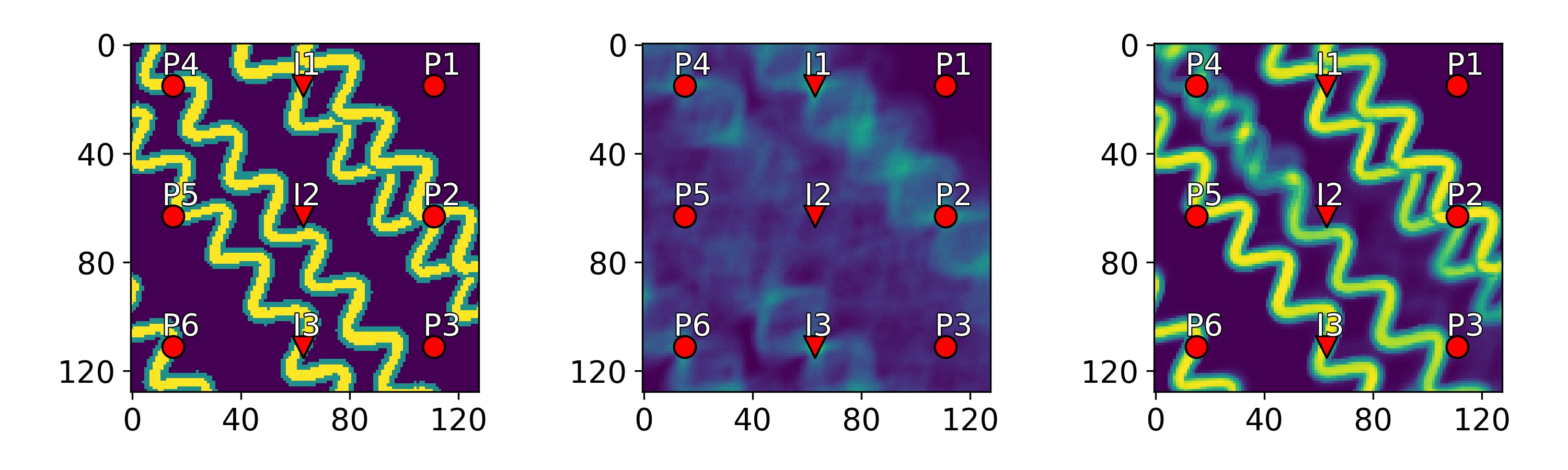}
        \caption{Layer~1}
        \label{fig:means_z1}
    \end{subfigure}
    \hfill
    \begin{subfigure}[b]{0.49\textwidth}
        \centering
        \includegraphics[width=\textwidth]{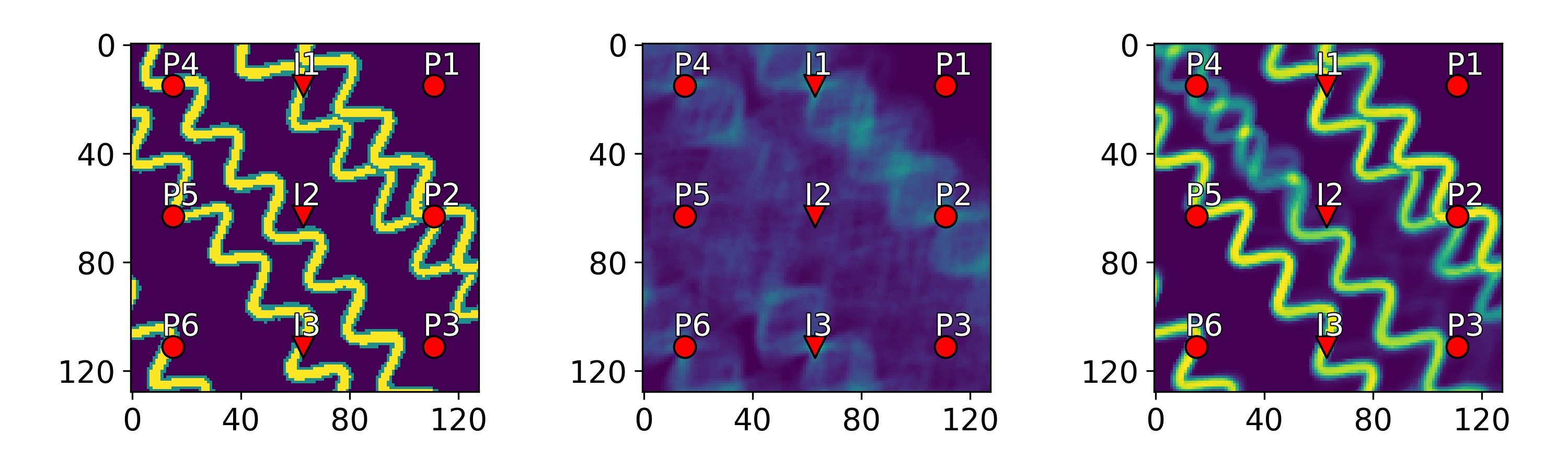}
        \caption{Layer~8}
        \label{fig:means_z8}
    \end{subfigure}

    \vspace{1em}

    \begin{subfigure}[b]{0.49\textwidth}
        \centering
        \includegraphics[width=\textwidth]{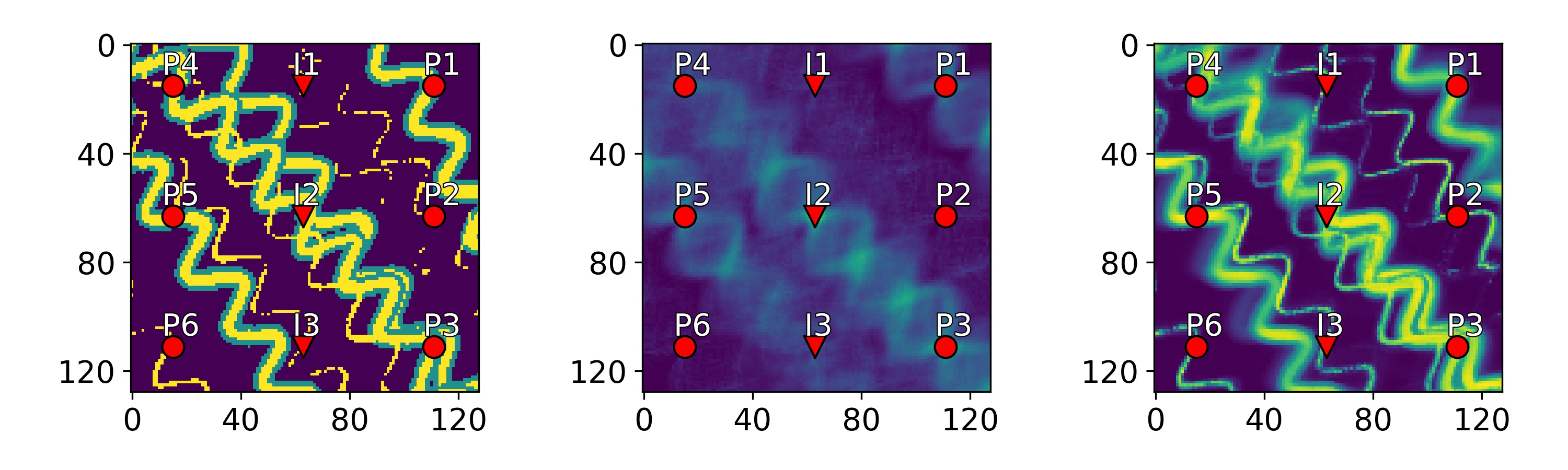}
        \caption{Layer~16}
        \label{fig:means_z18}
    \end{subfigure}
    \hfill
    \begin{subfigure}[b]{0.49\textwidth}
        \centering
        \includegraphics[width=\textwidth]{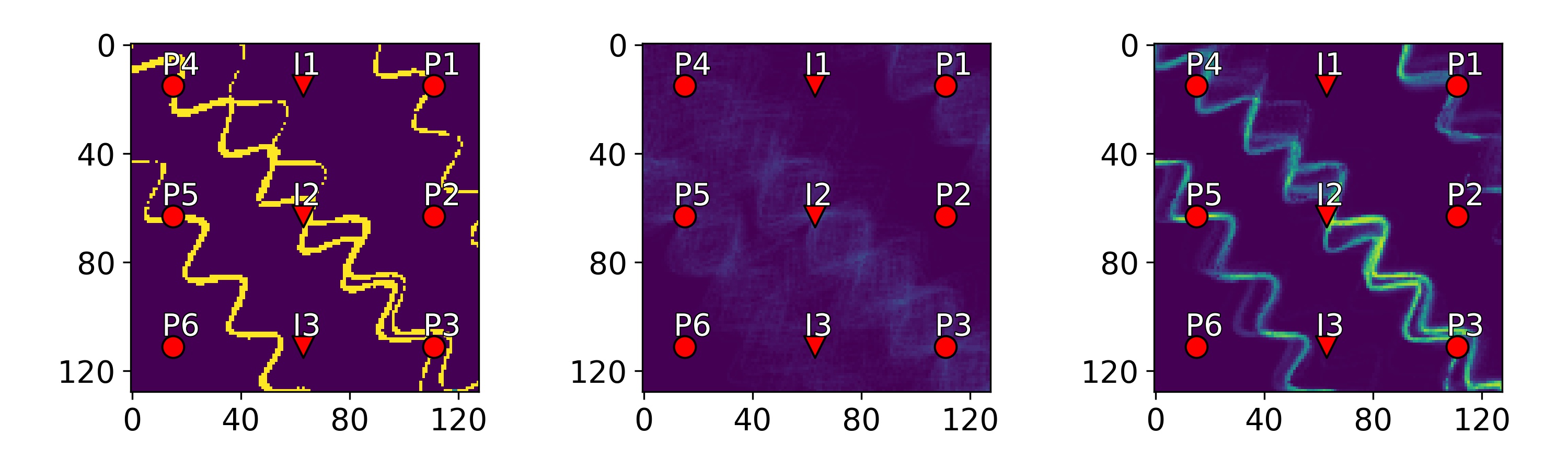}
        \caption{Layer~32}
        \label{fig:means_z32}
    \end{subfigure}

    \caption{Case 1: true (left), prior mean (center), and posterior mean (right) at layers~1, 8, 16 and 32 across the ensemble of realizations.}
    \label{fig:mean_layers}
\end{figure}

The true field, prior mean, and posterior mean are presented in Figure~\ref{fig:mean_layers} for four different layers. These areal views are for layers~1 and~8 in the top portion of the model, and layers~16 and~32 in the middle and bottom of the model. The layer-by-layer results in Figure~\ref{fig:mean_layers} are consistent with the observations for the 3D representations in Figure~\ref{fig:case1_models}, again illustrating that uncertainty reduction is achieved and geological realism is maintained.

\begin{figure}[h]
    \centering

    \begin{subfigure}[b]{0.23\textwidth}
        \includegraphics[width=\textwidth, trim=50 50 50 50, clip]{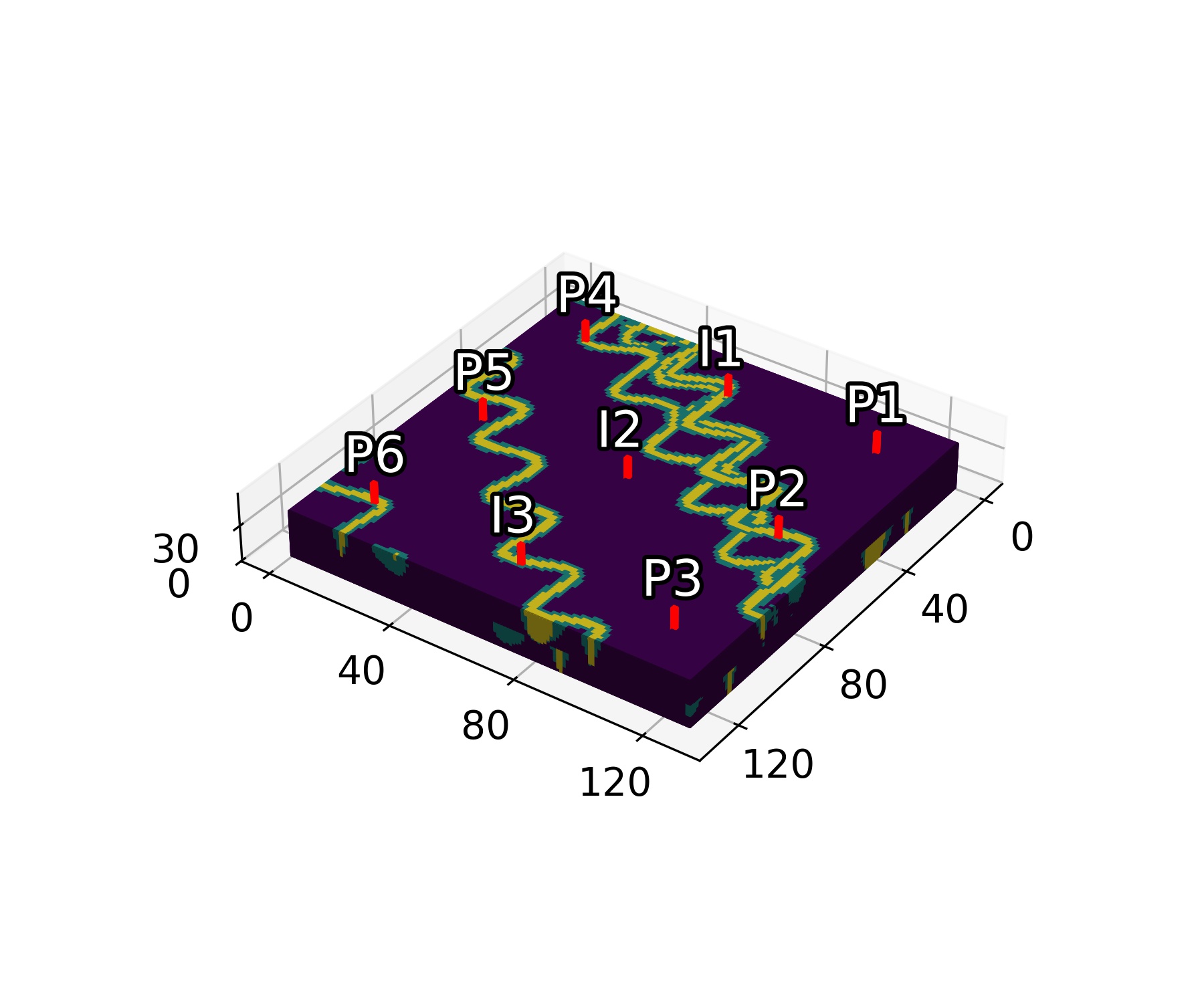}
        \caption{Prior 1}
    \end{subfigure}
    \hfill
    \begin{subfigure}[b]{0.23\textwidth}
        \includegraphics[width=\textwidth, trim=50 50 50 50, clip]{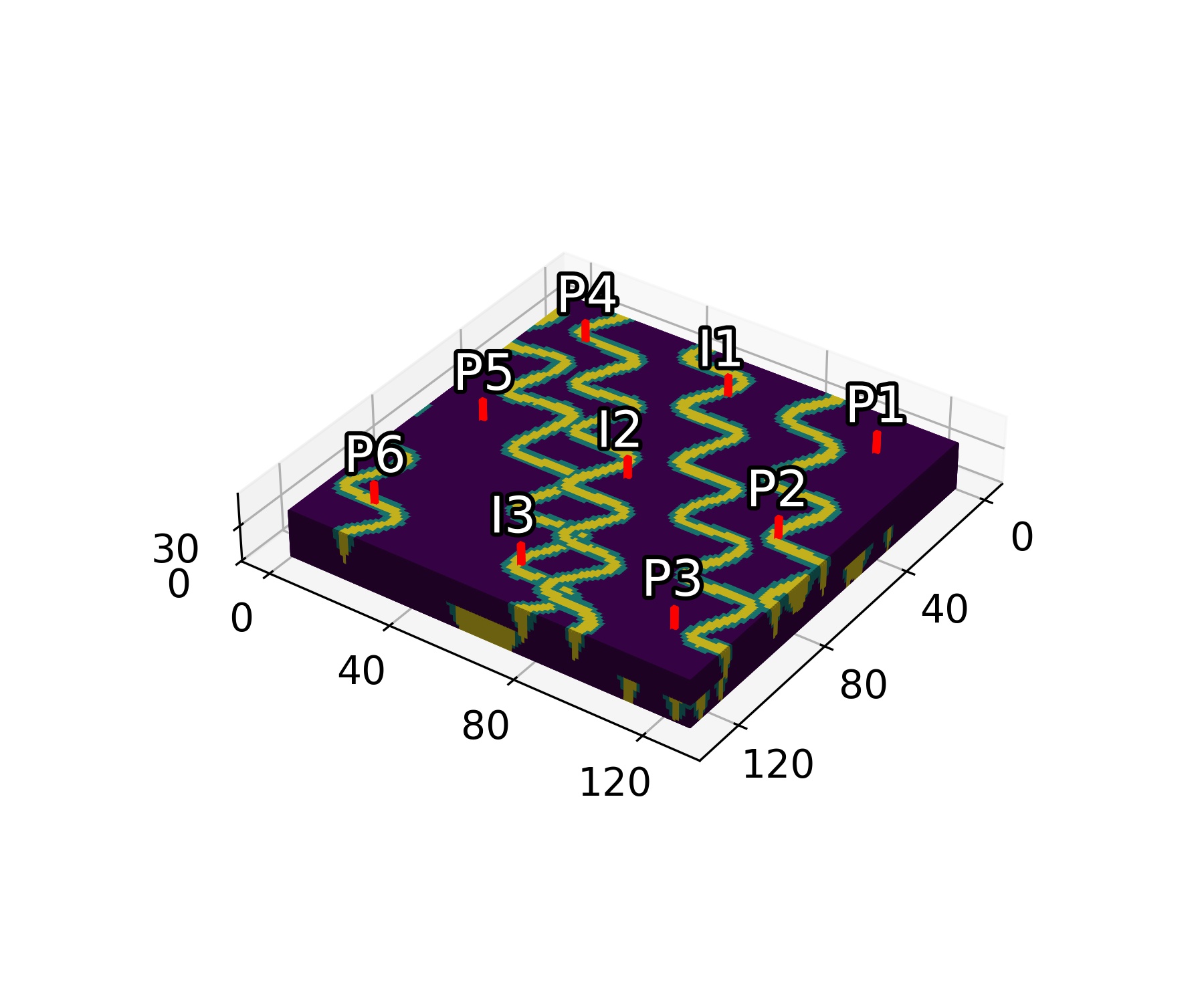}
        \caption{Prior 2}
    \end{subfigure}
    \hfill
    \begin{subfigure}[b]{0.23\textwidth}
        \includegraphics[width=\textwidth, trim=50 50 50 50, clip]{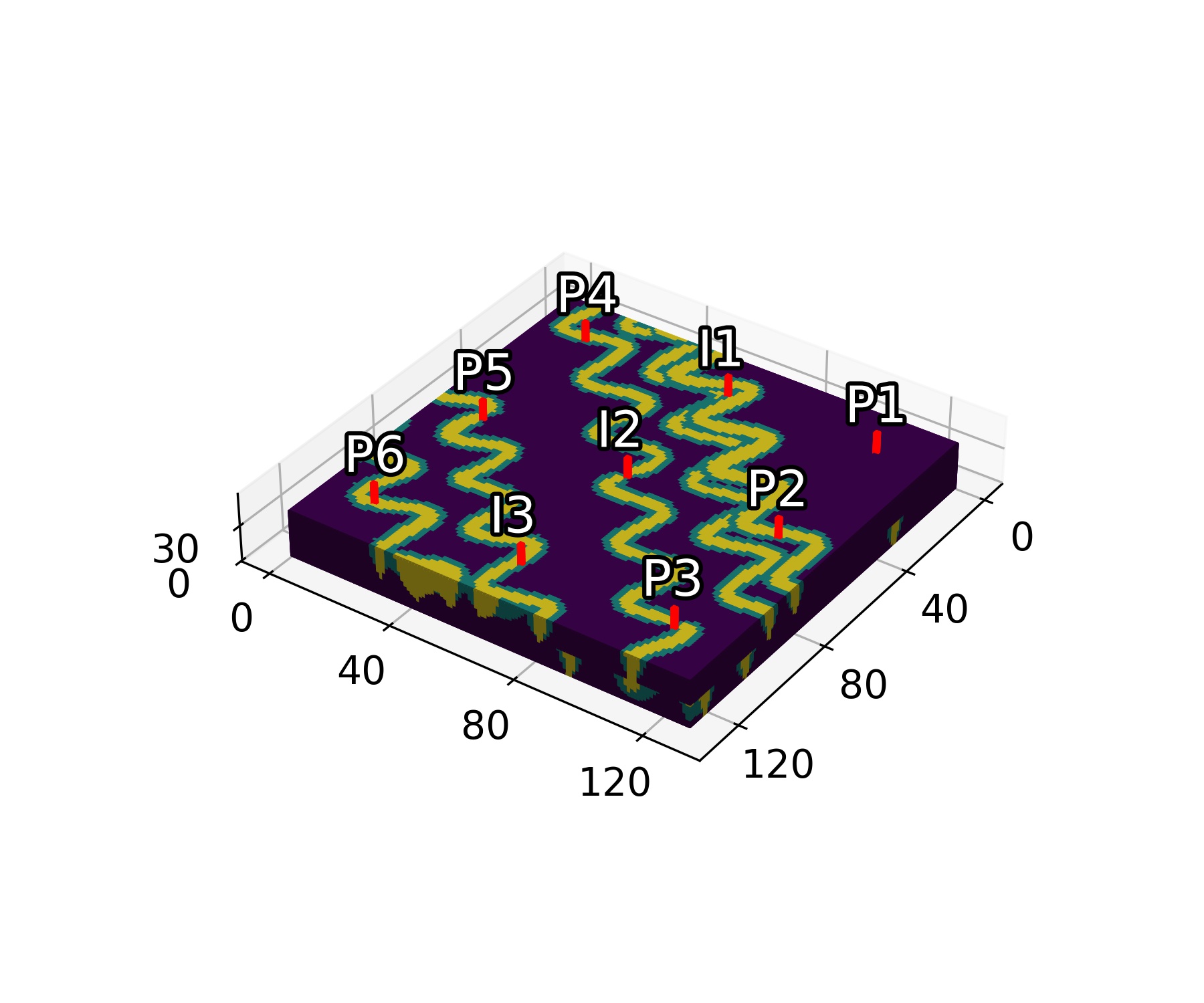}
        \caption{Prior 3}
    \end{subfigure}
    \hfill
    \begin{subfigure}[b]{0.23\textwidth}
        \includegraphics[width=\textwidth, trim=50 50 50 50, clip]{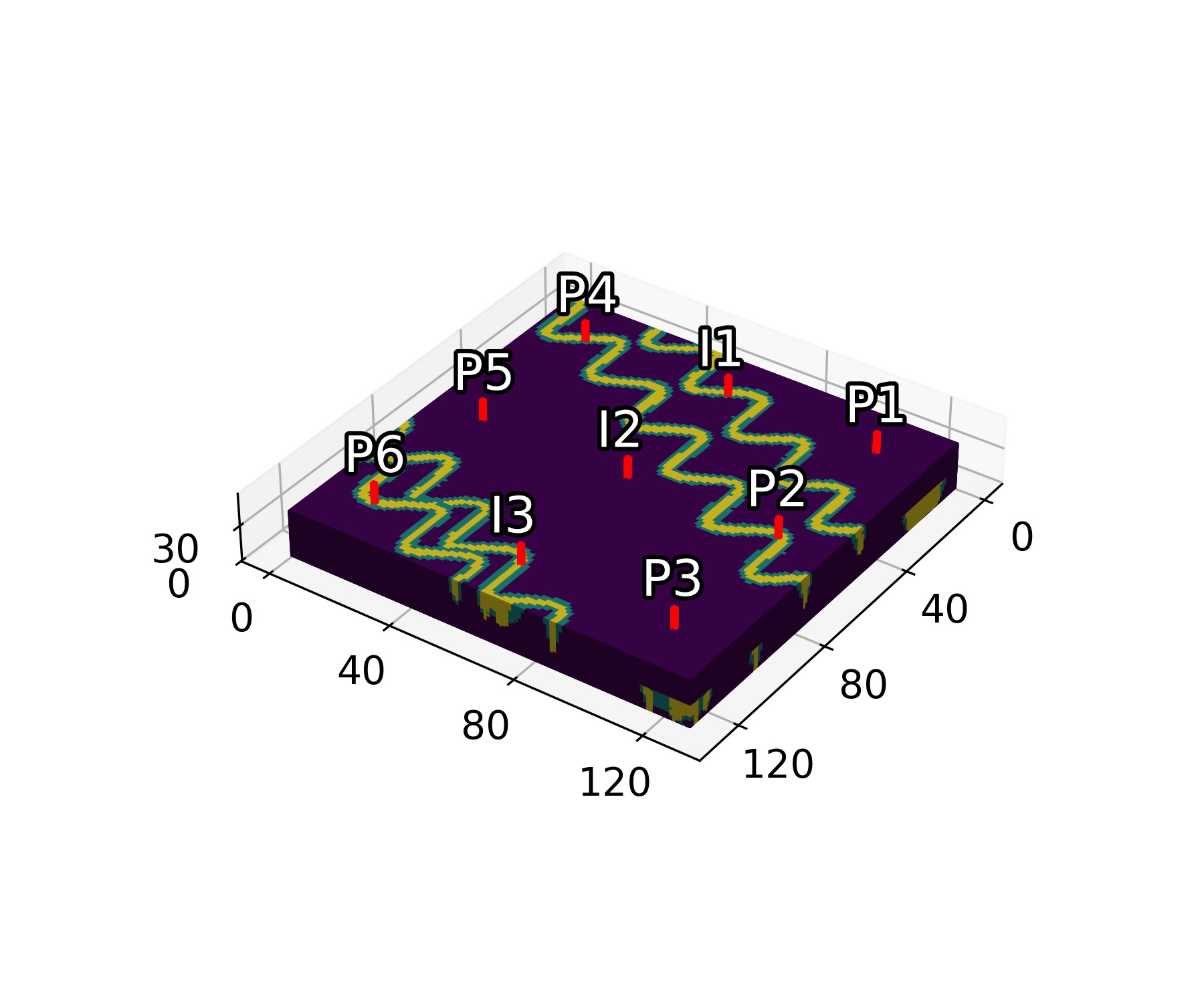}
        \caption{Prior 4}
    \end{subfigure}

    \vspace{1em}

    \begin{subfigure}[b]{0.23\textwidth}
        \includegraphics[width=\textwidth, trim=50 50 50 50, clip]{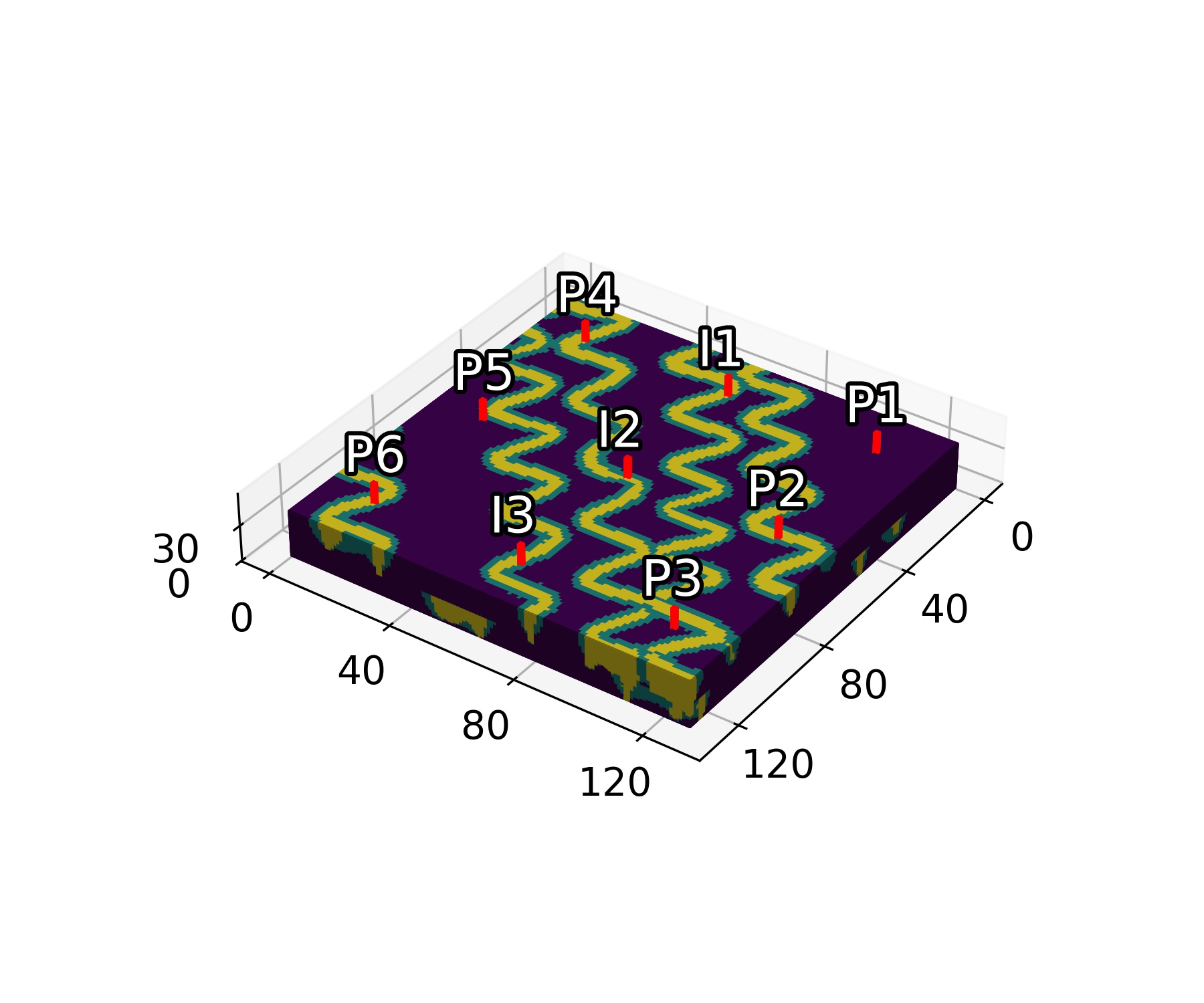}
        \caption{Post.~1}
    \end{subfigure}
    \hfill
    \begin{subfigure}[b]{0.23\textwidth}
        \includegraphics[width=\textwidth, trim=50 50 50 50, clip]{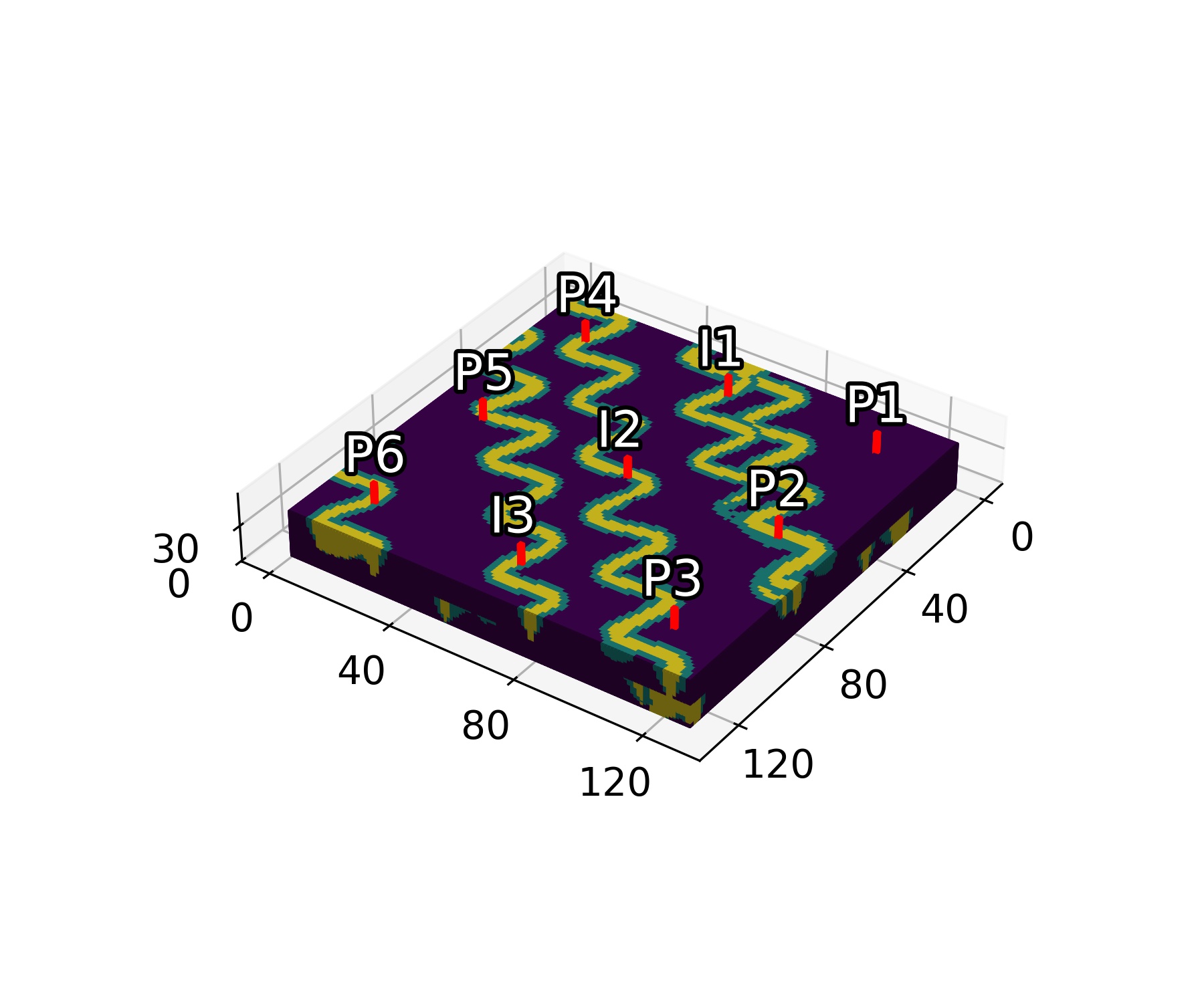}
        \caption{Post.~2}
    \end{subfigure}
    \hfill
    \begin{subfigure}[b]{0.23\textwidth}
        \includegraphics[width=\textwidth, trim=50 50 50 50, clip]{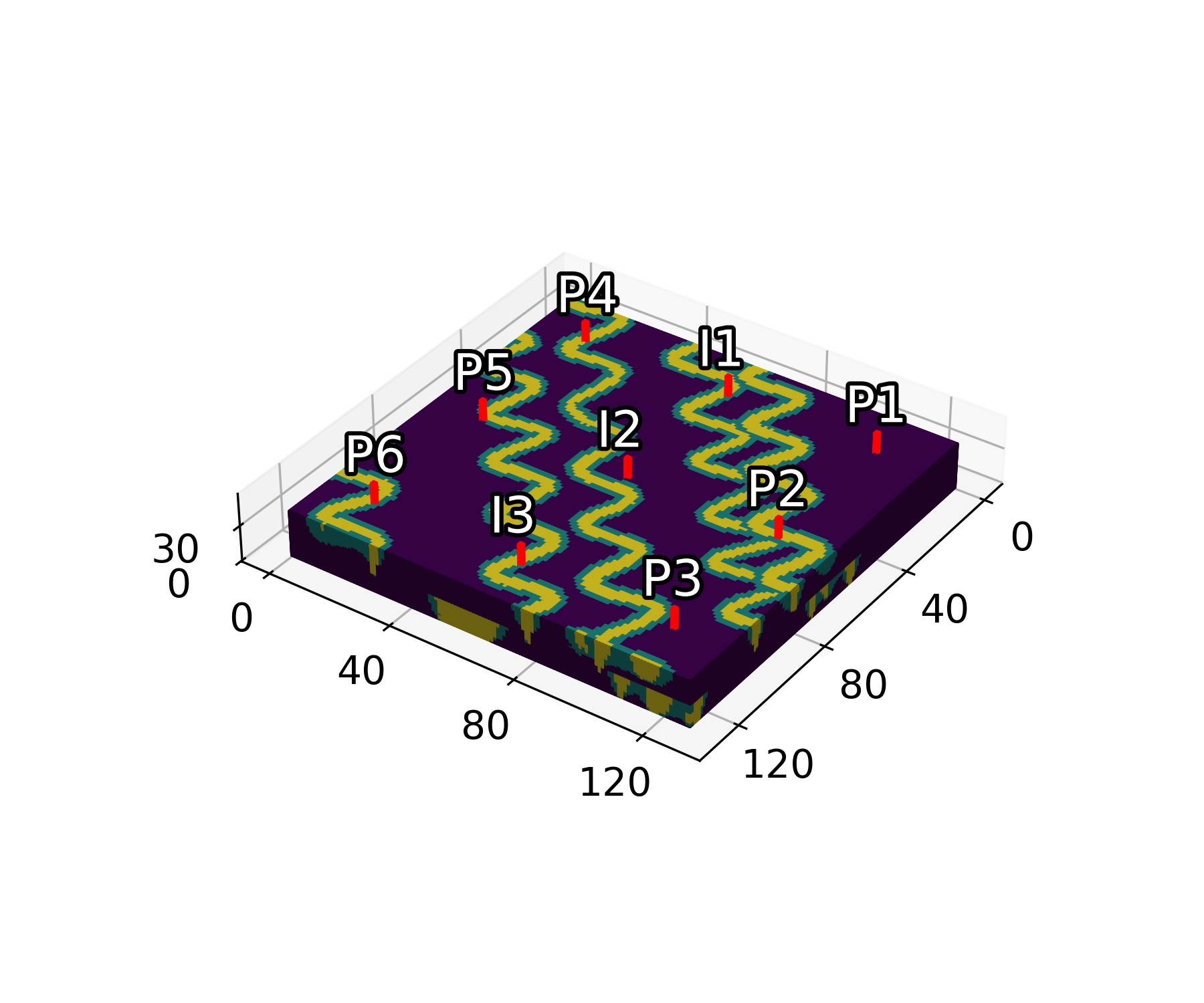}
        \caption{Post.~3}
    \end{subfigure}
    \hfill
    \begin{subfigure}[b]{0.23\textwidth}
        \includegraphics[width=\textwidth, trim=50 50 50 50, clip]{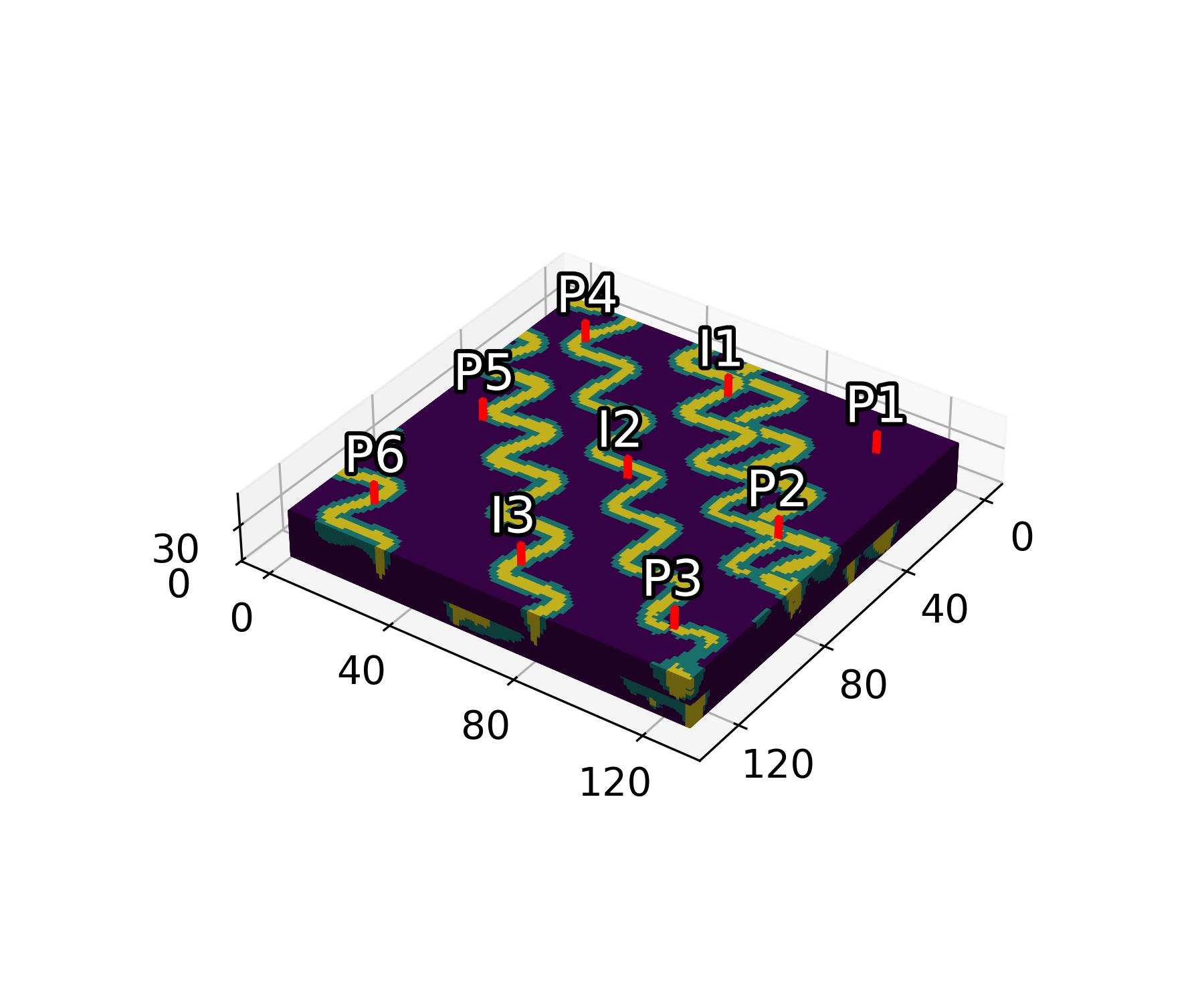}
        \caption{Post.~4}
    \end{subfigure}

    \caption{Case 1: randomly selected realizations from the prior (top row) and posterior (bottom row) ensembles.}
    \label{fig:prior_posterior_tops}
\end{figure}

\begin{figure}[h]
    \centering

    \begin{subfigure}[b]{0.23\textwidth}
        \includegraphics[width=\textwidth, trim=50 50 50 50, clip]{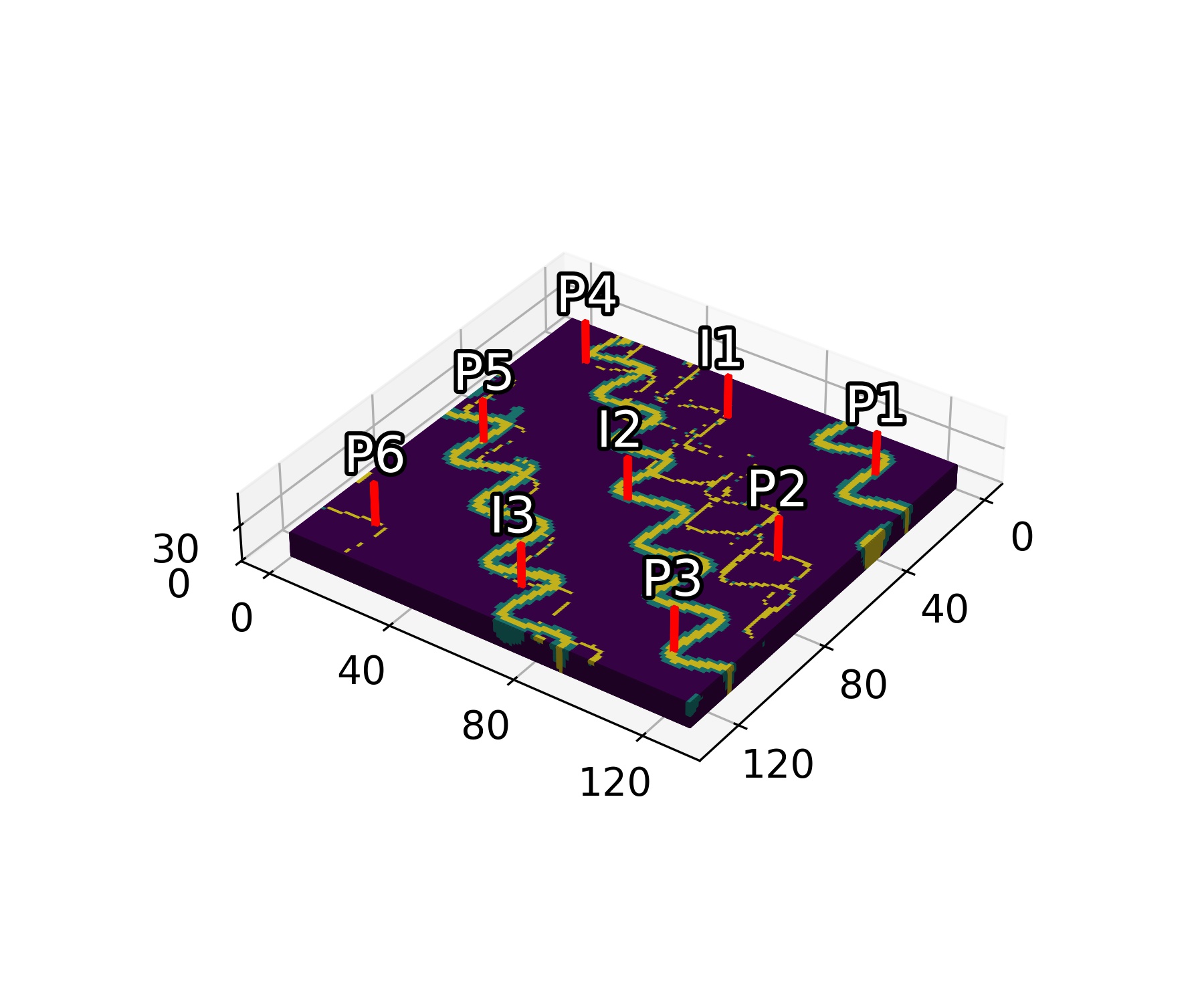}
        \caption{Prior 1}
    \end{subfigure}
    \hfill
    \begin{subfigure}[b]{0.23\textwidth}
        \includegraphics[width=\textwidth, trim=50 50 50 50, clip]{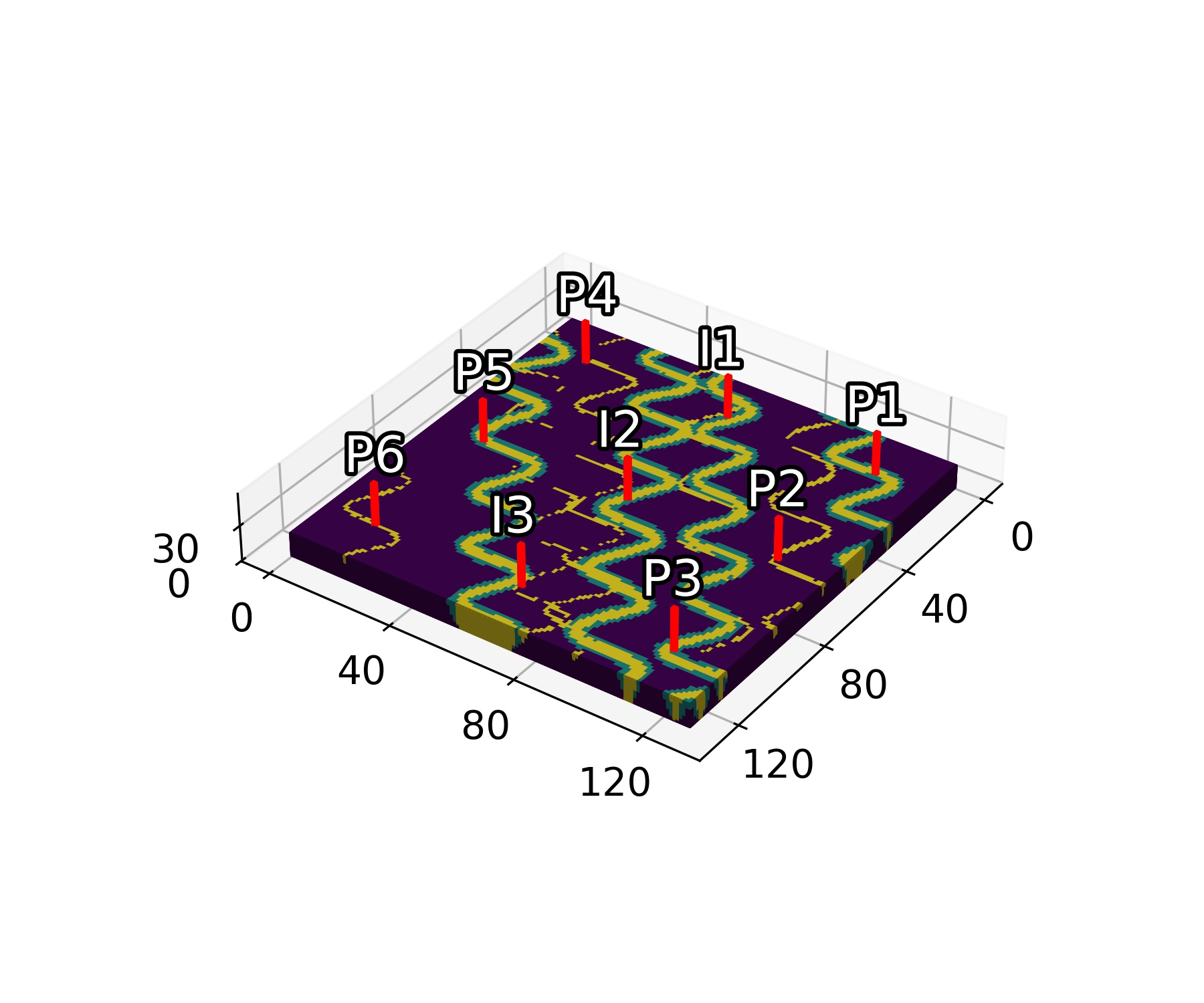}
        \caption{Prior 2}
    \end{subfigure}
    \hfill
    \begin{subfigure}[b]{0.23\textwidth}
        \includegraphics[width=\textwidth, trim=50 50 50 50, clip]{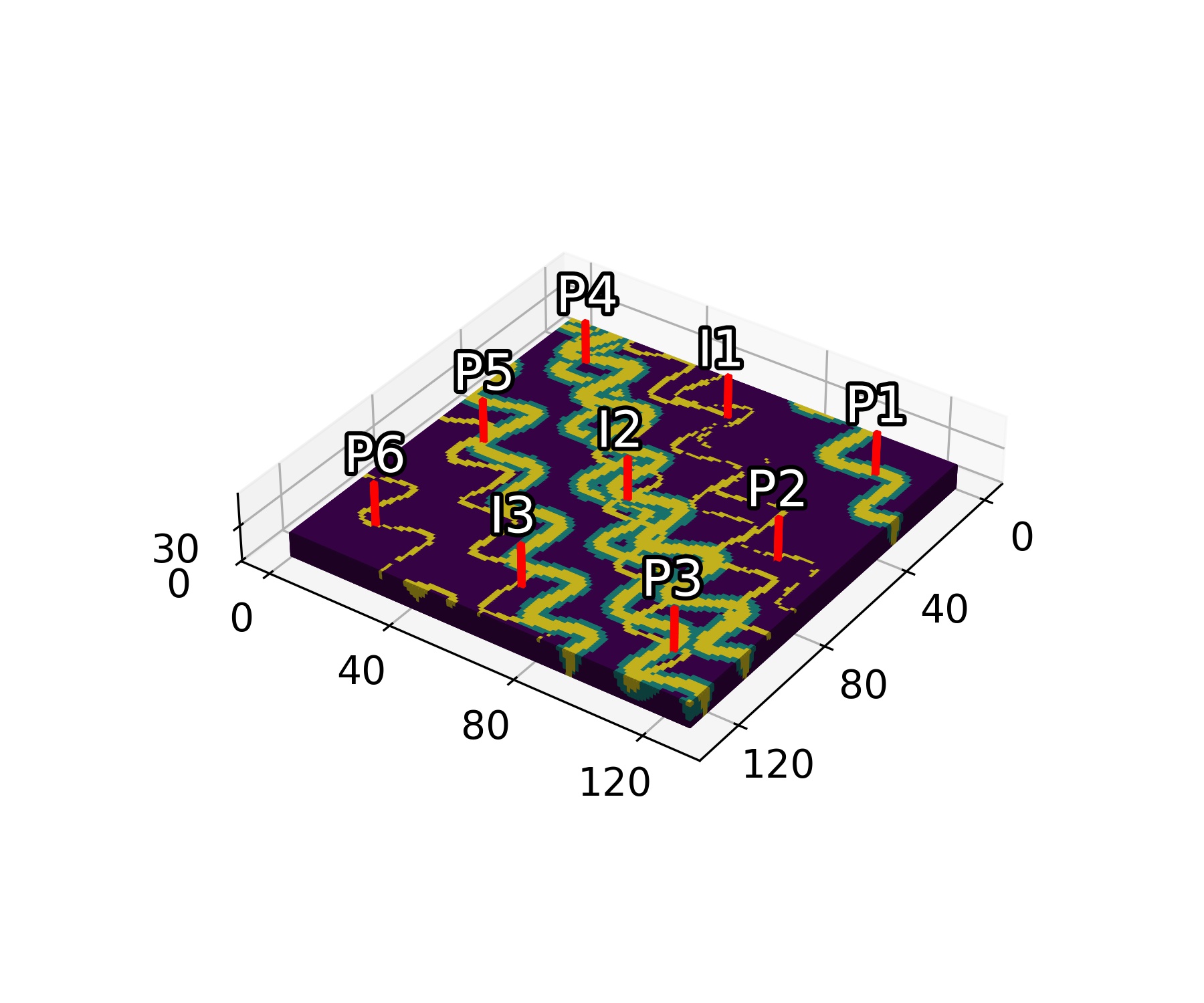}
        \caption{Prior 3}
    \end{subfigure}
    \hfill
    \begin{subfigure}[b]{0.23\textwidth}
        \includegraphics[width=\textwidth, trim=50 50 50 50, clip]{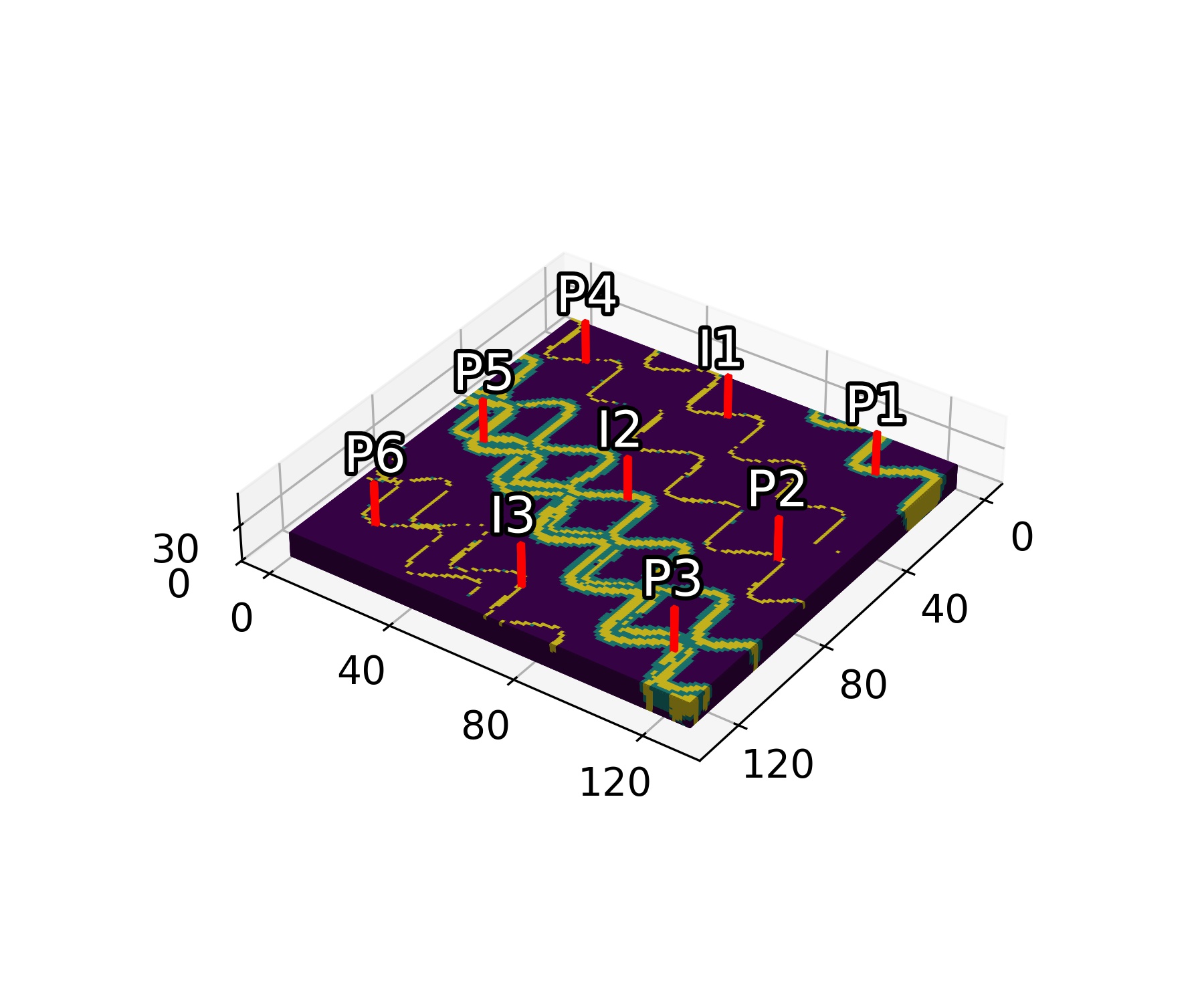}
        \caption{Prior 4}
    \end{subfigure}

    \vspace{1em}

    \begin{subfigure}[b]{0.23\textwidth}
        \includegraphics[width=\textwidth, trim=50 50 50 50, clip]{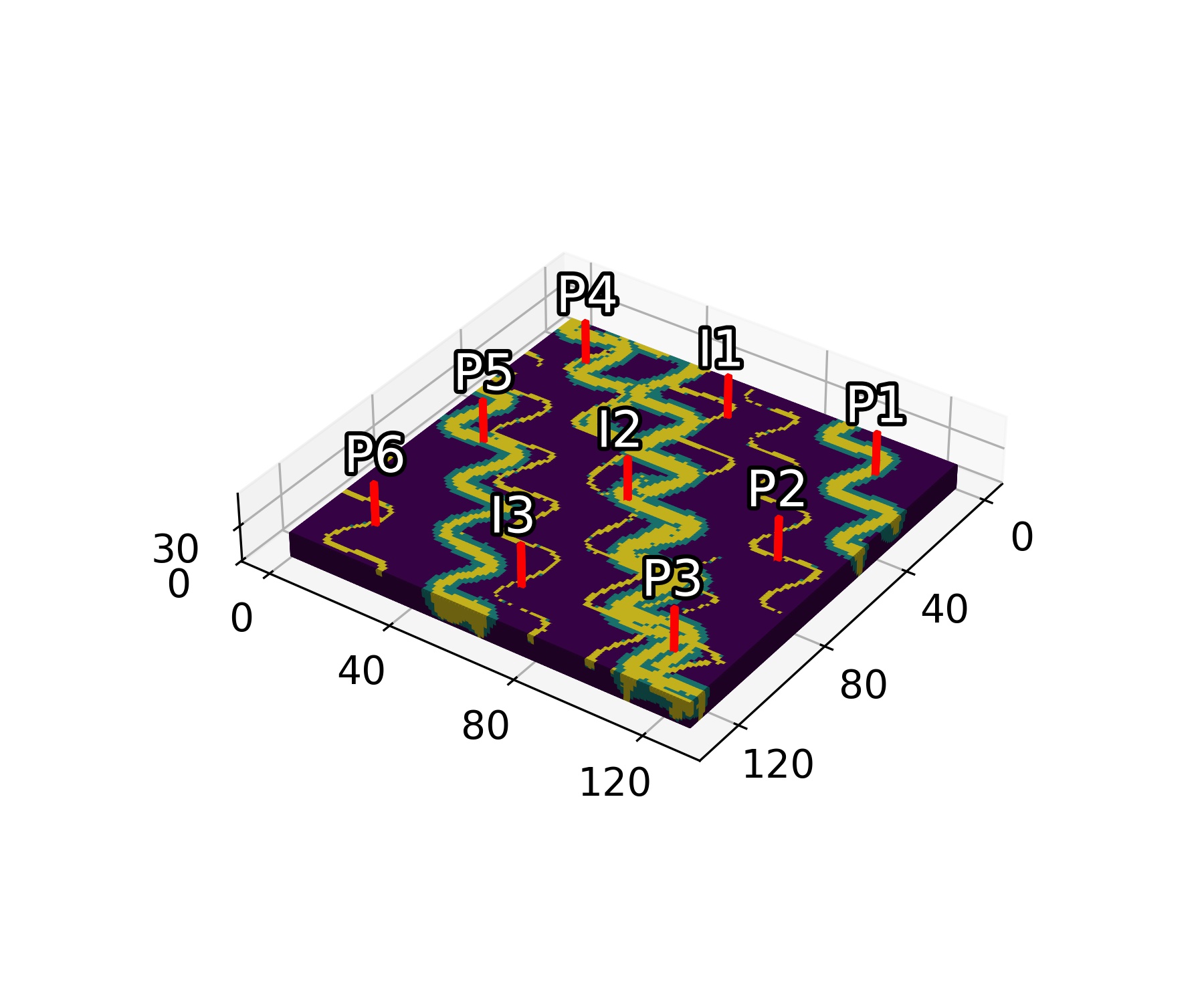}
        \caption{Post.~1}
    \end{subfigure}
    \hfill
    \begin{subfigure}[b]{0.23\textwidth}
        \includegraphics[width=\textwidth, trim=50 50 50 50, clip]{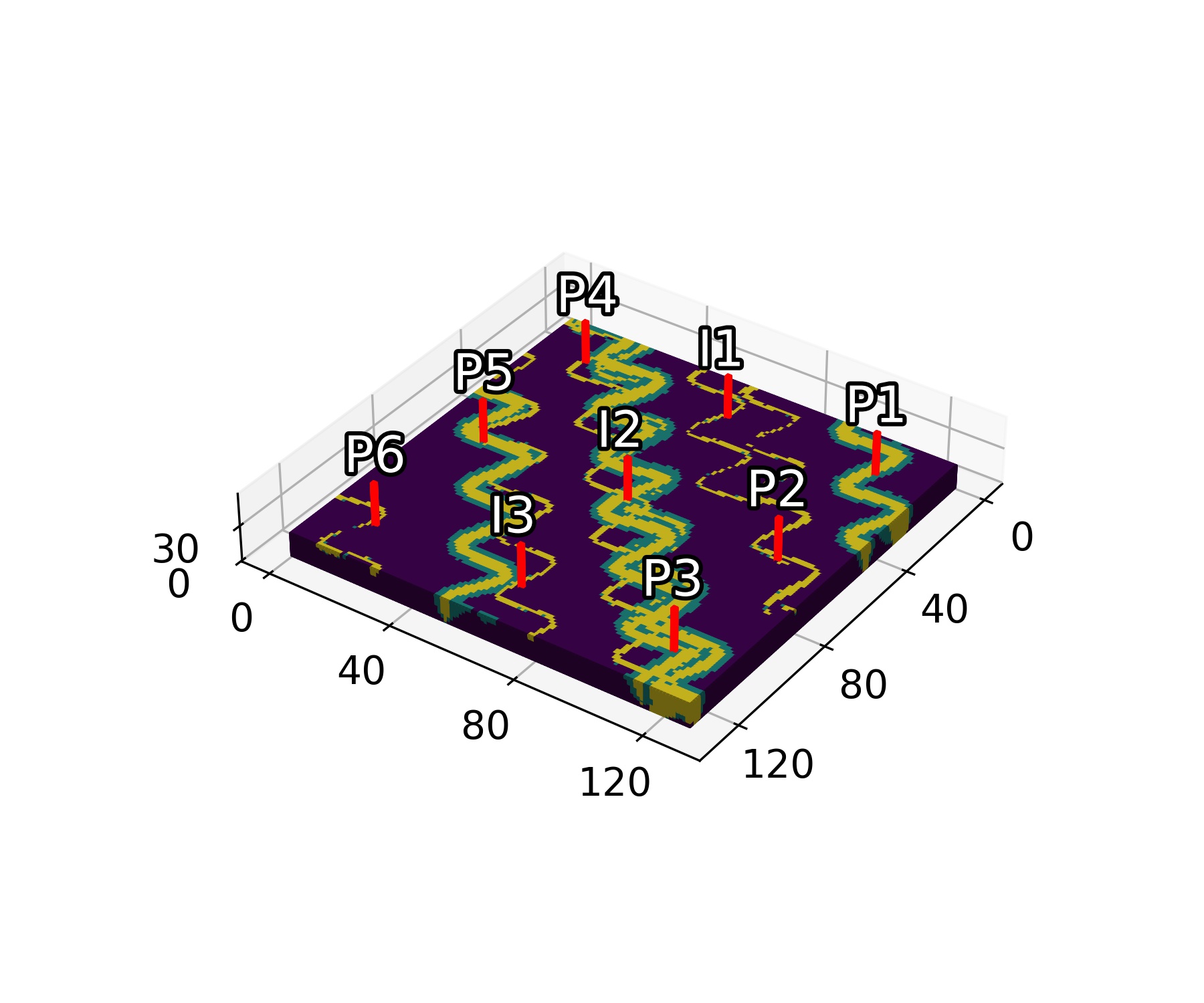}
        \caption{Post.~2}
    \end{subfigure}
    \hfill
    \begin{subfigure}[b]{0.23\textwidth}
        \includegraphics[width=\textwidth, trim=50 50 50 50, clip]{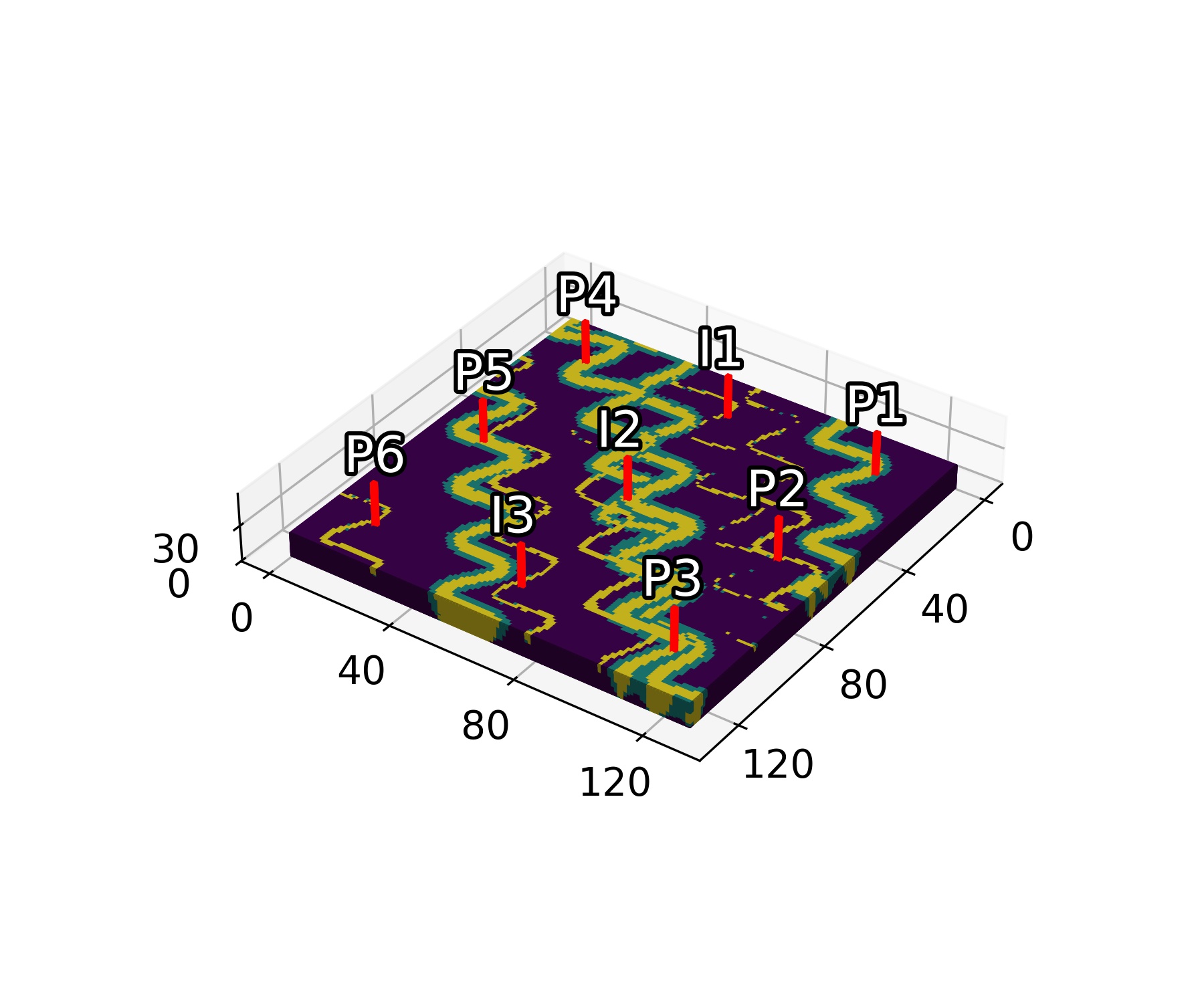}
        \caption{Post.~3}
    \end{subfigure}
    \hfill
    \begin{subfigure}[b]{0.23\textwidth}
        \includegraphics[width=\textwidth, trim=50 50 50 50, clip]{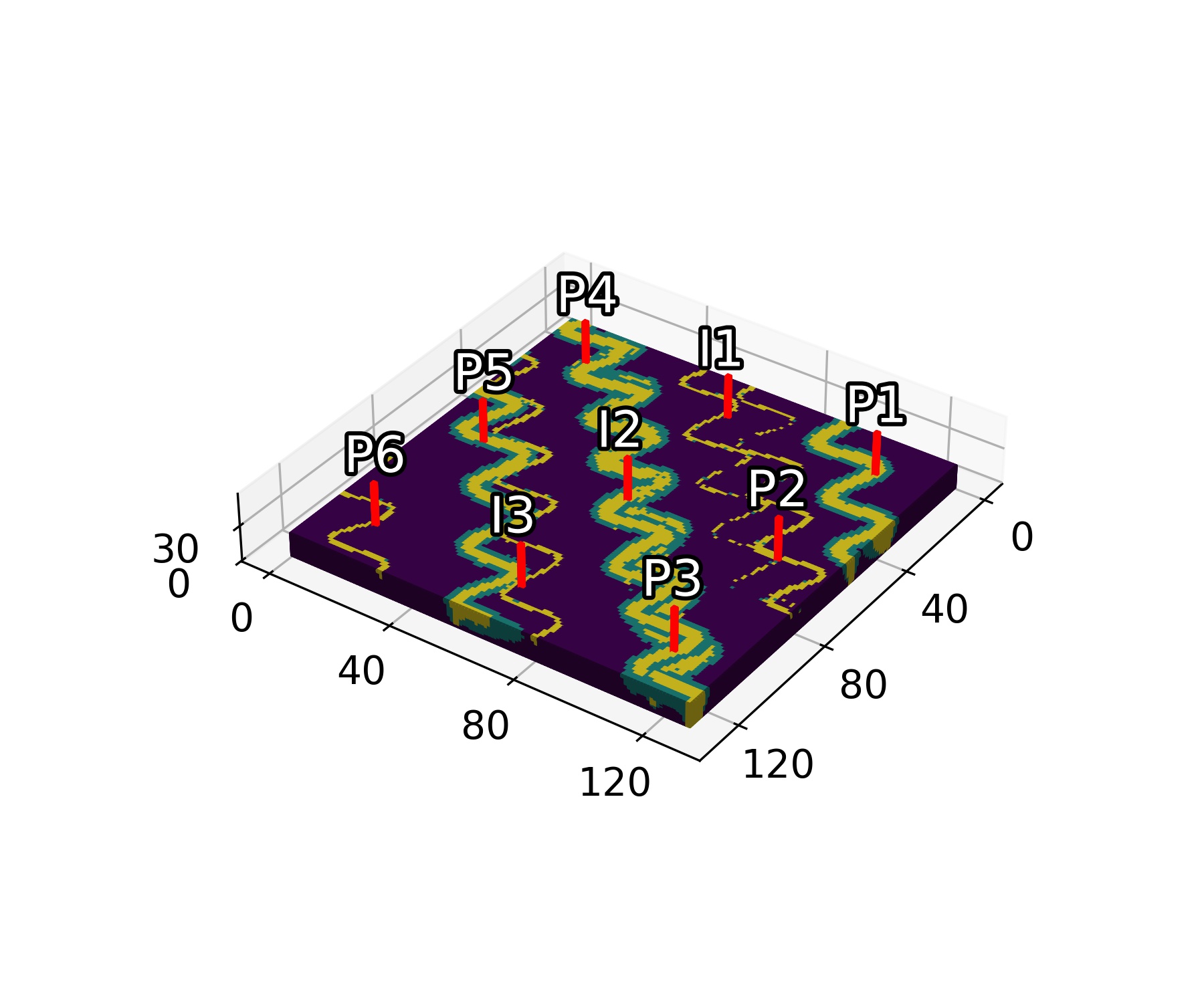}
        \caption{Post.~4}
    \end{subfigure}

    \caption{Case 1: lower halves of randomly selected realizations from the prior (top row) and posterior (bottom row) ensembles.}
    \label{fig:prior_posterior_bottoms}
\end{figure}

Randomly selected realizations from the prior (top row) and posterior (bottom row) ensembles are shown in Figure~\ref{fig:prior_posterior_tops} (full models) and  Figure~\ref{fig:prior_posterior_bottoms} (lower halves). The prior ensemble is characterized by a range of channel orientations and channel densities. The posterior realizations, by contrast, display features consistent with the synthetic true model. Specifically, the channel locations, density and orientation, are seen to be in alignment with those of the true model (Figure~\ref{fig:case1_models}a and Figure~\ref{fig:case1_models}d).

\begin{figure}[h]
    \centering

    \begin{subfigure}[b]{0.32\textwidth}
        \centering
        \includegraphics[width=\textwidth]{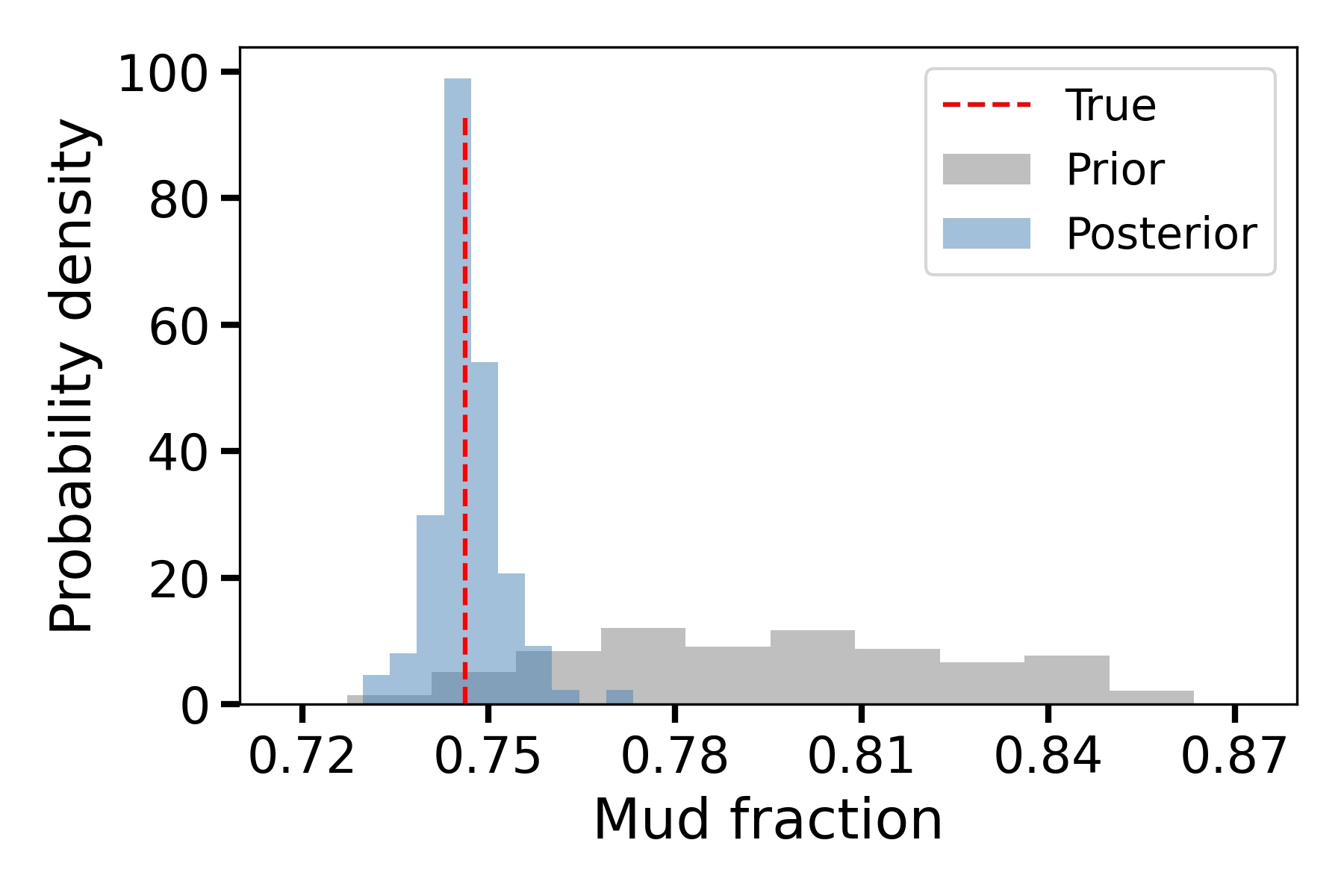}
        \caption{Mud fraction}
        \label{fig:mud_fraction1}
    \end{subfigure}
    \hfill
    \begin{subfigure}[b]{0.32\textwidth}
        \centering
        \includegraphics[width=\textwidth]{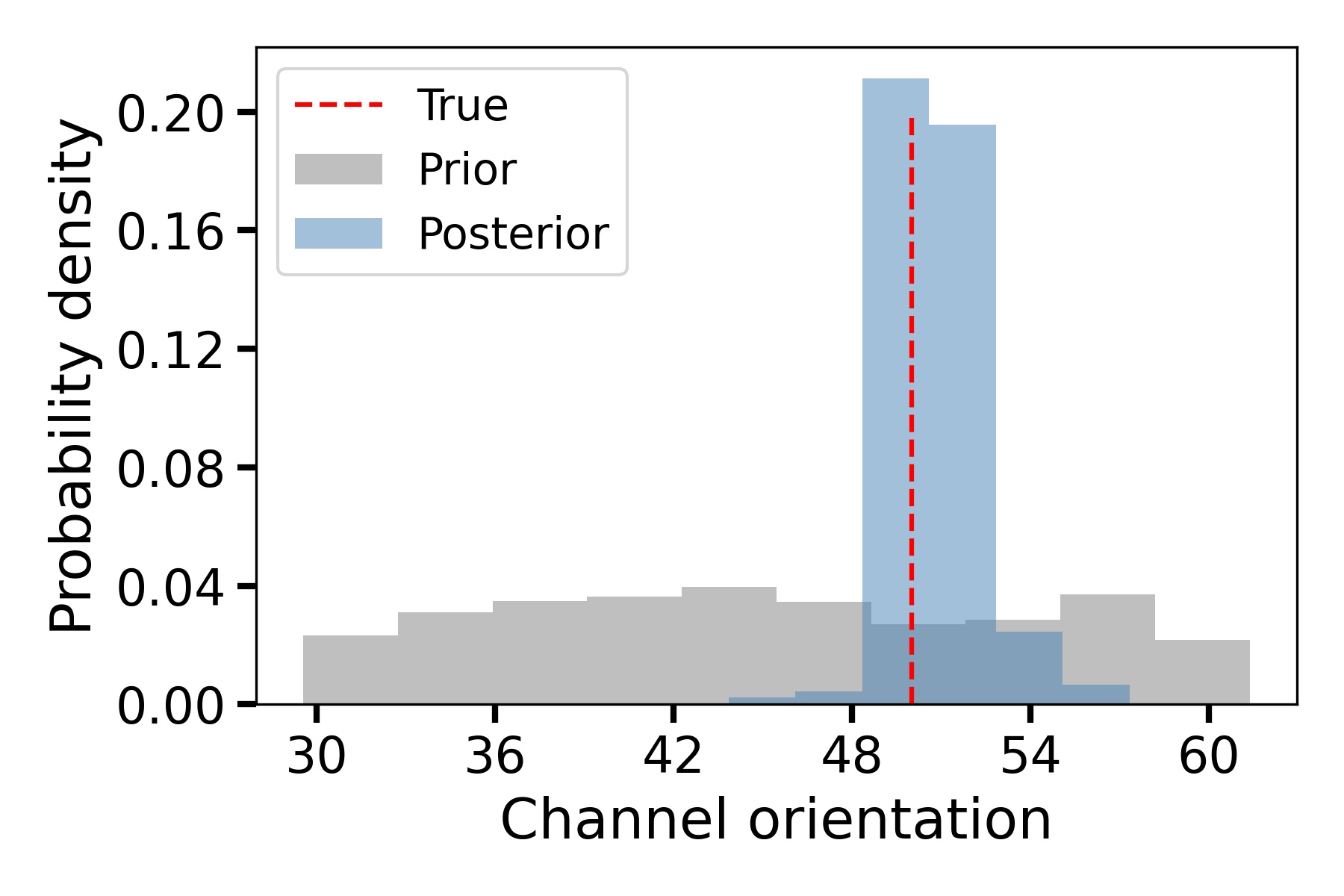}
        \caption{Channel orientation}
        \label{fig:channel_orientation1}
    \end{subfigure}
    \hfill
    \begin{subfigure}[b]{0.32\textwidth}
        \centering
        \includegraphics[width=\textwidth]{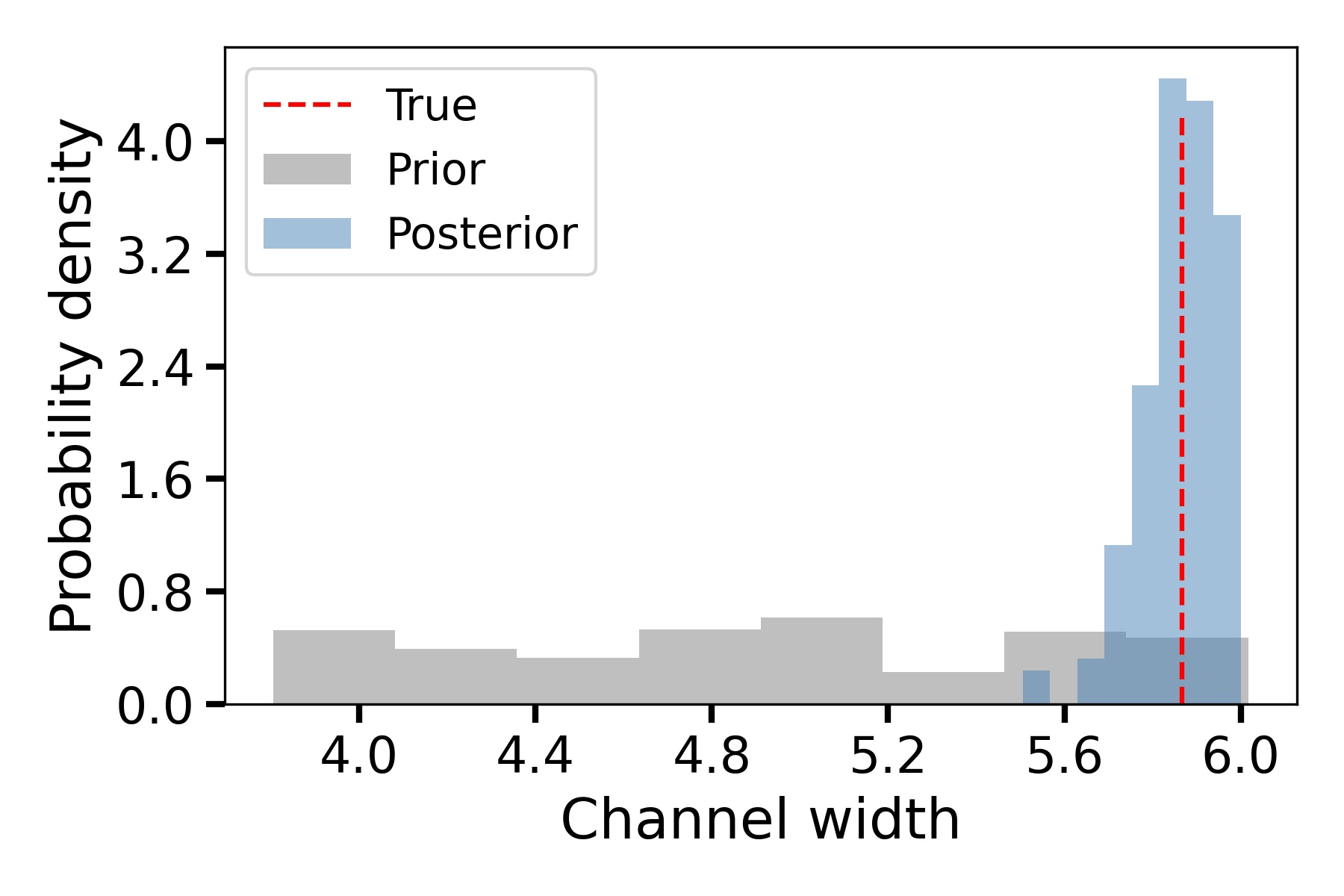}
        \caption{Channel width}
        \label{fig:channel_width1}
    \end{subfigure}
    \caption{\textcolor{black}{Case 1: history matching results for geological scenario parameters. Gray regions show the 
    prior distribution and blue histograms denote the posterior distribution. Vertical red dashed lines indicate true values.}}
    \label{fig:case1_params}
\end{figure}

Finally, we present prior and posterior distributions for the geological scenario parameters -- mud fraction, channel orientation, and channel width. \textcolor{black}{In Figure~\ref{fig:case1_params}, we display the histograms for the prior and the posterior distributions in gray and light blue, respectively. The vertical red lines indicate the true values, as given in Table~\ref{table:true_params}. Mud facies fractions (Figure~\ref{fig:case1_params}a) are computed directly from the (prior or posterior) geomodel realizations. For channel width and orientation, the CNN-based procedure ($\tilde{\mathbf{s}} = \text{CNN}(\mathbf{r})$) described in Section~\ref{results_models} is applied for each prior and posterior realization. Histograms of the scenario parameters are then constructed from the full set of $N_e=200$ prior and posterior geomodels. Note that the histograms for the prior models are slightly nonuniform, in contrast to the specifications of Table~\ref{table:orientation_width_number}. This is due to limitations and approximations in the OBM procedure applied to generate the training samples and to Monte Carlo error.}

Very close alignment between the posterior parameter distributions and the true values is observed in Figure~\ref{fig:case1_params}. Some amount of spread in the posterior histograms is evident, though the mode of each distribution is near the true scenario parameter in all cases. Consistent with the posterior geomodels discussed earlier, these results highlight the ability of the overall procedure to provide uncertainty reduction in the scenario parameters. We reiterate that this is achieved automatically within our workflow -- we do not need to perform any sort of two-level history matching where the scenario parameters are determined in an initial step. This represents a significant advantage of our unified parameterization.

\subsection{Aggregate results for Cases~2 and 3}

In the interest of brevity, we present a limited set of results for Cases~2 and~3. Our purpose here is to demonstrate that the workflow can provide appropriate uncertainty reduction for true models corresponding to a range of scenario parameters. The scenario parameters for these cases are provided in Table~\ref{table:true_params}. The prior models used in the history matching for Cases~2 and~3 are the same as those used for Case~1.

\begin{figure}[h]
    \centering
    \begin{subfigure}[b]{0.23\textwidth}
        \centering
        \includegraphics[width=\textwidth, trim=50 50 50 50, clip]{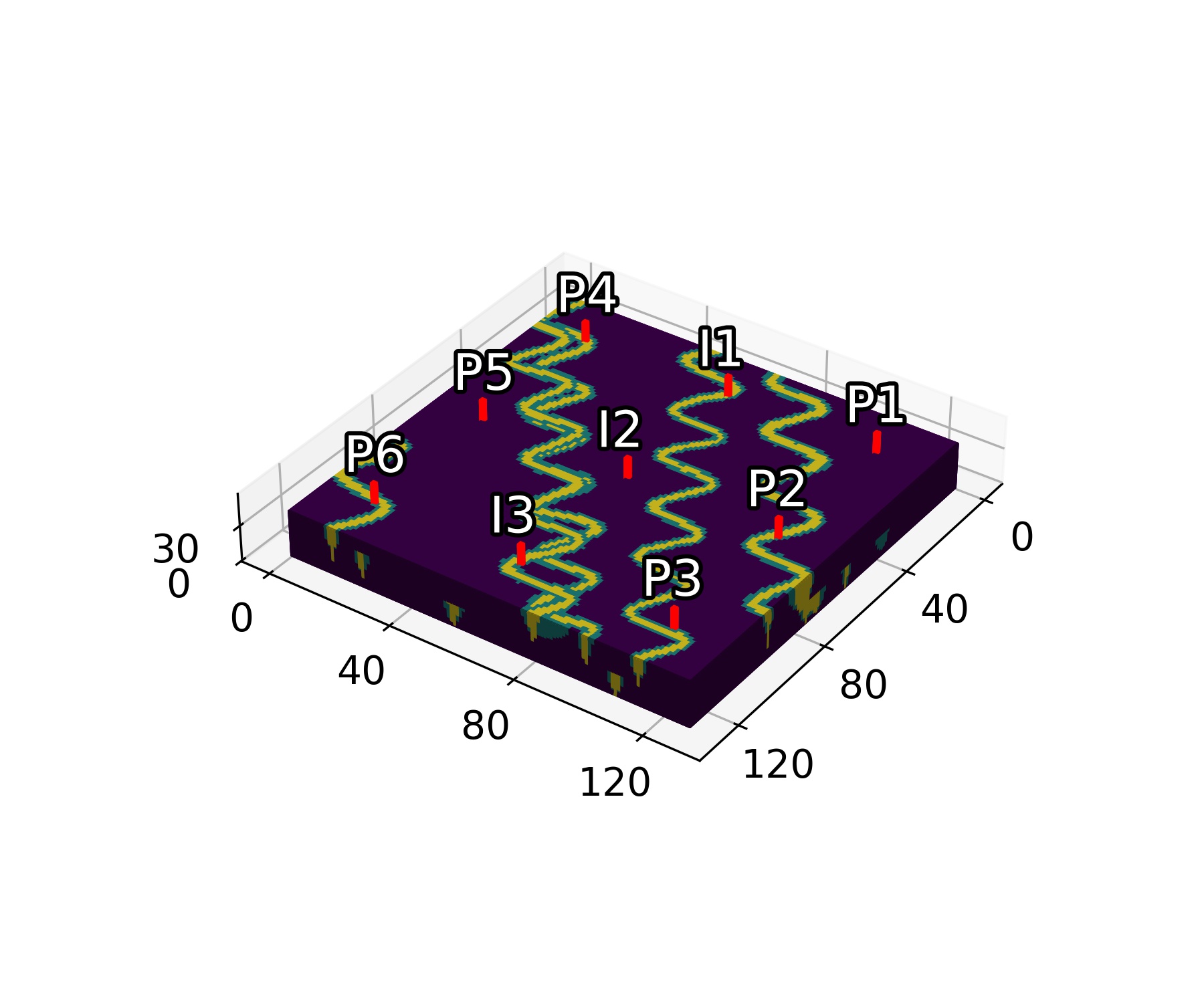}
        \caption{True 2}
        \label{fig:true_top_2}
    \end{subfigure}
    \hfill
    \begin{subfigure}[b]{0.23\textwidth}
        \centering
        \includegraphics[width=\textwidth, trim=50 50 50 50, clip]{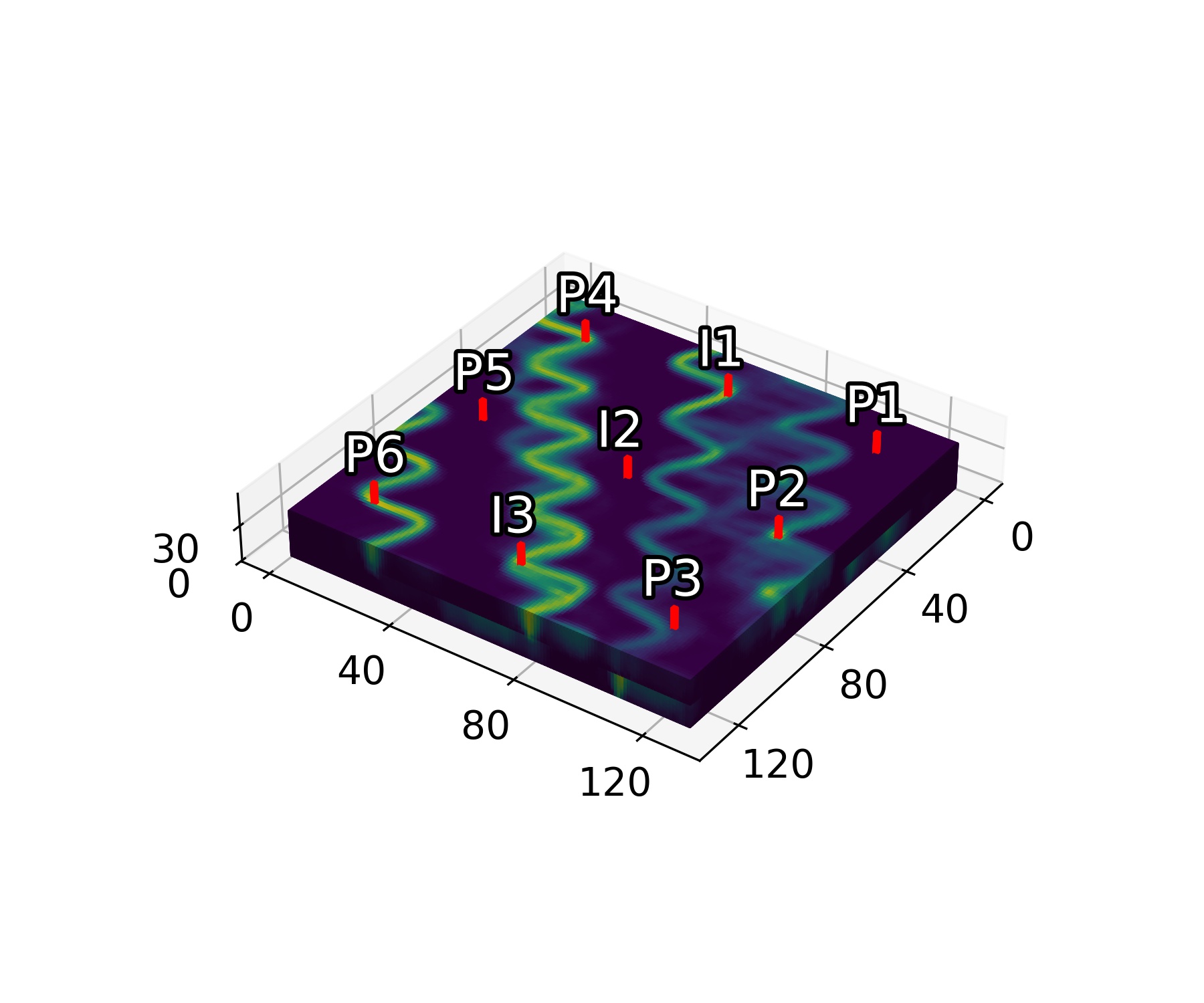}
        \caption{Post.~2}
        \label{fig:post_mean_top_2}
    \end{subfigure}
    \hfill
    \begin{subfigure}[b]{0.23\textwidth}
        \centering
        \includegraphics[width=\textwidth, trim=50 50 50 50, clip]{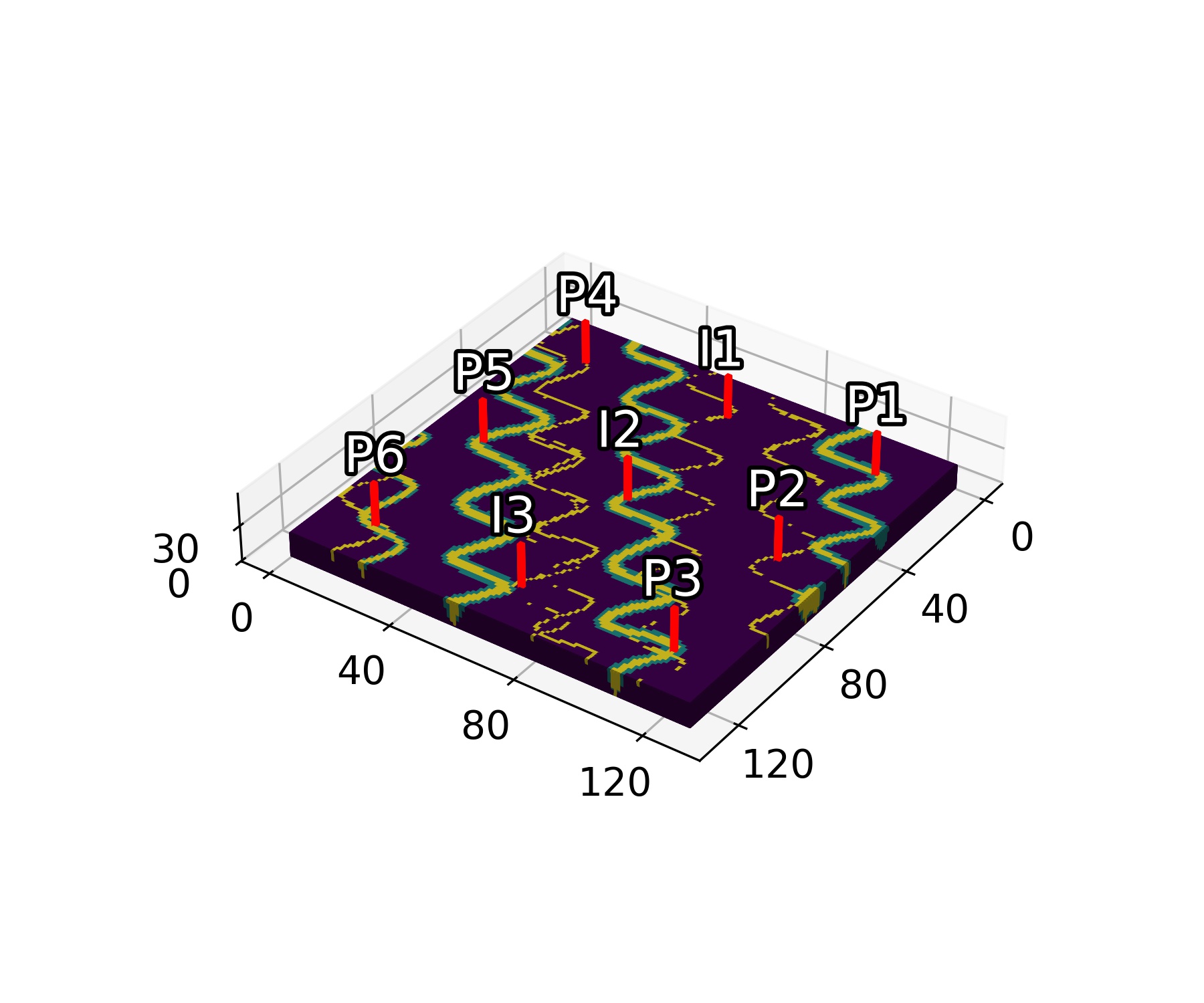}
        \caption{True 2 (lower)}
        \label{fig:true_bottom_2}
    \end{subfigure}
    \hfill
    \begin{subfigure}[b]{0.23\textwidth}
        \centering
        \includegraphics[width=\textwidth, trim=50 50 50 50, clip]{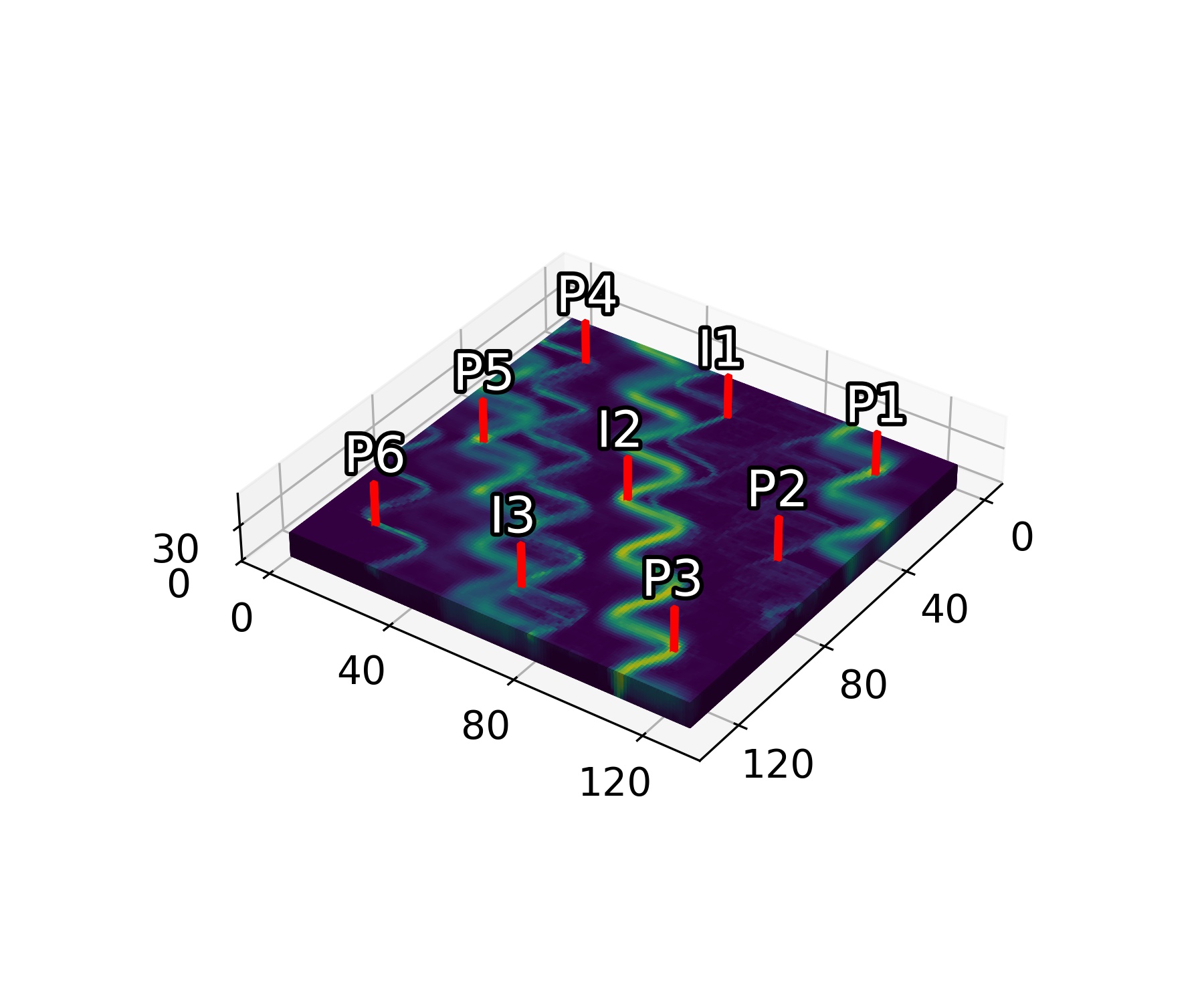}
        \caption{Post.~2 (lower)}
        \label{fig:post_mean_bottom_2}
    \end{subfigure}

    \vspace{1em} 

    \begin{subfigure}[b]{0.23\textwidth}
        \centering
        \includegraphics[width=\textwidth, trim=50 50 50 50, clip]{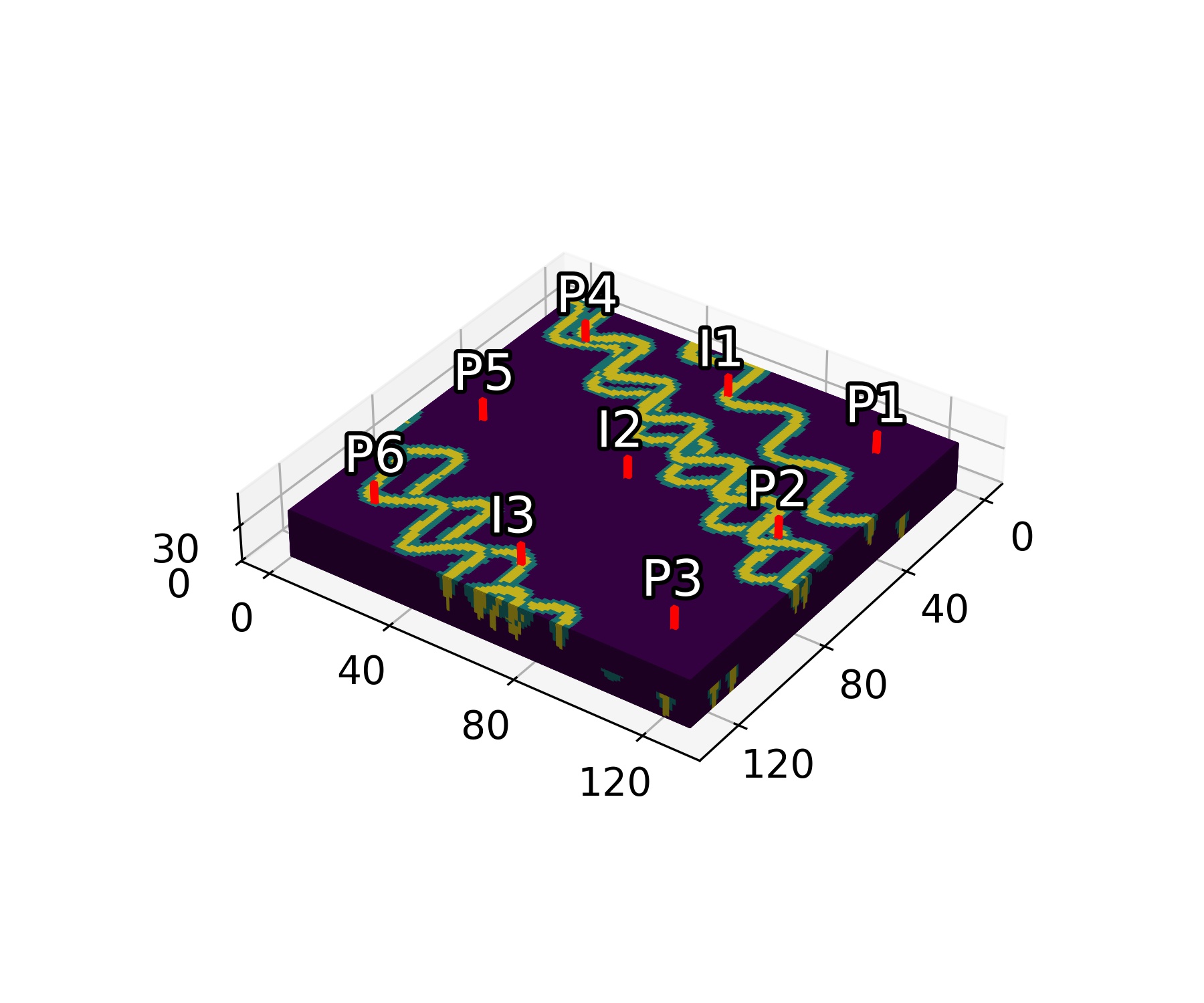}
        \caption{True 3}
        \label{fig:true_top_3}
    \end{subfigure}
    \hfill
    \begin{subfigure}[b]{0.23\textwidth}
        \centering
        \includegraphics[width=\textwidth, trim=50 50 50 50, clip]{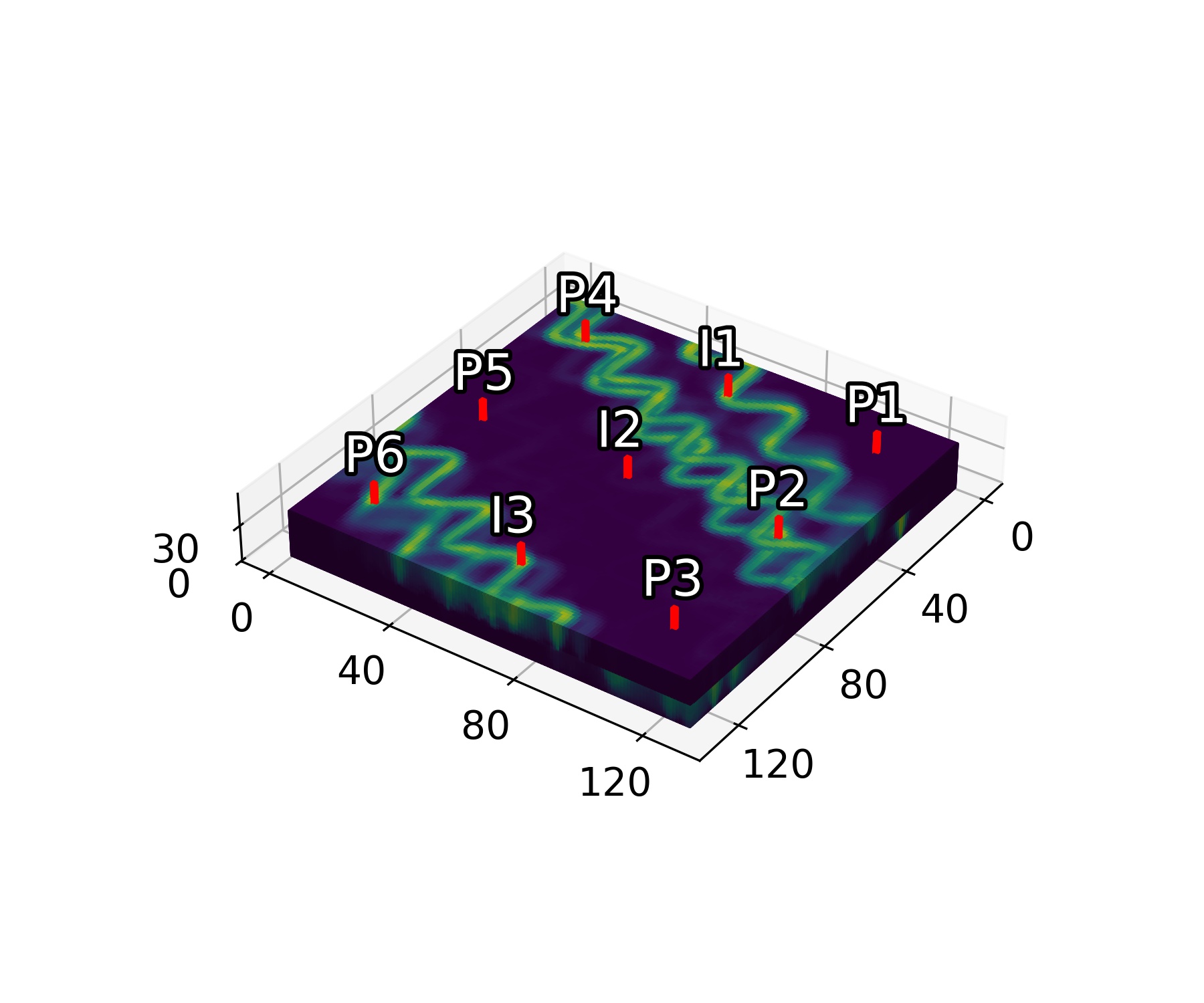}
        \caption{Post.~3}
        \label{fig:post_mean_top_3}
    \end{subfigure}
    \hfill
    \begin{subfigure}[b]{0.23\textwidth}
        \centering
        \includegraphics[width=\textwidth, trim=50 50 50 50, clip]{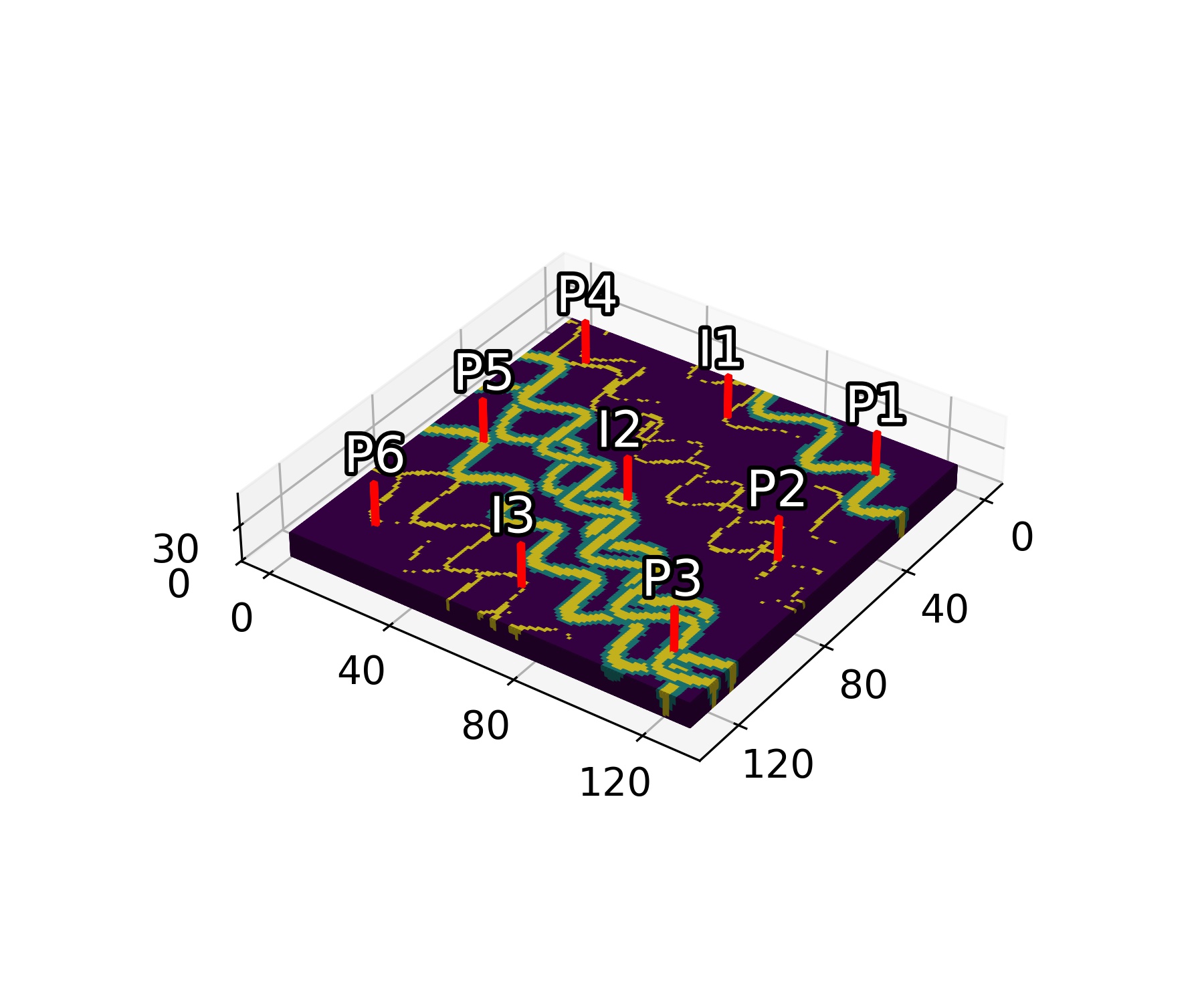}
        \caption{True 3 (lower)}
        \label{fig:true_bottom_3}
    \end{subfigure}
    \hfill
    \begin{subfigure}[b]{0.23\textwidth}
        \centering
        \includegraphics[width=\textwidth, trim=50 50 50 50, clip]{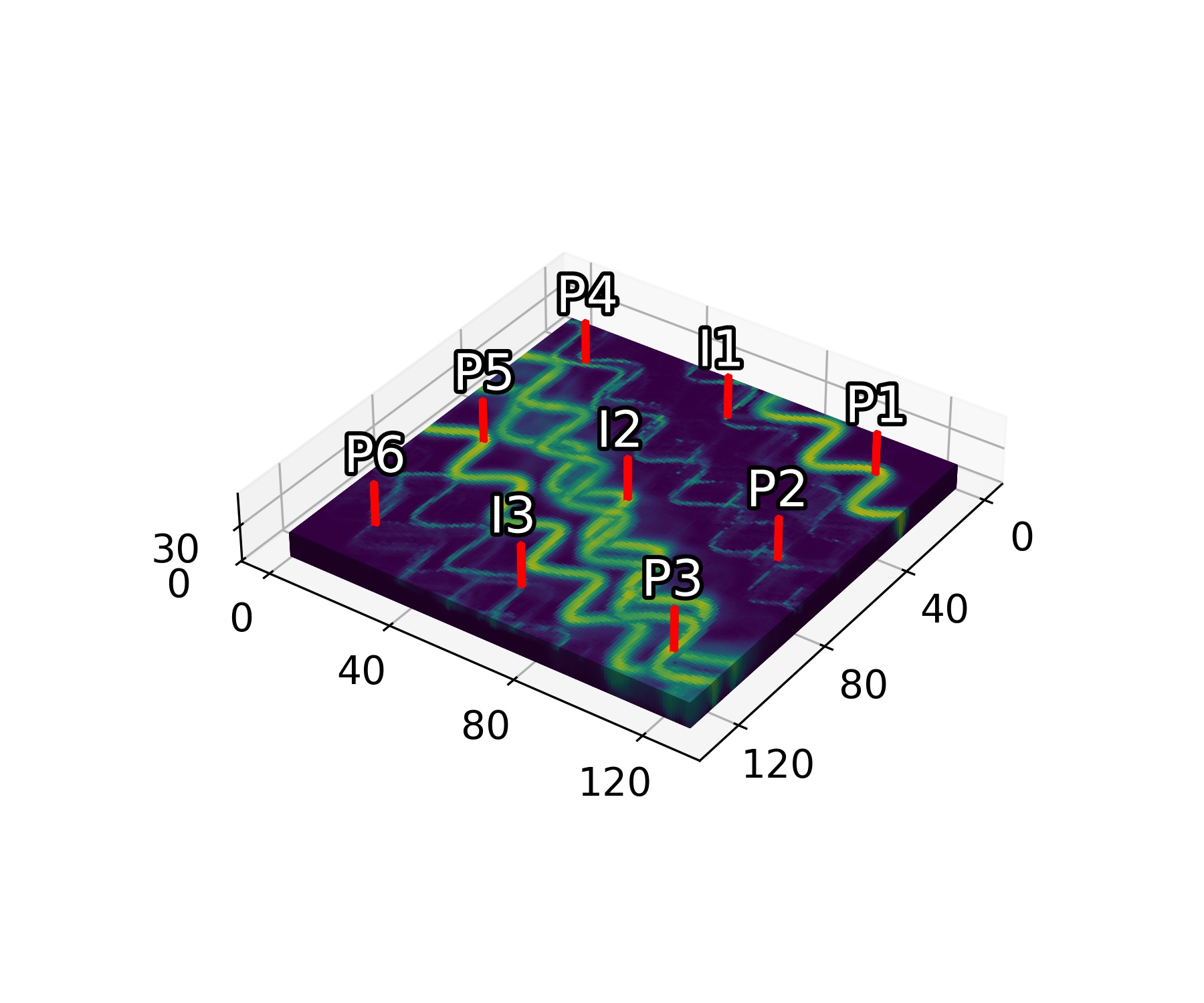}
        \caption{Post.~3 (lower)}
        \label{fig:post_mean_bottom_3}
    \end{subfigure}

    \caption{Top row: true model and posterior mean for Case~2. Bottom row: true model and posterior mean for Case~3. Full model and lower halves shown for both cases.}
    \label{fig:case23_models}
\end{figure}

The true models and posterior ensemble means (full model and lower half) for Cases~2 and~3 are displayed in Figure~\ref{fig:case23_models}. The prior means for the full model and for the lower half of the model were shown earlier in Figure~\ref{fig:case1_models}b and Figure~\ref{fig:case1_models}e. It is clear from Figure~\ref{fig:case23_models} that the posterior means for Case~2 and Case~3 closely resemble the true models in both cases. In particular, although the two true models correspond to quite different channel locations and orientations, both true models are accurately reflected (visually) in the posterior means. The tightly intertwining channels at various locations in True~3, for example, are well resolved in the posterior mean. This is evident both for the full model and for the lower half.

\begin{figure}[h]
    \centering

    \begin{subfigure}[b]{0.32\textwidth}
        \centering
        \includegraphics[width=\textwidth]{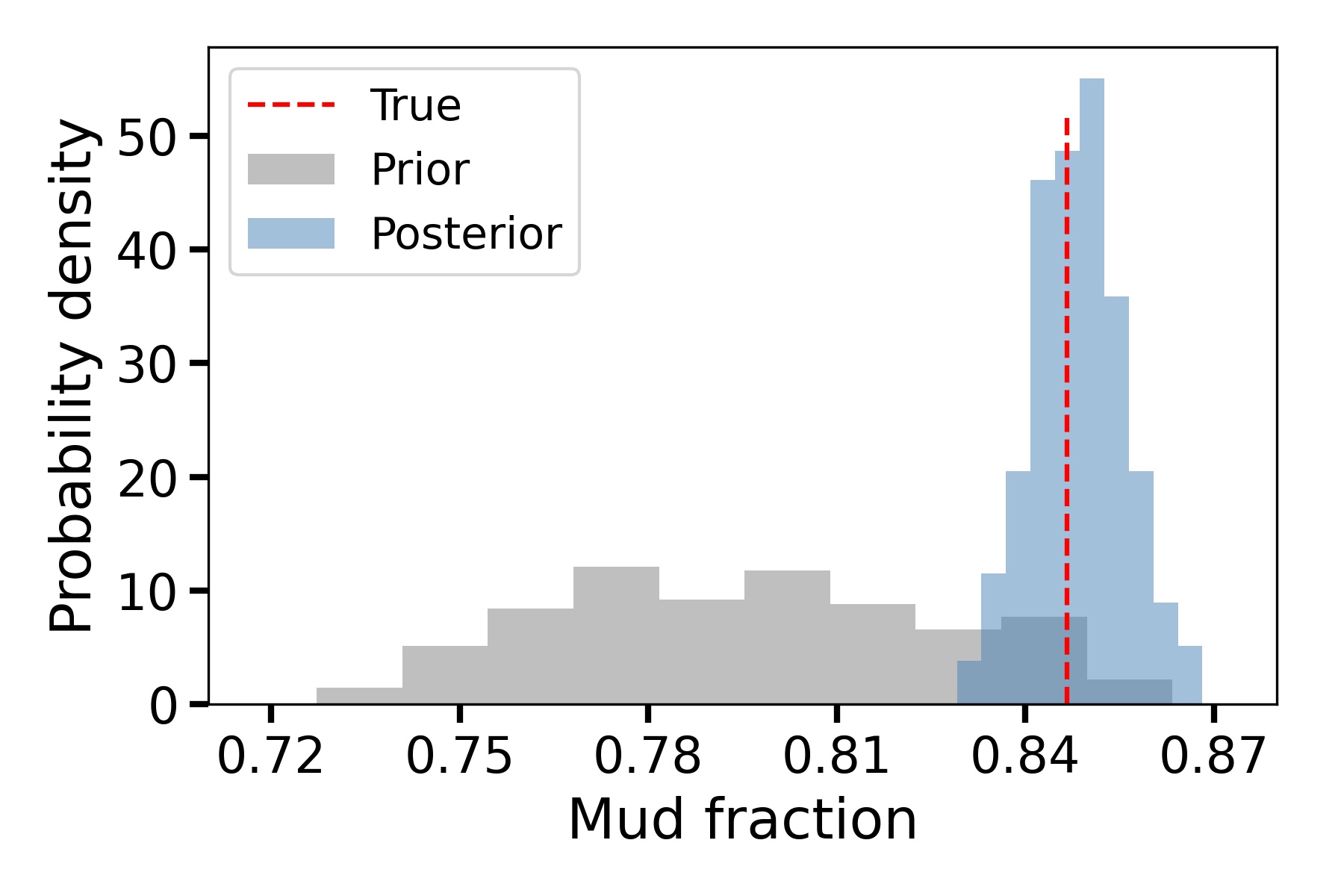}
        \caption{Mud fraction (Case~2)}
        \label{fig:mud_fraction2}
    \end{subfigure}
    \hfill
    \begin{subfigure}[b]{0.32\textwidth}
        \centering
        \includegraphics[width=\textwidth]{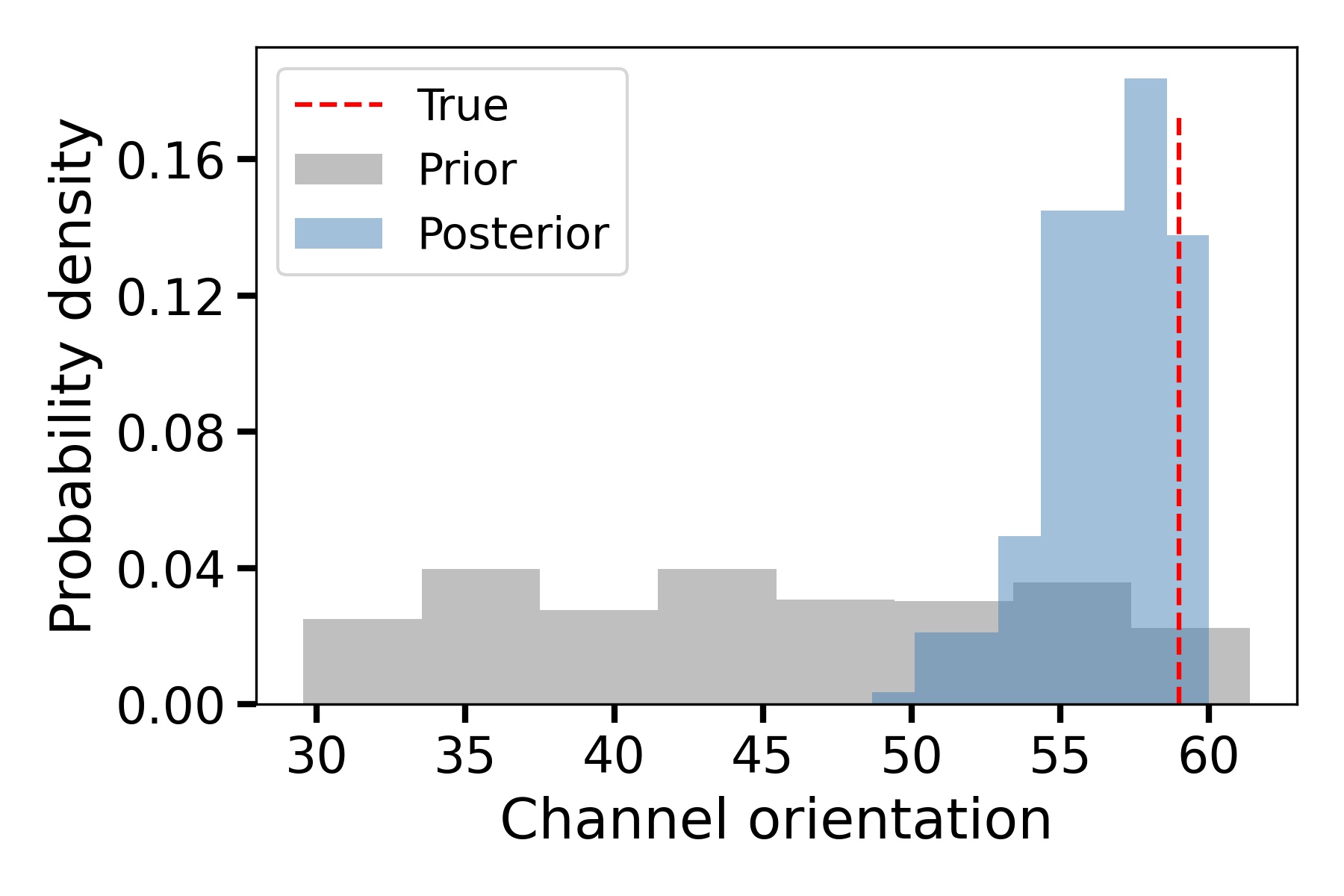}
        \caption{Channel orient.~(Case~2)}
        \label{fig:channel_orientation2}
    \end{subfigure}
    \hfill
    \begin{subfigure}[b]{0.32\textwidth}
        \centering
        \includegraphics[width=\textwidth]{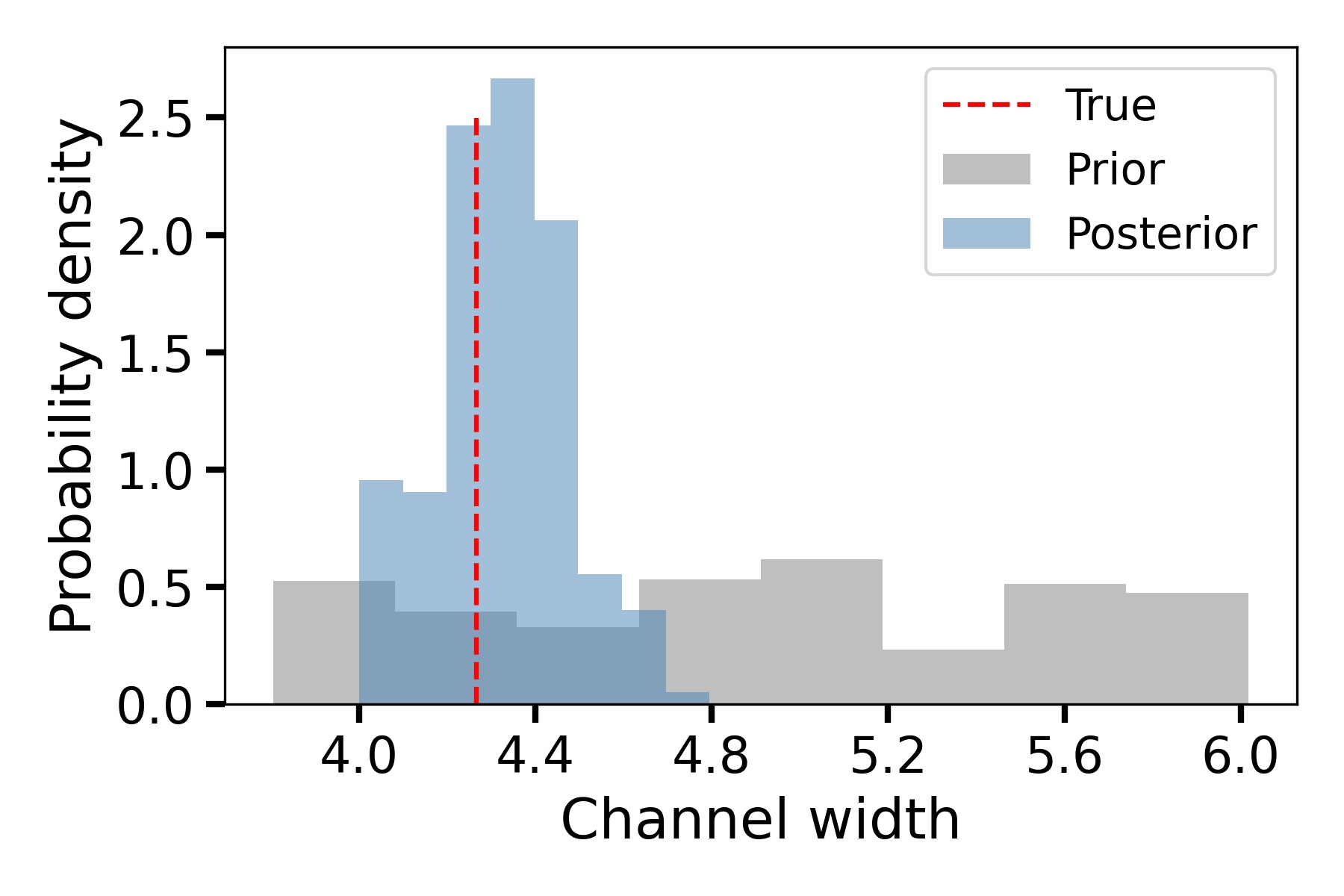}
        \caption{Channel width (Case~2)}
        \label{fig:channel_width2}
    \end{subfigure}

    \vspace{1em} 

    \begin{subfigure}[b]{0.32\textwidth}
        \centering
        \includegraphics[width=\textwidth]{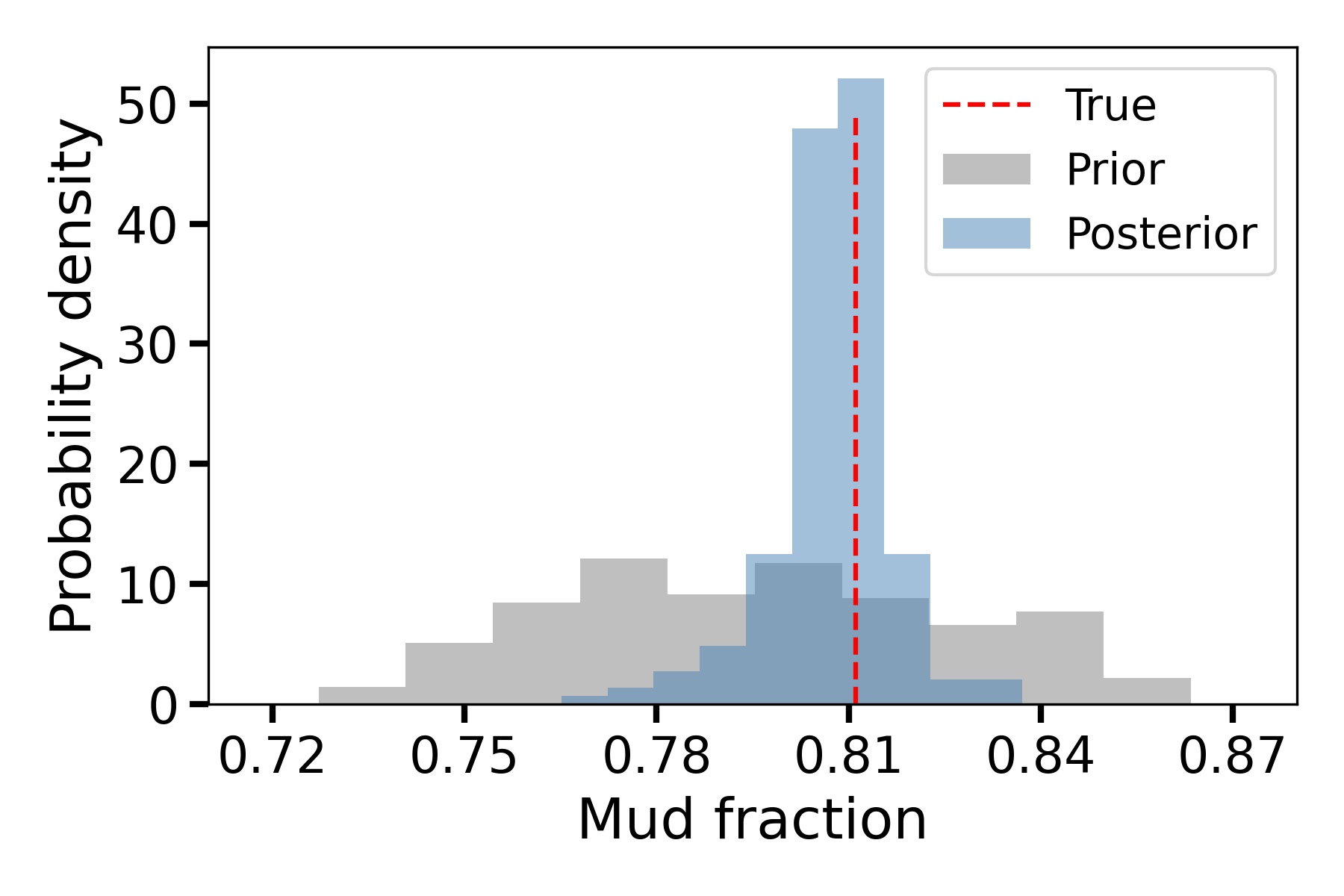}
        \caption{Mud fraction (Case~3)}
        \label{fig:mud_fraction3}
    \end{subfigure}
    \hfill
    \begin{subfigure}[b]{0.32\textwidth}
        \centering
        \includegraphics[width=\textwidth]{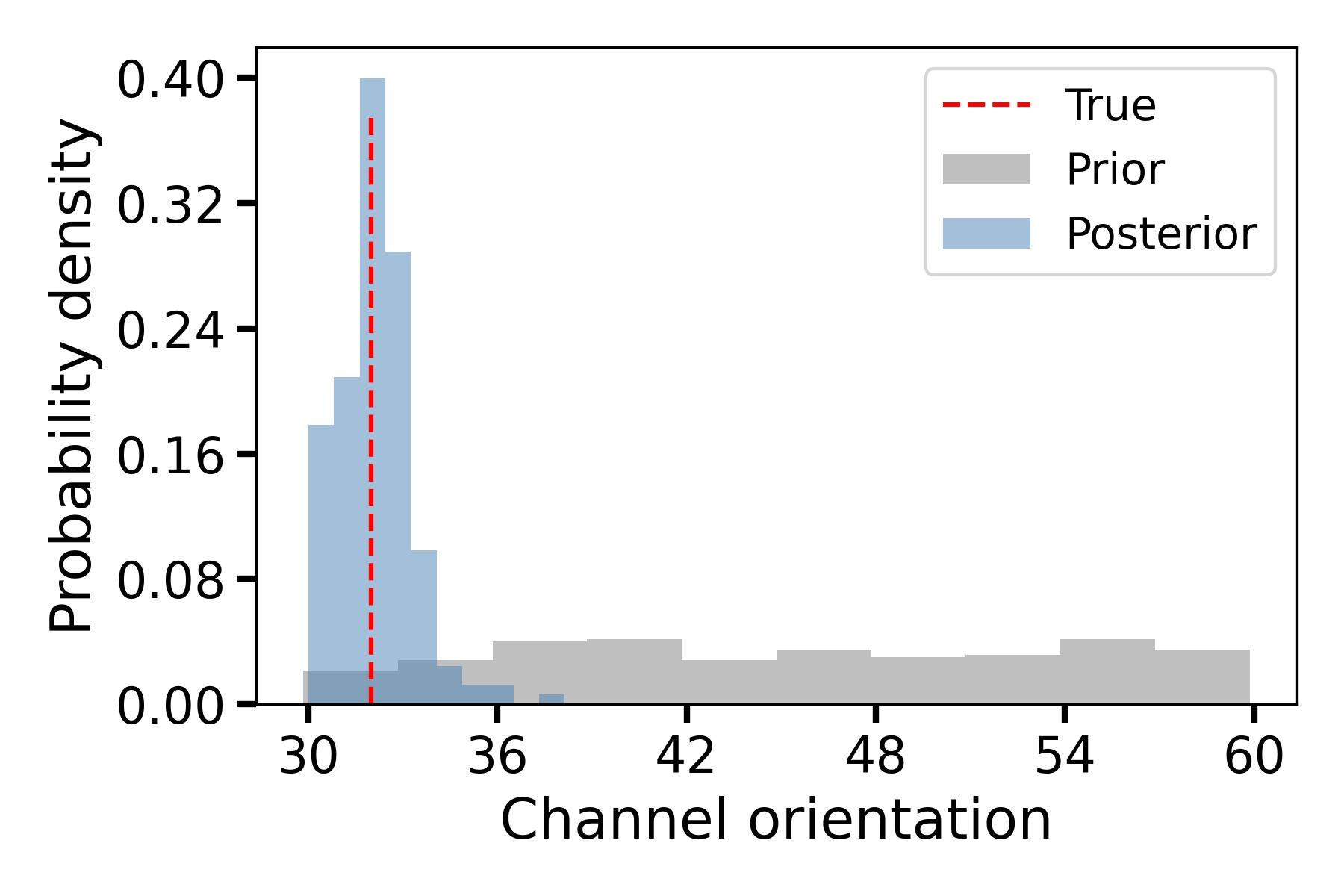}
        \caption{Channel orient.~(Case~3)}
        \label{fig:channel_orientation3}
    \end{subfigure}
    \hfill
    \begin{subfigure}[b]{0.32\textwidth}
        \centering
        \includegraphics[width=\textwidth]{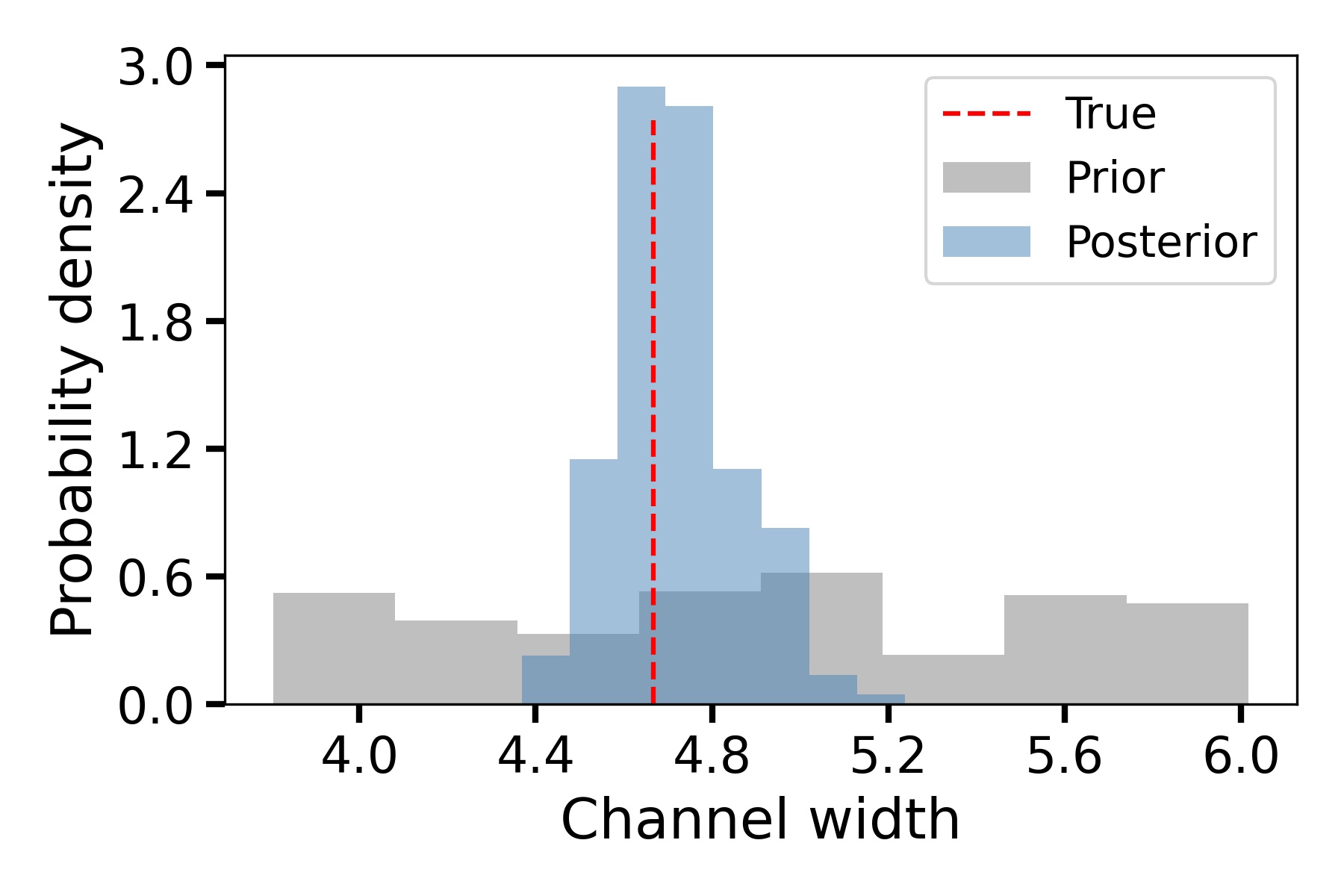}
        \caption{Channel width (Case~3)}
        \label{fig:channel_width3}
    \end{subfigure}

    \caption{\textcolor{black}{History matching results for geological scenario parameters for Case~2 (top row) and Case~3 (bottom row). Gray regions show the prior distribution and blue histograms denote the posterior distribution. Vertical red dashed lines indicate true values.}}
    \label{fig:case23_params}
\end{figure}

The prior and posterior scenario parameter distributions, along with the true values, appear in Figure~\ref{fig:case23_params}. \textcolor{black}{The priors again deviate from strict uniformity.} Consistent with Case~1 (Figure~\ref{fig:case1_params}), we see substantial uncertainty reduction in the scenario parameters for Cases~2 and~3. The posterior histograms include the true values in all cases, though slight deviation between the true values and the posterior modes is evident in Figure~\ref{fig:case23_params}a,b,c. We do not view this as a concern since the modes of the marginal distributions need not coincide with the mode of the joint distribution in general multivariate problems.

For both Cases~2 and 3, we observe a level of uncertainty reduction and forecast accuracy in well rates that is consistent with that for Case~1. Similarly, individual posterior realizations appear geologically realistic and resemble the true models, as would be expected from the means shown in Figure~\ref{fig:case23_models}. These results attest to the robustness of the workflow across a range of geological scenarios.

\clearpage
\section{Concluding remarks}
\label{conclusion}
In this study, we developed a latent diffusion model for the parameterization of large 3D systems with variable geological scenario parameters. Our 3D-LDM procedure provides a direct mapping from geomodel space to latent space (and vice versa), along with substantial dimension reduction. The implementation combines a 3D variational autoencoder and a 3D U-net, trained sequentially, and it includes a novel perceptual loss term that acts to improve the realism of the generated geomodels. An additional loss term enables conditioning to hard data. We applied the 3D-LDM for the parameterization of three-facies fluvial channel systems with variable (uncertain) scenario parameters. These included mud facies fraction, channel orientation, and channel width. For any set of scenario parameters, an essentially infinite number of realizations can be generated, so the range of uncertainty considered with the 3D-LDM procedure can be very wide.

The training dataset consisted of 3000 conditional realizations, generated with geostatistical software, of dimensions 128 $\times$ 128 $\times$ 32 (524,288 grid blocks). The trained 3D-LDM was characterized by a latent variable of overall dimension 1024, which corresponds to a dimension reduction ratio of 512 relative to the physical geomodels. New LDM-generated realizations exhibited the key geological features present in the reference samples, and they displayed the correct degree of variability across geological scenarios. The quality of the generated models was assessed by means of visual inspection and quantitative metrics, including facies fractions distributions, variograms, and flow response statistics. Accuracy in all these metrics was demonstrated.

We then performed history matching for an immiscible, two-phase flow problem involving three injection wells and six production wells, all operating under BHP control. Approximately uniform priors on the uncertain scenario parameters were applied. The history matching procedure entailed adjusting the latent variables, using ESMDA, to achieve low mismatch between simulation results and (synthetic) field observations. Three `true' models characterized by different geological scenario parameters were tested. For all cases, we observed significant uncertainty reduction and a high degree of consistency between the posterior scenario parameters and the true parameters. Posterior geomodels were also shown to resemble the true model much more closely than the prior models. Predictions for well rates were presented for Case~1, and these showed uncertainty reduction and correspondence with the true results.

Future research directions include testing our parameterization method with more rigorous history matching algorithms, such as Markov chain Monte Carlo (MCMC), to assess the accuracy of the ESMDA procedure used here. The stochastic character of Monte Carlo-based methods may, in some respects, be more compatible with the 3D-LDM latent space, so MCMC can be used to provide benchmark results. Because MCMC typically requires ${\mathcal O}(10^4-10^5)$ forward runs, a fast surrogate model for flow simulation will be needed for this assessment. \textcolor{black}{This surrogate will take the geomodel as input, and provide as output the injection and production rates. Research in this direction, using convolutional layers to extract spatial features and recurrent layers to capture temporal dynamics, is underway.}

Extending the framework to handle larger metaparameter ranges, other scenario parameters, and additional geological settings, such as 3D naturally fractured and deltaic fan systems, would represent very useful developments. Conditioning to other data types or to ``soft'' geological information is also of interest. Finally, the application (and extension as required) of the overall methodology to real cases should be performed.

\section*{Declaration of competing interest}
The authors declare that they have no known competing financial interests or personal relationships that could have appeared to influence the work reported in this paper. The second author is on the editorial board of this journal.

\section*{Acknowledgments}
\label{cknowledge}
We are grateful to the Stanford Doerr School of Sustainability and to the industrial affiliates of the Stanford Smart Fields Consortium for financial support. We also thank the SDSS Center for Computation for providing the computational resources used in this work.\\

\section*{Code availability}
The source codes and datasets used in this work are available for download at \url{https://github.com/guidodf09/ldm_3d_geomodel}.
The codes are based on the implementations provided in the \url{https://github.com/huggingface/diffusers} and \url{https://github.com/Project-MONAI/GenerativeModels} repositories. The Python libraries \texttt{diffusers} and \texttt{monai} are used in our implementation.

\newpage
\vfill\eject

\newpage
\bibliographystyle{cas-model2-names.bst}
\bibliography{bibliography.bib}

\begin{section}*{Appendix A}
\section*{3D-LDM algorithms}
\label{appendix_a}

\begin{minipage}[t]{0.85\textwidth}
\begin{algorithm}[H]
\caption{VAE training}\label{alg:vae_train}
\begin{algorithmic}[1]
\While {\text{not converged}}
\State Sample from dataset $\mathbf{m}_0 \sim q(\mathbf{m}_0)$
\State Encode and decode sample $\boldsymbol{\xi}_0 = \mathcal{E}(\mathbf{m}_0), \hat{\mathbf{m}}_0 = \mathcal{D}(\boldsymbol{\xi}_0)$
\State Compute $L_{\text{VAE}}$ (Eq.~\ref{eqn:loss_vae}), with:
\Statex \hspace{\algorithmicindent} $L_{\text{recon}} = \| \mathbf{m}_0 - \hat{\mathbf{m}}_0 \|_2^2$
\Statex \hspace{\algorithmicindent} $L_{\text{KL}} = D_{\text{KL}}\left[\mathcal{N}(\mathcal{E}_{\mu}, \mathcal{E}_{\sigma}^2) \,\|\, \mathcal{N}(\mathbf{0}, \mathbf{I}_{n_c}) \right]$
\Statex \hspace{\algorithmicindent} $L_{\text{h}} = \frac{1}{N_{\text{h}}} \| \mathbf{H}(\mathbf{m}_0 - \hat{\mathbf{m}}_0) \|_2^2$
\Statex \hspace{\algorithmicindent} $L_{\text{perc}} = \sum_{d \in \{x,y,z\}} \frac{1}{N_d^{\text{sub}}} \sum_{i=1}^{N_d^{\text{sub}}} \| \phi(\mathbf{m}_{0,i}^{(d)}) - \phi(\hat{\mathbf{m}}_{0,i}^{(d)}) \|_2^2$
\State Gradient descent step on $\nabla_{\theta} L_{\text{VAE}}(\theta)$ 
\EndWhile 
\end{algorithmic}
\end{algorithm}
\end{minipage}
\\
\begin{minipage}[t]{0.85\textwidth}
\begin{algorithm}[H]
\caption{U-net training}\label{alg:unet_train}
\begin{algorithmic}[1]
\While {\text{not converged}}
\State Sample from dataset $\mathbf{m}_0 \sim q(\mathbf{m}_0)$
\State Encode sample $\boldsymbol{\xi}_0 = \mathcal{E}(\mathbf{m}_0)$
\State Sample discrete step $t \sim U({1, \ldots, T})$ and noise $\boldsymbol{\epsilon}_t \sim \mathcal{N}(\mathbf{0}, \mathbf{I}_{n_{c}})$ 
\State Compute $\bar{\alpha}_t = \prod_{s=1}^{t} \alpha_s$
\State Perform noising step to compute $\boldsymbol{\xi}_{t}$ (Eq.~\ref{eqn:noising_step_ldm})
\State Predict $\boldsymbol{\epsilon}_{\theta}(\boldsymbol{\xi}_t,t)$ with U-net
\State Compute $L_{\text{U-net}}$ (Eq.~\ref{eqn:loss_ldm})
\State Gradient descent step on $\nabla_{\theta} L_{\text{U-net}}(\theta)$
\EndWhile
\end{algorithmic}
\end{algorithm}
\end{minipage}
\\
\begin{minipage}[t]{0.85\textwidth}
\begin{algorithm}[H]
\caption{Inference (generation)}\label{alg:ldm_generate}
\begin{algorithmic}[1]
\State Sample noise $\boldsymbol{\xi}_T \sim \mathcal{N}(\mathbf{0}, \mathbf{I}_{n_c})$
\For{${t = T, \ldots, 1}$}
    \State Compute $\bar{\alpha}_t = \prod_{s=1}^{t} \alpha_s$
    \State Predict $\boldsymbol{\epsilon}_{\theta}(\boldsymbol{\xi}_t,t)$ with U-net
    \State Perform denoising step to compute $\boldsymbol{\xi}_{t-1}$ (Eq.~\ref{eqn:sampling_ddim_ldm})
\EndFor
\State Decode denoised latent ${\mathbf{m}}_0^{\text{LDM}} = \mathcal{D}(\boldsymbol{\xi}_0)$
\State \textbf{return} generated sample ${\mathbf{m}}_0^{\text{LDM}}$
\end{algorithmic}
\end{algorithm}
\end{minipage}
\end{section}

\newpage

\begin{section}*{Appendix B}

\section*{3D-LDM model architecture}
\label{appendix_b}
Tables~\ref{tab:vae_architecture} and \ref{tab:unet_architecture} define the VAE and U-net architectures used in our 3D-LDM implementation.

\begin{table}[h]
\centering
\caption{
Variational autoencoder architecture used in this work, with input shape $(N_x, N_y, N_z, 1)$. Each ResBlock consists of GroupNorm → SiLU → Conv3D. An AttentionBlock is applied at the bottom level. Downsampling uses Conv3D; upsampling uses ConvTranspose3D. 
}
\footnotesize
\renewcommand{\arraystretch}{1.5}
\begin{tabular}{lllcc}
\hline
\textbf{Stage} & \textbf{Operation} & \textbf{Filters size, stride, padding} & \textbf{Output} & \textbf{Attn.} \\
\hline
Input & Input & -- & $(N_x, N_y, N_z, 1)$ & -- \\
\hline
\textbf{Encoder} & Conv3D & 64 filters, $3\times3\times3$, s=1, p=1 & $(N_x, N_y, N_z, 64)$ & No \\
\hline
\multirow{2}{*}{Down Block \#1} & ResBlock & 64 filters, $3\times3\times3$, s=1, p=1 & $(N_x, N_y, N_z, 64)$ & No \\
& Downsample & 64 filters, $3\times3\times3$, s=2, p=1 & $\left(\frac{N_x}{2}, \frac{N_y}{2}, \frac{N_z}{2}, 64\right)$ & \\
\hline
\multirow{2}{*}{Down Block \#2} & ResBlock & 128 filters, $3\times3\times3$, s=1, p=1 & $\left(\frac{N_x}{2}, \frac{N_y}{2}, \frac{N_z}{2}, 128\right)$ & No \\
& Downsample & 128 filters, $3\times3\times3$, s=2, p=1 & $\left(\frac{N_x}{4}, \frac{N_y}{4}, \frac{N_z}{4}, 128\right)$ & \\
\hline
\multirow{2}{*}{Down Block \#3} & ResBlock & 256 filters, $3\times3\times3$, s=1, p=1 & $\left(\frac{N_x}{4}, \frac{N_y}{4}, \frac{N_z}{4}, 256\right)$ & No \\
& Downsample & 256 filters, $3\times3\times3$, s=2, p=1 & $\left(\frac{N_x}{8}, \frac{N_y}{8}, \frac{N_z}{8}, 256\right)$ & \\
\hline
\multirow{4}{*}{Bottom Block} & ResBlock & 512 filters, $3\times3\times3$, s=1, p=1 & $\left(\frac{N_x}{8}, \frac{N_y}{8}, \frac{N_z}{8}, 512\right)$ & Yes \\
& AttentionBlock & Multi-head Self-Attn, 512 filters & $\left(\frac{N_x}{8}, \frac{N_y}{8}, \frac{N_z}{8}, 512\right)$ & Yes \\
& ResBlock & 512 filters, $3\times3\times3$, s=1, p=1 & $\left(\frac{N_x}{8}, \frac{N_y}{8}, \frac{N_z}{8}, 512\right)$ & Yes \\
& Conv3D & 1 filter, $1\times1\times1$, s=1, p=0 & $\left(\frac{N_x}{8}, \frac{N_y}{8}, \frac{N_z}{8}, 1\right)$ & No \\
\hline
\textbf{Latent variable} & Sampling $\sim \mathcal{N}(\mathcal{E}_{\mu}, \mathcal{E}_{\sigma}^2)$ & -- & $\left(\frac{N_x}{8}, \frac{N_y}{8}, \frac{N_z}{8}, 1\right)$ & -- \\
\hline
\textbf{Decoder} & Conv3D & 512 filters, $3\times3\times3$, s=1, p=1 & $\left(\frac{N_x}{8}, \frac{N_y}{8}, \frac{N_z}{8}, 512\right)$ & No \\
\hline
\multirow{4}{*}{Up Block \#1} & ResBlock & 512 filters, $3\times3\times3$, s=1, p=1 & $\left(\frac{N_x}{8}, \frac{N_y}{8}, \frac{N_z}{8}, 512\right)$ & Yes \\
& AttentionBlock & Multi-head Self-Attn, 512 filters & $\left(\frac{N_x}{8}, \frac{N_y}{8}, \frac{N_z}{8}, 512\right)$ & Yes \\
& ResBlock & 512 filters, $3\times3\times3$, s=1, p=1 & $\left(\frac{N_x}{8}, \frac{N_y}{8}, \frac{N_z}{8}, 512\right)$ & Yes \\
& Upsample & 512 filters, $3\times3\times3$, s=2, p=1 & $\left(\frac{N_x}{4}, \frac{N_y}{4}, \frac{N_z}{4}, 512\right)$ & \\
\hline
\multirow{2}{*}{Up Block \#2} & ResBlock & 256 filters, $3\times3\times3$, s=1, p=1 & $\left(\frac{N_x}{4}, \frac{N_y}{4}, \frac{N_z}{4}, 256\right)$ & No \\
& Upsample & 256 filters, $3\times3\times3$, s=2, p=1 & $\left(\frac{N_x}{2}, \frac{N_y}{2}, \frac{N_z}{2}, 256\right)$ & \\
\hline
\multirow{2}{*}{Up Block \#3} & ResBlock & 128 filters, $3\times3\times3$, s=1, p=1 & $\left(\frac{N_x}{2}, \frac{N_y}{2}, \frac{N_z}{2}, 128\right)$ & No \\
& Upsample & 128 filters, $3\times3\times3$, s=2, p=1 & $(N_x, N_y, N_z, 128)$ & \\
\hline
\multirow{2}{*}{Output Block} & ResBlock & 64 filters, $3\times3\times3$, s=1, p=1 & $(N_x, N_y, N_z, 64)$ & No \\
& Conv3D & 1 filter, $3\times3\times3$, s=1, p=1 & $(N_x, N_y, N_z, 1)$ & No \\
\hline
\end{tabular}
\label{tab:vae_architecture}
\end{table}

\begin{table}[h]
\centering
\caption{Diffusion U-net architecture used in this work, with input shape $(n_x, n_y, n_z, 1)$, i.e., the VAE latent space. The discrete step $t$ embedding is applied inside ResBlocks as: Linear → SiLU → Linear → 256-dimensional vector. Blocks have the same meaning as for the VAE.
}
\footnotesize	
\renewcommand{\arraystretch}{1.8} 
\begin{tabular}{lllcc}
\hline
\textbf{Stage} & \textbf{Operation} & \textbf{Filters size, stride, padding} & \textbf{Output} & \textbf{$\boldsymbol{t}$ emb.} \\
\hline
Input         & Input                     & --                     & $(n_x, n_y, n_z, 1)$              & No \\
Initial Conv  & Conv3D                         & 64 filters, 3×3×3, s=1, p=1 & $(n_x, n_y, n_z, 64)$             & No \\
\hline
Down Block \#1 & ResBlock                      & 64 filters, 3×3×3, s=1, p=1 & $(n_x, n_y, n_z, 64)$             & Yes \\
              & Downsample                    & 128 filters, 3×3×3, s=2, p=1 & $\left(\frac{n_x}{2}, \frac{n_y}{2}, \frac{n_z}{2}, 128\right)$ &  \\
\hline
Down Block \#2 & ResBlock + Attn.          & 128 filters, 3×3×3, s=1, p=1 & $\left(\frac{n_x}{2}, \frac{n_y}{2}, \frac{n_z}{2}, 128\right)$ & Yes \\
              & Downsample                    & 256 filters, 3×3×3, s=2, p=1 & $\left(\frac{n_x}{4}, \frac{n_y}{4}, \frac{n_z}{4}, 256\right)$ &  \\
\hline
Down Block \#3 & ResBlock + Attn.          & 256 filters, 3×3×3, s=1, p=1 & $\left(\frac{n_x}{4}, \frac{n_y}{4}, \frac{n_z}{4}, 256\right)$ & Yes \\
\hline
Mid Block     & ResBlock + Attn.           & 256 filters, 3×3×3, s=1, p=1 & $\left(\frac{n_x}{4}, \frac{n_y}{4}, \frac{n_z}{4}, 256\right)$ & Yes \\
\hline
Up Block \#1  & ResBlock + Attn.           & 256 filters, 3×3×3, s=1, p=1 & $\left(\frac{n_x}{4}, \frac{n_y}{4}, \frac{n_z}{4}, 256\right)$ & Yes \\
              & Upsample                     & 128 filters, 4×4×4, s=2, p=1  & $\left(\frac{n_x}{2}, \frac{n_y}{2}, \frac{n_z}{2}, 128\right)$ &  \\
\hline
Up Block \#2  & ResBlock + Attn.           & 128 filters, 3×3×3, s=1, p=1 & $\left(\frac{n_x}{2}, \frac{n_y}{2}, \frac{n_z}{2}, 128\right)$ & Yes \\
              & Upsample                     & 64 filters, 4×4×4, s=2, p=1  & $(n_x, n_y, n_z, 64)$             &  \\
\hline
Up Block \#3  & ResBlock                      & 64 filters, 3×3×3, s=1, p=1  & $(n_x, n_y, n_z, 64)$             & Yes \\
\hline
Output       & GroupNorm → SiLU → Conv3D       & 1 filter, 3×3×3, s=1, p=1    & $(n_x, n_y, n_z, 1)$              & No \\
\hline
\end{tabular}
\label{tab:unet_architecture}
\end{table}

\end{section}

\newpage

\end{document}